\DocumentMetadata{testphase=new-or-1}
\documentclass{natureJLH}

\usepackage{graphicx}
\usepackage{ragged2e}
\usepackage[none]{hyphenat}
\usepackage{tikz}
\usepackage[normalem]{ulem} 
\usepackage{amssymb}
\usepackage{amsmath}
\usepackage{caption}
\usepackage{booktabs,multirow}
\usepackage{multirow}
\usepackage{pdflscape}
\usepackage{xcolor}
\usepackage{hyperref}
\usepackage{url}
\usepackage{bm}
\usepackage{txfonts}
\usepackage[]{graphicx}
\usepackage{booktabs}
\usepackage{subcaption}
\usepackage{xcolor}
\usepackage{ulem}
\usepackage{color}
\usepackage{xurl}
\usepackage{listings}
\usepackage{pifont}
\usepackage{newtxtext}
\usepackage{longtable}
\usepackage{makecell}
\usepackage{marvosym}
\hypersetup{
    colorlinks=true,
    linkcolor=black,     
    citecolor=black,   
    urlcolor=black,       
    linkbordercolor=red, 
    citebordercolor=green,
    urlbordercolor=cyan,
    pdfborder={0 1 1}
}

\newcommand{\araa}{Annu. Rev. Astron. Astrophys.}   
\newcommand{\aj}{Astron. J.}   
\newcommand{\apj}{Astrophys. J.}   
\newcommand{\apjs}{Astrophys. J. Suppl. Ser.}   
\newcommand{\apss}{Astrophys. Space Sci.}   
\newcommand{\aap}{Astron. Astrophys.}   
\newcommand{\aapr}{Astron. Astrophys. Rev.}   
\newcommand{\aaps}{Astron. Astrophys. Suppl.}   
\newcommand{\mnras}{Mon. Not. R. Astron. Soc.}   
\newcommand{\nar}{New Astron. Rev.}   
\newcommand{\pasp}{Publ. Astron. Soc. Pac.}   
\newcommand{\raa}{Res. Astron. Astrophys.} 

\title{A Be star-black hole binary with a wide orbit from LAMOST time-domain survey}
\author{Qian-Yu An$^{\dagger,1}$, Yang Huang$^{\dagger,2,3}$, Wei-Min Gu$^{\textsuperscript{\Letter},1}$, Yong Shao$^{4,5}$, Zhi-Xiang Zhang$^6$, Tuan Yi$^{7}$, B. D. Lailey$^{8}$, T. A. A. Sigut$^{8,9}$, Kyle Akira Rocha$^{10,11}$, Meng Sun$^{2,10}$, Seth Gossage$^{10}$, Shi-Jie Gao$^{4,5}$, Shan-Shan Weng$^{12}$, Song Wang$^{2,3,13,14}$, Bowen Zhang$^{3,13}$, Xinlin Zhao$^{3,13}$, Senyu Qi$^{15}$, Shilong Liao$^{3,16}$, Jianghui Ji$^{17}$, Junfeng Wang$^1$, Jianfeng Wu$^1$, Mouyuan Sun$^1$, Xiang-Dong Li$^{4,5}$, Jifeng Liu$^{\textsuperscript{\Letter},2,3,13,14}$. }

\begin{document}
\maketitle

\begin{affiliations}

\item Department of Astronomy, Xiamen University, Xiamen, Fujian 361005, People's Republic of China
\item National Astronomical Observatories, Chinese Academy of Sciences, Beijing 100012, People’s Republic of China
\item School of Astronomy and Space Science, University of Chinese Academy of Sciences, Beijing 100049,  People's Republic of China
\item Department of Astronomy, Nanjing University, Nanjing 210023, People's Republic of China
\item Key Laboratory of Modern Astronomy and Astrophysics, Nanjing University, Ministry of Education, Nanjing 210023, People's Republic of China
\item College of Physics and Information Engineering, Quanzhou Normal University, Quanzhou 362000, People's Republic of China
\item Department of Astronomy, School of Physics, Peking University, Beijing 100871, People's Republic of China
\item Department of Physics and Astronomy, The University of Western Ontario, 1151 Richmond Street, London N6A 3K7, Canada
\item Institute for Earth and Space Exploration (IESX), The University of Western Ontario, 7134 Perth Drive, London N6A 5B7, Canada
\item Center for Interdisciplinary Exploration and Research in Astrophysics (CIERA), Northwestern University, 1800 Sherman Ave, Evanston, IL 60201, USA
\item Department of Physics \& Astronomy, Northwestern University, 2145 Sheridan Road, Evanston, IL 60208, USA
\item School of Physics and Technology, Nanjing Normal University, Nanjing 210023, People's Republic of China
\item Key Lab of Optical Astronomy, National Astronomical Observatories, Chinese Academy of Sciences, Beijing 100101, People's Republic of China
\item Institute for Frontiers in Astronomy and Astrophysics, Beijing Normal University, Beijing, 102206, People’s Republic of China
\item Physics Department, Tsinghua University, Beijing 100084, People’s Republic of China
\item Shanghai Astronomical Observatory, Chinese Academy of Sciences, Shanghai 200030, People's Republic of China
\item Key Laboratory of Planetary Sciences, Purple Mountain Observatory, Chinese Academy of Sciences, Nanjing 210008, People's Republic of China\\
$\dagger$ These authors contributed equally;\\
$\textsuperscript{\Letter}$ Corresponding authors: guwm@xmu.edu.cn; jfliu@nao.cas.cn.

\end{affiliations}

\begin{abstract}
Binary systems consisting of an early type star and a black hole (BH) are crucial for understanding various astrophysical phenomena, particularly the origins of detected gravitational wave sources\cite{2020ApJ...905L..15B,2022PhR...955....1M}. 
Be binary systems are expected to represent a key evolutionary stage in hosting BHs. However, while hundreds of Be X-ray binaries are known\cite{2023A&A...671A.149F}, the only confirmed BH candidate in a Be binary remains highly controversial\cite{2014Natur.505..378C,2022arXiv220812315R,2023A&A...677L...9J}. We report the discovery of ALS\,8814, a Be star-BH binary with a moderately eccentric ($e = 0.23$) and wide orbit ($P = 176.6$ days), revealed by the radial velocity (RV) measurement of the visible Be star. Our analysis, combining flux-calibrated spectra in the Balmer discontinuity region and spectral template matching, yields a mass of $11.2^{+1.4}_{-1.2}$ $M_\odot$ for the Be star.
The minimum mass of the unseen companion, assuming an edge-on inclination ($i = 90^{\circ}$), is $9.8\pm 0.7\,M_\odot$. We rule out the presence of non-degenerate companions in ALS\,8814, indicating that it can only be a BH. This discovery represents a robust case of a Be-BH binary, identified purely through precise RV measurements from a single set of lines. The extremely low peculiar velocity of ALS\,8814 suggests that the BH is formed via a direct core-collapse with a negligible natal kick, implying an almost perfect alignment between the Be star's spin and the orbital plane. In this context, the binary's inclination angle is estimated to be 22$^{\circ}$–49$^{\circ}$ by analyzing the shallow double-peaked profile of the H$\alpha$ emission line\cite{2024MNRAS.527.2585L}. This inclination range corresponds to a BH mass estimate between $15\,M_\odot$ and $58\,M_\odot$. As the only unambiguous Be-BH binary system known to date, ALS\,8814 provides valuable constraints on the BH formation in a binary system with a high-mass companion.
\end{abstract}

\par To discover early type star -- BH binary systems, the RV measurements were systematically analyzed for approximately 900 O/B-type stars with at least five observations targeted by the Large Aperture Multi-Object Spectroscopic Telescope\cite{2012RAA....12.1197C} (LAMOST) time-domain survey\cite{2021RAA....21..292W}, a major component of the LAMOST Phase II spectroscopic surveys conducted from December 2017 to June 2022.
One B-type star (hereafter ALS\,8814) exhibits clear periodic RV variation from red observations, including six exposures from the low-resolution survey (LRS) and 30 exposures from the medium-resolution survey (MRS), respectively. 
Four follow-up observations conducted by High-Resolution Spectrograph mounted on Lijiang 2.4-m Telescope\cite{2015RAA....15..918F,2019RAA....19..149W,2023RAA....23k5008Y} (2.4-m Telescope/HiRES), and Double Spectrograph mounted on Palomar’s 200-inch telescope (P200/ DBSP) between December 2022 and September 2023 confirmed the orbital period. The strong double-peaked H$\alpha$ emission profile, moving in the same direction as absorption lines from the stellar photosphere, along with the light curve from All-Sky Automated Survey for Supernovae\cite{2014ApJ...788...48S,2017PASP..129j4502K} (ASAS-SN; see Methods), suggests ALS\,8814 is a typical Be-type star\cite{1981BeSN....4....9J}. 
It is a very bright star ($V = 10.2$\,mag) towards Galactic Anti-Centre with (Galactic longitude $l$, Galactic latitude $b$) = (188.46557$^{\circ}$, 2.64901$^{\circ}$). 

\par
\begin{figure*}
\begin{center}
\includegraphics[width=1\textwidth]{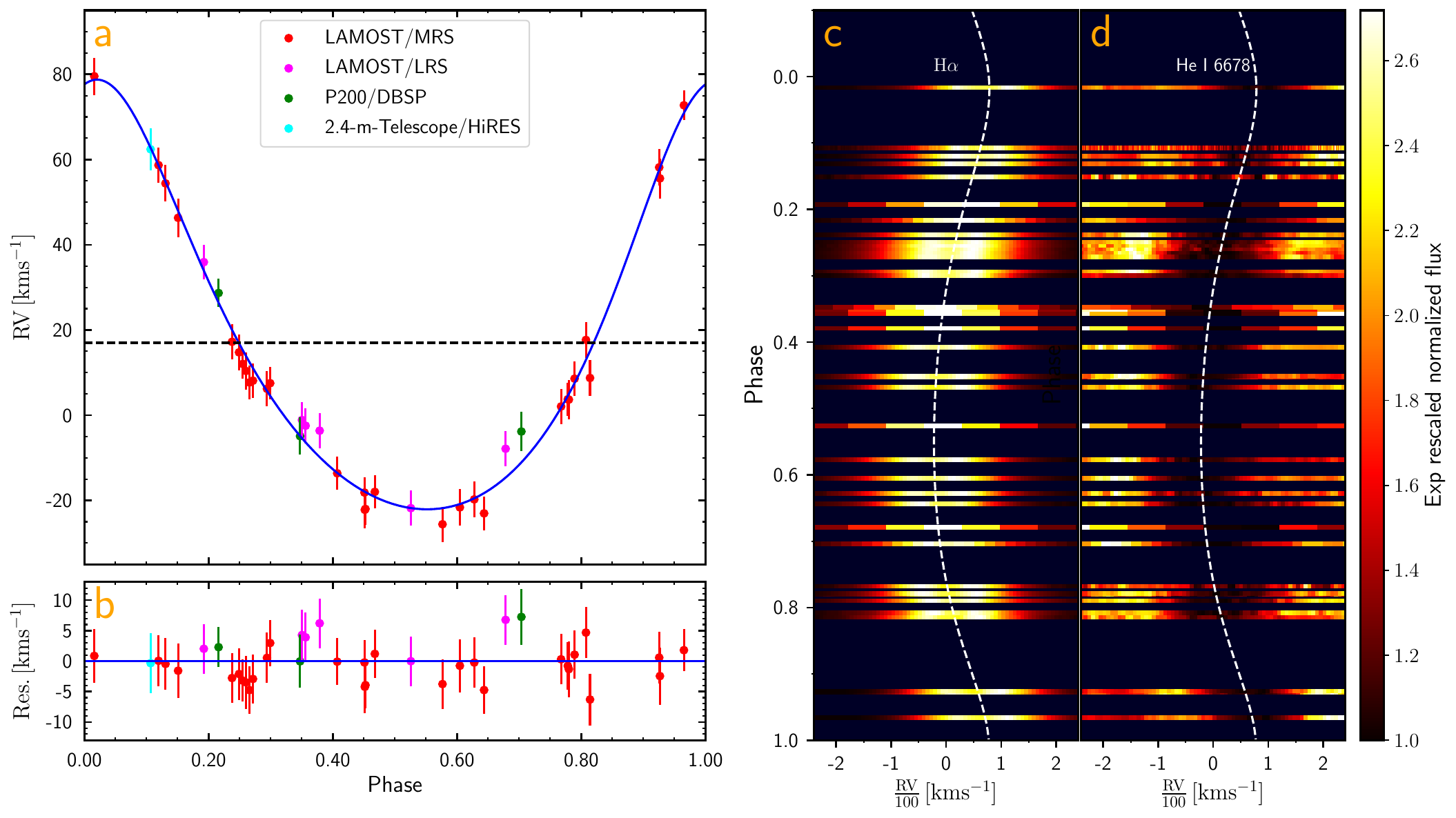}
\caption{{\bf RV of the Be star in ALS\,8814.} {\bf a,} Phase-folded RV curve of the Be star. Data points are folded using $T_0$ = 2,458,042.291 and red$P = 176.55$\,days. The RVs are measured from spectra obtained by LAMOST/MRS (red), LAMOST/LRS (magenta), P200/DBSP (green) and 2.4-m Telescope/HiRES (light green). 
The blue line represents the best-fit binary orbit model with parameters $K_{\rm Be, 0} = 50.4\pm1.3$\,km\,s$^{-1}$, $e = 0.23 \pm 0.02$ and $V_{\rm Be, 0} = 17.0 \pm 0.7$\,km\,s$^{-1}$. The black dashed line indicates the centre-of-mass velocity $V_{\rm Be, 0} = 17.0$\,km\,s$^{-1}$.
{\bf b,} Residuals for the orbital fit to the Be star. {\bf c,} The dynamical spectra of H$\alpha$ emission line, phased with the orbital period of $176.55$ days. Different colours represent normalized fluxes at distinct wavelengths for each spectrum. To enhance colour contrast, we use e-exponential rescaled normalized fluxes rather than normalized fluxes themselves. The white dashed curve is RV curve of Be star taken from the best-fit model, i.e. the blue line in panel {\bf a}). {\bf d,} Same with panel {\bf c}, but for He I 6678 line.}
\label{fig:RV_curve}
\end{center}
\end{figure*}

 A Lomb-Scargle analysis\cite{1976Ap&SS..39..447L,1982ApJ...263..835S} of ALS\,8814's RV, as measured from the wing of H$\alpha$ emission line in 30 LAMOST/MRS, six LAMOST/LRS, three P200/DBSP and one 2.4-m Telescope/HiRES observations collected between 2017 and 2023, reveals a significant period of 177 days.
The motions of absorption lines (e.g., He I 6678 shown in Supplementary Fig.~\ref{fig:spectra}(b)) detected in these spectra are all consistent with that of wing of H$\alpha$ emission. Performing binary orbit fitting on the RV curve yields an orbital period $P = 176.6\pm0.1$\,days, a semi-amplitude $K_{\rm Be} = 50.4\pm1.3$\,km~s$^{-1}$, an orbital eccentricity $e = 0.23\pm0.02$ and a centre-of-mass velocity $V_{0} = 17.0\pm0.7$\,km~s$^{-1}$. Figure~\ref{fig:RV_curve} shows the phase-folded RV curve. The mass function of the unseen companion is
\begin{equation}
f(M_2) = \frac{M_2^3 \sin^3 i}{(M_1 + M_2)^2} = \frac{K^{3} P (1-e^2)^\frac{3}{2}}{2 \pi G} = 2.16\pm0.17~M_\odot,
\end{equation}
where $M_{1}$ and $M_{2}$ are the masses of the Be star and the unseen companion, respectively, $i$ is the orbital inclination, and $G$ is the gravitational constant.

\par
\begin{figure*}
\begin{center}
\includegraphics[width=0.87\textwidth]{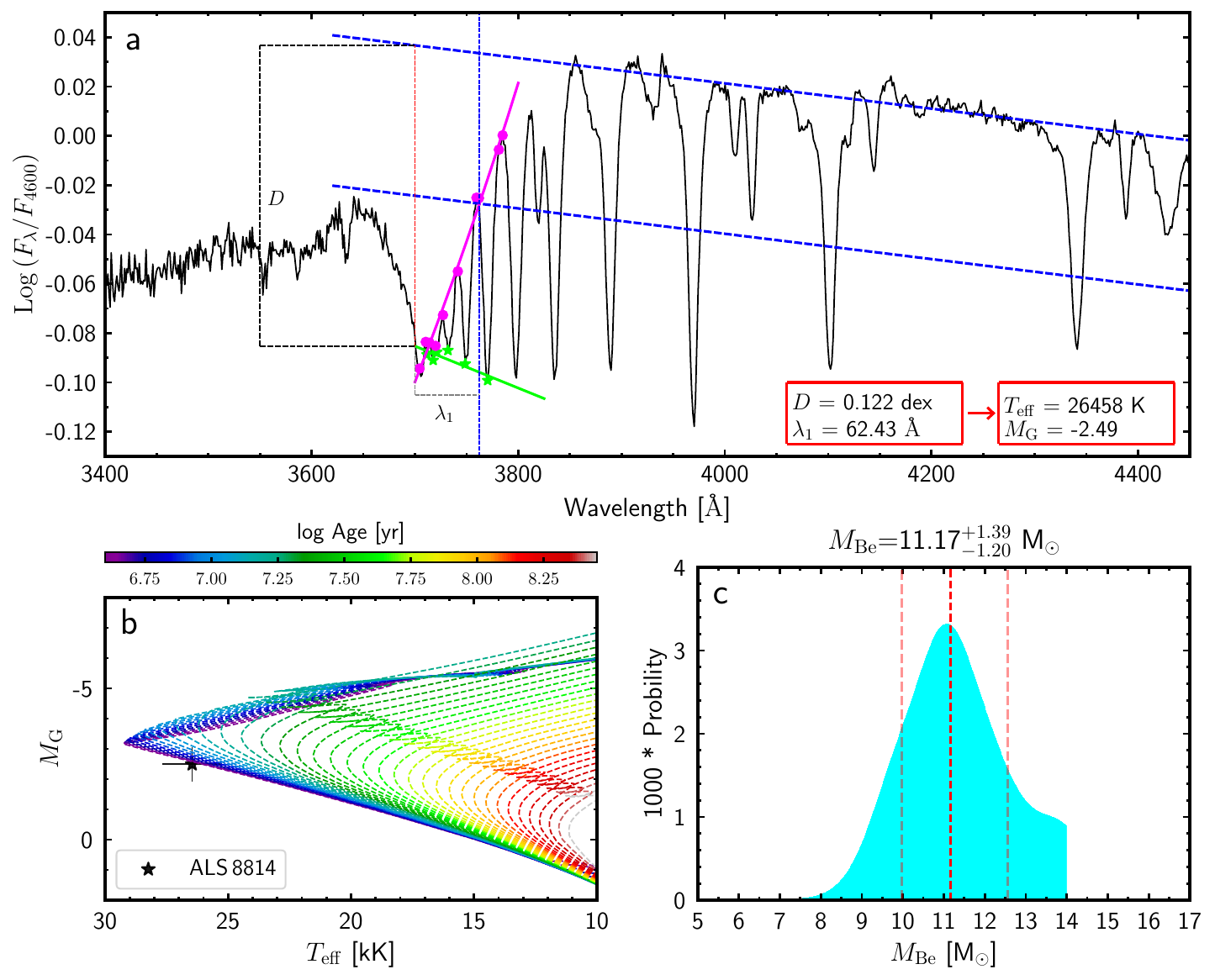}
\caption{{\bf Stellar parameters of Be star in ALS\,8814.} {\bf a,} Flux-calibrated spectrum (normalized by the flux at 4600$\AA$) at Balmer discontinuity (BD) region of ALS\,8814. The upper blue dashed line represents the linear fit of the Paschen continuum near the BD region.
The green line connects the bottom of the
higher Balmer lines.
$D$ denotes the Balmer jump depth (in logarithmic space) measured from flux difference extrapolated from the two lines.
The lower blue line is obtained by shifting the upper one by $\frac{1}{2}D$.
This line intersects with the pseudo-continuum (determined by the magenta dots and curve) and the crossing-point minus $3700~\AA$\, gives the position of the Balmer jump $\lambda_1$.
$D$ and $\lambda_1$ are, respectively, indicators of effective temperature $T_{\rm eff}$ ans luminosity and their comparisons to an empirical grid can measure $T_{\rm eff}$ and $G$-band absolute magnitude of the Be star in ALS\,8814 (Methods).
 {\bf b,} The position of Be star on the effective temperature versus the absolute $G$ magnitude diagram. Overplotted lines colour coded by stellar ages are isochrones from \texttt{Parsec} stellar evolution model\cite{2022A&A...665A.126N} with solar metallicity and $\Omega/\Omega_{\rm cr} = 0.6$ (Methods).
{\bf c,} The probability distribution function of the mass of the Be star in ALS\,8814 derived by comparing observed parameters to those from  \texttt{Parsec} isochrones in a Bayesian framework (Methods).}
\label{fig:BCD1}
\end{center}
\end{figure*}

\begin{figure*}
\begin{center}
\includegraphics[width=1\textwidth]{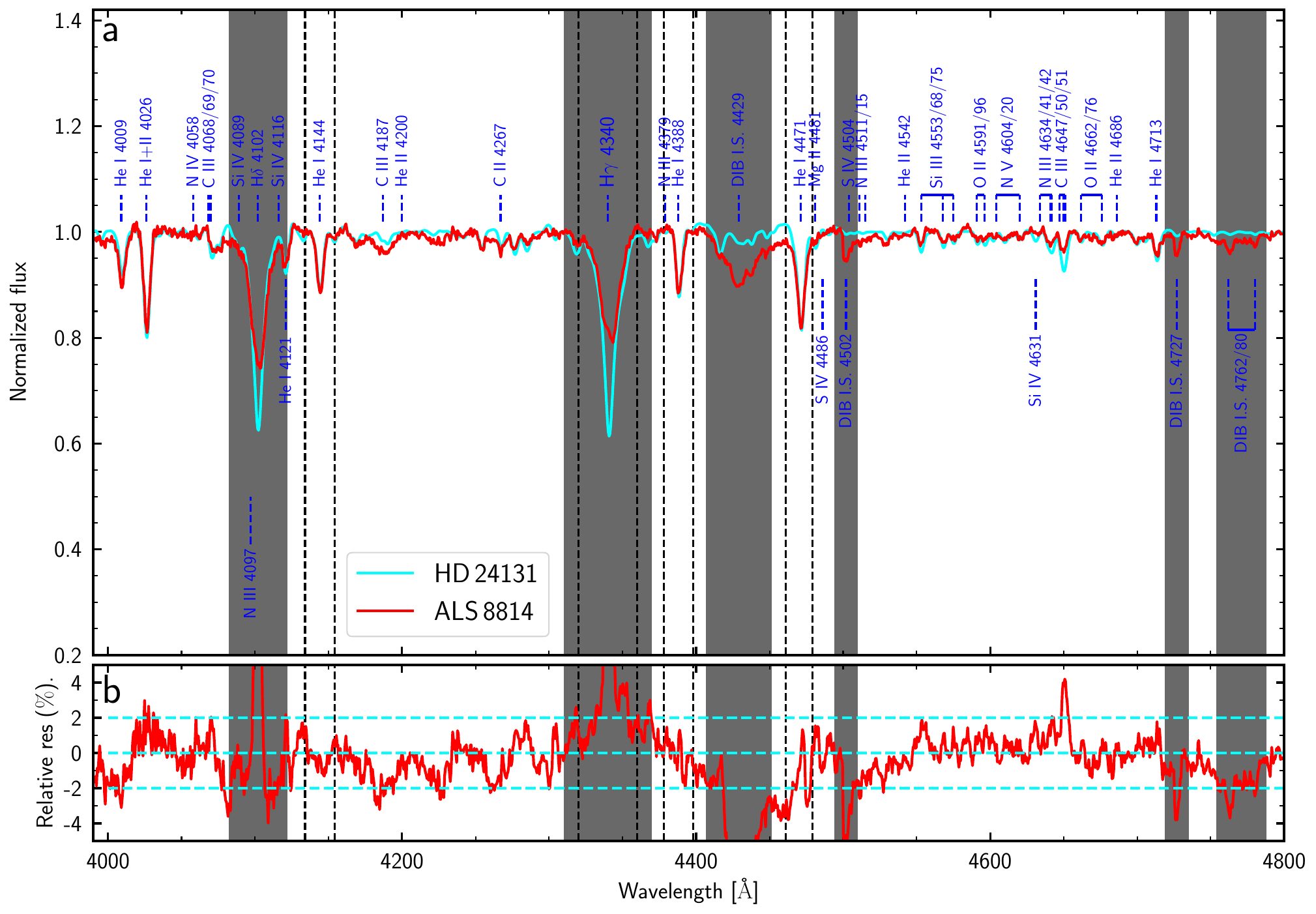}
\caption{{\bf Template-matching of the Be star in ALS\,8814.} {\bf a,} P200/DBSP spectrum of the
wavelength range of 3,990 $\AA$ to 4,800 $\AA$ (red) with the best observed spectrum (cyan). The Balmer lines (affected by the decretion disk) and diffuse interstellar bands, masked by these gray shadows, are excluded from chi-square calculations. {\bf b,} the relative fitting residuals from spectral template matching.}
\label{fig:spectra_match}
\end{center}
\end{figure*}

 
We apply the BCD (Barbier–Chalonge–Divan) spectrophotometric classification system\cite{1941AnAp....4...30B,1952AnAp...15..201C,2009A&A...501..297Z,2018A&A...609A.108S} to derive stellar parameters for ALS\,8814 based on measurements of the Balmer discontinuity region.
By using the BCD parameters (the mean spectral position of the Balmer jump $\lambda_1 = 62$\,$\mathrm{\AA}$\, and the size of the Balmer jump $D = 0.12$\,dex) measured from the flux-calibrated low-resolution spectrum observed by Xinglong 2.16-m Telescope\cite{2016PASP..128k5005F} (Figure~\ref{fig:BCD1}(a)), ALS\,8814 is found to be a B1V star with an effective temperature $T_{\rm eff} = 26,500$\,K and a $G$-band absolute magnitude $M_{G} = -2.49$ (see Methods).
Using high-quality, moderate-resolution spectra obtained with P200/DBSP for ALS\,8814 and a newly constructed spectral classification grid for of B-type stars\cite{2024A&A...690A.176N}, the standard template-matching technique identifies ALS\,8814 as a B1V type star (Figure~\ref{fig:spectra_match}; see Methods), which is in excellent agreement with the classification from the BCD method. Fitting the values of $T_{\mathrm{eff}}$ and $M_{\mathrm{G}}$ corresponding to a B1V star to the \texttt{PARSEC}\cite{2022A&A...665A.126N} stellar isochrones (assuming solar metallicity and $\omega_{\rm i} = 0.60$) leads to a mass of $M_{\mathrm{Be}} = 11.2^{+1.4}_{-1.2}\,M_\odot$, a radius of $R_{\mathrm{Be}} = 4.9^{+1.6}_{-1.0}\,R_\odot$  and an age of $\tau_{\mathrm{Be}} = 7.9^{+5.2}_{-2.9}$ Myr for the Be star (Figure~\ref{fig:BCD1}(b-c); see Methods).
The distance and reddening along the line-of-sight of ALS\,8814 are further found to be $1.1^{+0.3}_{-0.2}$\,kpc and $A_{V} = 2.68 \pm 0.04$\,mag, where $V$ is $V$-band magnitude, based on the broadband spectral energy distribution (SED) fitting (see Methods). 

\par Based on the measured Be star mass and the mass function of the unseen companion, we derive a minimum companion mass of $9.8 \pm 0.7 ~M_\odot$ (assuming $i = 90^\circ$). If the companion were a non-degenerate object $-$ even in the least luminous scenario, such as an unevolved main-sequence star $-$ the luminosity contribution of the companion would exceed $\sim$50\% of that of the primary Be star. Such a luminous companion should leave clear imprints in the optical spectra. However, spectral disentangling analysis (Methods) reveals no detectable signals from the companion, ruling out the possibility of single main-sequence star companion. Furthermore, an equal-mass inner binary scenario is also excluded: such a configuration would produce significant spectral signatures at the RV extrema (phases of the maximum/minimum RV) and should also be detectable in the SED. Yet, both the optical spectra of ALS\,8814 and its SED (Supplementary Fig.~\ref{fig:sed_plot}) are fully consistent with a single-star model. Finally, we evaluate the possibility of ALS\,8814 being a stripped star formed via post-mass transfer, as proposed for systems like LB-1\cite{2020A&A...639L...6S} and HR\,6819\cite{2022A&A...659L...3F}. However, ALS\,8814 exhibits classical Be star characteristics $-$ coherent motion of emission and absorption lines $-$ which strongly disfavors post-interaction evolution (Methods). Collectively, these constraints leave the BH as the most plausible explanation for the unseen companion.

\par The BH mass can be estimated if the binary inclination angle can be estimated. This angle is equivalent to that of the Be star rotation axis (denoted as $i_{\rm Be}$) if assuming spin-orbit alignment, which is expected and indeed supported by the observational evidence\cite{1994AJ....107..306H}. Using the machine learning method conducted by the algorithm of neural networks tasked with regression\cite{2024MNRAS.527.2585L}, we analyze the H$\alpha$ emission line to estimate the orbital inclination to be $36.0^{+8.5}_{-9.7}$ degrees, corresponding to a black hole mass of $23.5^{+16.8}_{-6.4} \,M_\odot$ (see Methods). For validation, we also model the absorption line profiles of He I 6678, He I 5048 (carefully selected from LAMOST/MRS to minimize contamination from emission lines) and one blue arm of the P200 spectrum which has relatively high resolving power and high signal-to-noise ratio (SNR) using the {\tt TLUSTY}\cite{1995ApJ...439..875H,2007ApJS..169...83L} model spectrum. The projected rotational velocity of the Be star, $V\sin i_{\rm Be}$, is measured to be $248.3^{+2.1}_{-2.2}$\,km~s$^{-1}$ (see Methods).
As fast rotators, the actual velocities $V$ of the Be stars were found to be as high as 0.7-0.8 times of the critical rotation speeds $V_{\rm c}$, which can be properly predicted by \texttt{PHOEBE}\cite{2020ApJS..250...34C} with a given stellar mass and equivalent radius. The inclination angle is then estimated to be $i_{\rm Be} = 36.2^{+13.5}_{-7.7}$ degrees by adopting a distribution of $V/V_{\rm c} = \mathcal{N}(0.73, 0.17)$ from a recent measurement\cite{2016A&A...595A.132Z} for Be stars, which agrees with that derived by the machine learning method. Finally, we consider the potential spin-orbit misalignment angle, estimated to be approximately 10 degrees from our binary population synthesis simulations (see Methods and Supplementary Fig.~\ref{fig:kick}). This misalignment is treated as a systematic uncertainty in the orbital inclination measured via the machine learning method. Accounting for this uncertainty, the inclination is estimated to range from 22 to 49 degrees, corresponding to a BH mass between $15\,M_{\odot}$ and $58\,M_{\odot}$. Given this mass range, the unseen companion is significantly more massive than the visible star, strongly indicating that it is a BH, with few viable alternative explanations.

\par ALS 8814 stands out as the most reliably identified Be-BH system, established through precise determinations of both the Be star's mass and the binary mass function, along with a reasonable estimate of the orbital inclination angle.
The only known Be-BH candidate MWC 656\cite{2014Natur.505..378C} is now being questioned due to the mass determination of the unseen companion, which relies on the mass ratio derived from RVs of both the Be star and the accretion disc powered by the unseen companion. 
However, the motions of the former are controversial, thus making an accurate measurement of its RVs challenging.
The most recent high-resolution spectroscopic observations of MWC\,656 provided another solution for mass ratio, preferring a neutron star/white dwarf/helium star with mass of 1-2\,$M_{\odot}$\cite{2022arXiv220812315R,2023A&A...677L...9J} rather than a BH with mass of 4-7\,$M_{\odot}$ as claimed in the previous study\cite{2014Natur.505..378C}. In the case of ALS\,8814, the RVs are derived solely from one set of structurally stable lines of the Be star (see Methods).
Therefore, our discovery of ALS\,8814 undoubtedly provides the most compelling case for the existence of a Be-BH system, as long predicted by binary evolution models\cite{1999A&A...349..505R,2004ApJ...603..663Z,2014ApJ...796...37S}.
ALS\,8814 is an X-ray quiescent source with an upper limit of X-ray luminosity of 3.3\,$\times 10^{31}$ erg\,s$^{-1}$, as revealed by the 680-second observation from  {\it Swift\cite{2004ApJ...611.1005G}} X-ray Telescope (\emph{Swift}/XRT) and data from the eROSITA-DE Data Release 1\cite{2022A&A...661A...1B,2024A&A...682A..34M}.
The faint X-ray nature is further supported by the lack of detection of motion from the accretion disc in all collected optical spectra.
The upper limit of X-ray luminosity for ALS\,8814 is one to two orders of magnitude fainter than that of Be X-ray binaries with similar orbital properties (see Supplementary Fig.~\ref{fig:X-ray}), which is naturally expected since a BH event horizon exists (see Methods).

\par
The discovery of a red{$>$15 $M_\odot$} BH at a young binary system, largely with (super-)solar metallicity, is not preferred by the stellar evolution models\cite{2010ApJ...714.1217B}. The same case is found for the massive BH ($\sim 21~M_{\odot}$) within high-mass X-ray binary (HMXB) Cygnus X$-$1\cite{2021Sci...371.1046M}.
The commonly used mass-loss rates of Wolf–Rayet stars\cite{2010ApJ...714.1217B} (hereafter $f_{\rm WR}$) are required to be substantially reduced to form such massive stellar BHs at solar metallicity.
However, our binary population synthesis simulations\cite{2020ApJ...898..143S} (Methods) show that the maximum mass of the BH formed in Be-BH systems is only $\sim 18~M_{\odot}$ (see Supplementary Fig.~\ref{fig:model}(a)), even when adopting a reduced mass loss-rate of $0.3f_{\rm WR}$, which is sufficient for producing a BH in Cygnus X$-$1\cite{2021Sci...371.1046M}.
In the simulations, the BH mass can only reach as high as the medium estimate of that of ALS\,8814 by at least an order of magnitude reduction on the mass loss rate, i.e. $0.1f_{\rm WR}$ (see Supplementary Fig.~\ref{fig:model}(b)). Besides adopting a similar mass-loss rate of $\sim$0.1$-$0.2$f_{\rm WR}$ from Ref.\cite{2000A&A...360..227N}, the \texttt{POSYDON} framework\cite{2023ApJS..264...45F,2024arXiv241102376A}, a \texttt{MESA}-based\cite{2011ApJS..192....3P} binary population synthesis code, includes more significant considerations such as self-consistent mass transfer rate calculations, solving for stellar structure, and accounting for rotation. Consequently, the \texttt{POSYDON} code also predicts that the BH mass at solar metallicity can exceed $20~M_{\odot}$\cite{2023NatAs...7.1090B,2024ApJ...971..133R} (see Supplementary Fig.~\ref{fig:model}(c)).
The high-mass BH of ALS\,8814 thus provides a stringent constraint on the stellar-wind mass-loss prescriptions that largely determine the maximum BH mass at solar metallicity.

\par
ALS\,8814 falls on the low-eccentricity sequence on the period-eccentricity diagram defined by the sample of Be HMXBs\cite{2023A&A...671A.149F} (see Supplementary Fig.~\ref{fig:periodecc}).
The high-eccentricity sequence Be HMXBs are believed to receive substantial natal kicks of $\sim 300$\,km\,s$^{-1}$, while the low eccentricity sequence ones may experience much smaller/negligible kicks from supernova (SN) explosions\cite{2002ApJ...574..364P}.
This indicates that a small or negligible impulse was posed to ALS\,8814 system when its BH was formed.
Another strong independent evidence is its extremely low 3D peculiar velocities $v_{\rm pec}^{z =0} = 8.0 \pm 6.8$\,km\,s$^{-1}$ (at Galactic plane crossing; see Methods) relative to local Galactic rotation.
This finding suggests ALS\,8814 BH was formed through fallback SN or direct core-collapse.
This has been predicted by stellar evolution theories for the formation of massive BHs, and ALS\,8814 can thus serve as clear evidence for this process.

\par
ALS\,8814 is a BH binary found by the LAMOST time-domain survey, one of the leading large-scale spectroscopic surveys to catch non-accreting compact binary systems\cite{2021RAA....21..292W} relied on long-term RV monitoring.
The major population of dormant BHs under the iceberg is disclosing by this survey, together with the similar campaigns like the Sloan Digital Sky Survey\cite{2019BAAS...51g.274K}/The Apache Point Observatory Galactic Evolution Experiment\cite{2017AJ....154...94M} and the high-precision astrometric one from the European Space Agency (ESA) \emph{Gaia} satellite\cite{2023A&A...674A...1G}.
The upcoming samples thus allow one to explore the upper limit of BH mass under different parameters such as metallicity and type of visible companion, the physics of core-collapse and even the orbit evolution.

\bibliography{auriga.bib}
\clearpage
\clearpage
\clearpage
\clearpage

\begin{longtable}{lll}
    \caption{\textbf{Summary of parameters for ALS\,8814.}} \\
    \hline\hline
    \multicolumn{3}{l}{\bf{Astrometric data of the Be star}}   \\ 
    Right ascension & $\alpha$\_J2000\,[deg] & 93.64398 \\
    Declination & $\delta$\_J2000\,[deg] & 22.90951 \\
    Apparent magnitude & $V$\,[mag] &  10.23 \\
    Parallax & $\varpi$\,[mas] & $ 0.44\pm0.29 $ \\
    Proper motion in RA & $\mu_{\alpha}$\,[$\rm mas\,yr^{-1}$] & $ -1.03\pm0.33 $ \\
    Proper motion in Dec & $\mu_{\delta}$\,[$\rm mas\,yr^{-1}$] & $ -1.52\pm0.26 $ \\
    RUWE &  & $ 18.58 $ \\
    Astrometric\_excess\_noise & [mas] & $ 2.32 $ \\
    Astrometric\_excess\_noise\_sig &  & $ 6,271.03 $ \\
    Position toward the Galactic centre & $X\left[{\rm kpc}\right]$&$-9.28^{+0.27}_{-0.29}$\\ 
    Position in the direction of Galactic rotation & $Y\left[{\rm kpc}\right]$&$-0.16^{+0.04}_{-0.04}$\\ 
    Position toward the North Galactic Pole & $Z\left[{\rm kpc}\right]$&$0.08^{+0.01}_{-0.01}$\\ 
    Velocity toward the Galactic centre & $U\,\left[{\rm km\,s^{-1}}\right]$ & $-9.51^{+0.66}_{-0.62}$ \\ 
    Velocity in the direction of Galactic rotation & $V\,\left[{\rm km\,s^{-1}}\right]$ & $3.19^{+1.72}_{-2.14}$ \\ 
    Velocity toward the North Galactic Pole & $W\,\left[{\rm km\,s^{-1}}\right]$ & $-2.88^{+2.50}_{-3.20}$\\ 
    \hline
    \multicolumn{3}{l}{\bf{Parameters of the Be star}}  \\ 
    Effective temperature & $T_{\rm eff}$\,[K] & $26,458\pm1,204 $ \\
    Absolute $G$-band magnitude & $M_G$\,[mag] & $-2.49\pm0.56 $ \\
    Surface gravity  & $\log g ~[\rm cm\,s^{-2}]$  & $4.09^{+0.09}_{-0.16}$  \\
    Projected rotation velocity & $V\sin i\,\left[{\rm km\,s^{-1}}\right]$ &  $248.3^{+2.1}_{-2.2}$ \\    
    Bolometric luminosity & $L\,[L_{\odot}]$ & $10,544^{+7,445}_{-3,689}$ \\ 
    Age & $\tau_{\mathrm{Be}}\,[\rm{Myr}]$ & $7.9^{+5.2}_{-2.9}$ \\
    Mass &  $M_1\,[M_{\odot}]$ & $11.17^{+1.39}_{-1.20}$ \\
    Distance &  $D\,[\rm{pc}]$ & $1,115.32^{+341.03}_{-240.26}$ \\
    Extinction & $A_{V}$\,[mag] & $ 2.68 \pm 0.04$ \\
    \hline
    \multicolumn{3}{l}{\bf{Orbital parameters}}  \\ 
    \hline
    \\ \\
    \hline
    Semi-amplitude & $K\,[\rm{km~s^{-1}}]$ & $50.41^{+1.34}_{-1.28}$ \\
    Period  & $P$\,[day]  & $176.55\pm0.11$  \\
    Eccentricity & $e$ &  $0.23\pm0.02$ \\
    Systemic velocity & $V_0\,[\rm{km~s^{-1}}]$  & $16.99^{+0.70}_{-0.74}$  \\
    Mass function & $f(M_2)\,[M_\odot]$  & $2.16\pm0.17$  \\
    Inclination by $V\sin i$ measurement & $i\,[\rm{deg}]$ & $36.2^{+13.5}_{-7.7}$ \\ 
    Inclination by machine learning method & $i\,[\rm{deg}]$ & $36.0^{+8.5}_{-9.7}$ \\  
    Time of periastron & $T_0\,[\rm{HJD-2,458,042}]$ & $0.291^{+2.565}_{-2.789}$ \\ 
    Argument of periastron & $\omega\,[\rm{Rad}]$ & $-0.17\pm0.08$ \\ 
    Mass of the black hole by assuming spin-orbit alignment & $M_{2,~36.0^{+8.5}_{-9.7}\,^{\circ}}\,[M_\odot]$ & $23.5^{+16.8}_{-6.4}$ \\
    Mass of the black hole (assuming $i=49.1^{\circ}$) & $M_{2,~49.1^{\circ}}\,[M_\odot]$ & $15.1\pm1.0$ \\ 
    Mass of the black hole (assuming
    $i=22.1^{\circ}$) & $M_{2,~22.1^{\circ}}\,[M_\odot]$ & $57.8\pm3.9$ \\
    \hline
    \label{tab:summary}
\end{longtable}

\clearpage
\begin{methods}
\subsection{Spectroscopy observations and data reduction.}
ALS\,8814 is one of the targets in the LAMOST time-domain survey, a key project of the LAMOST Phase II survey\cite{Zong2020}. This project aims to obtain time series spectra using both the LAMOST low-resolution spectrograph (with a resolving power of $R \sim $1,800 over the wavelength range 3,700 to 9,000 $\mathrm{\AA}$) and the LAMOST medium-resolution spectrograph (with a resolving power of $R \sim$\,7,500 over the wavelength range 4,950 to 5,350 $\mathrm{\AA}$ in the blue arm and 6,300 to 6,800 $\mathrm{\AA}$ in the red arm) across nearly 100 plates, each covering 20 square degrees.
About 10\% of the time-domain plates, selected from the Kepler/K2 Campaigns, have top priority to be observed over 10 times, including the plate containing ALS\,8814. This particular plate was observed on 37 different nights (nine nights with low-resolution and 30 with medium-resolution) from November 5, 2017 to March 7, 2023. Notably, three low-resolution spectra are excluded from the final analysis due to issues with wavelength calibrations or bad pixels.
Details of the observations are listed in Extended Data Table~\ref{tab:single-epoch RVs}. All spectra were reduced using the LAMOST 2D pipeline\cite{Bai2017,Han2023}.

\par
One main goal of this time-domain survey is to discover early-type binary systems that potentially host compact objects (whatever neutron stars or BHs) via monitoring the variations of RV measured from time-series spectra.
In total, 665 O/B-type stars were targeted with at least 5 times during the entire LAMOST Phase II survey.
Amongst them, ALS\,8814 exhibits significant periodic RV variations with $P \sim 177$\,days.

\par 
To confirm the periodic variations, we conduct four follow-up optical spectroscopic observations of ALS\,8814 at different phases, using the P200/DBSP and the 2.4-m Telescope/HiRES.
The observation details are once again presented in Extended Data Table~\ref{tab:single-epoch RVs}.
The spectra are processed using standard spectroscopic procedures with \texttt{IRAF} \texttt{v2.17} (Image Reduction and Analysis Facility), including bias and flat field corrections, 1D spectrum extraction, and wavelength calibration. All the available spectra of ALS\,8814 are shown in Supplementary Fig.~\ref{fig:spectra}. The RVs measured from these new observations confirm the periodic signal detected in the LAMOST observations.

\subsection{Stellar properties from flux-calibrated spectrum.}
The BCD (Barbier–Chalonge–Divan) spectro- photometric classification system\cite{1941AnAp....4...30B,1952AnAp...15..201C,2009A&A...501..297Z,2018A&A...609A.108S} is adopted to derive the effective temperature ($T_{\rm eff}$) and $G$-band absolute magnitude ($M_{\rm G}$) of ALS\,8814. Although not commonly used, this method offers an efficient and robust approach for determining the fundamental parameters of young massive stars. It is particularly effective for Be stars, as the classification system is independent of interstellar extinction and circumstellar contributions, whether from extinction or emission (see discussions in Refs.\cite{Zorec1991,Cochetti2020} and a recent review by Ref.\cite{Zorec2023}). Using this technique, several research groups have successfully derived precise fundamental parameters for over 200 field Be and B[e] stars\cite{2018A&A...609A.108S,Zorec2005,Gkouvelis2016,Cochetti2020,Aidelman2023}.

In doing so, low-resolution spectra of ALS\,8814 and seven BCD standard stars (two of which are Be type stars) are obtained on October 22, 2022, using the  Xinglong 2.16-m Telescope/BFOSC.
The G6 grism and $1\farcs6$ long slit, installed on BFOSC, are selected to obtain spectra in the optical wavelength range 3,300 $\AA$ to 5,450 $\AA$ with a resolving power of approximately 7.5\,$\AA$, similar to the resolving power required in the original BCD system\cite{2009A&A...501..297Z}.
Meanwhile, spectra are also taken using a slit width of $7\farcs0$, to correct for flux loss due to the narrow slit width.
Furthermore, spectra of three spectroscopic standard stars are observed using a $7\farcs0$ width slit at various time intervals throughout the evening to derive the spectral response curve and assess its potential variations over time.
The observational log is presented in Extended Data Table~\ref{tab:BCD_obs}.
Eventually, we achieve a very high precision of flux calibration (better than 2\%-3\%) through above strategy and observations.
The spectra are processed using a specific pipeline designed for BFOSC/G6, written in {\tt IDL}. This pipeline includes bias and flat field corrections, 1D spectrum extraction, wavelength calibration, and flux calibration, with proper correction for flux loss due to the narrow slit width.

The Balmer jump depth $D$ and the position of the Balmer discontinuity $\lambda_1$ are then measured from the flux-calibrated, low-resolution spectra.
The value of $D$ is calculated at $\lambda = 3,700\,\AA$ as $D = \mathrm{Log}~(F_{3700+}/F_{3700-})$.
As shown in Figure~\ref{fig:BCD1}(a), $F_{3700+}$ represents the flux extrapolated from the Paschen continuum.
According to the standard procedure, $F_{3700-}$ represents the flux extrapolated from the ultraviolet continuum.
However, this procedure is not applicable to Be type stars, as a second Balmer jump in the ultraviolet region is excited due to the lower pressure in the circumstellar disc.
As examined by Ref.\cite{Gkouvelis2016}, $F_{3700-}$ can also be extrapolated from the higher Balmer lines (see the green line in Figure~\ref{fig:BCD1}(a)), which are not affected by the presence of circumstellar disc.
As shown in Supplementary Fig.~\ref{fig:bcd_comp}, the measured values of $D$ and $\lambda_1$ are in great consistency with those in Refs.~\cite{Gkouvelis2016,2009A&A...501..297Z} for the seven BCD standard stars.
The standard deviations of the differences between this work and the literature are only  0.024\,dex and 6.2\,$\AA$ for $D$ and $\lambda_1$, respectively, which are adopted as the measured uncertainties in the following analysis.

To derive the stellar properties of ALS\,8814, we collect 217 BCD standard stars from Ref.\cite{2009A&A...501..297Z} with well-measured BCD parameters, $T_{\rm eff}$, and extinction $E (B-V)$.
By cross-matching them with {\it Gaia} DR3\cite{2023A&A...674A...1G} and requiring a parallax error smaller than 15\%, 156 stars are finally left.
We then derive their $G$-band absolute magnitudes, $M_{\rm G} = G - A_{\rm G} - 5\log_{10}(d) + 5$, where $A_{G} = k_{G} \times E (B-V)$, with the extinction coefficient $k_{G}$ taken from Ref.\cite{Huang2021} and $d$ taken from Ref.\cite{Jones2021}. The $T_{\rm eff}$ and $M_{G}$ of ALS\,8814 are then derived using the mean and scatter of these values of standard stars with similar BCD parameters to ALS\,8814, i.e., $|D - 0.122| \le 0.024$~dex and $|\lambda_1 - 62.43| \le 6.2~\AA$ (see Supplementary Fig.~\ref{fig:BCD2}).
In this manner, $T_{\rm eff} = 26,458 \pm 1,204$\,K and $M_{G} = -2.49 \pm 0.56$ are found for ALS\,8814.

We further derive the physical properties of ALS\,8814 by matching the observed stellar parameters with theoretical isochrones from {\tt PARSEC}.
In doing so, the maximum-likelihood approach is adopted:
\begin{equation}
L (\tau, M, R) =  \prod_{i = 1}^{n}\frac{1}{\sqrt{2\pi}\sigma_i}\exp\left(\frac{-\chi^2}{2}\right)\text{,}
\end{equation}
where the $\chi^2$ is defined as:
\begin{equation}
\chi^2 = \sum_{i =1}^{n} \left(\frac{O_i - P_i(\tau, M, R)}{\sigma_i}\right)^2\text{.}
\end{equation}
Here $O$ represents the observed stellar parameters $T_{\rm eff}$ and $M_{G}$, and $\sigma$ denotes their uncertainties.
$P$ are the theoretical values from isochrones at a given $\tau$, $M$ and $R$.
$n$ is the number of observed parameters.
We note that here we only consider  {\tt PARSEC} isochrones with solar metallicity and rotation speed $\Omega/\Omega_{\rm cr} = 0.6$ (similar to the mean value found for early-type Be star).
In this way, we then obtain probability distribution function (PDF) for $\tau$, $M$ and $R$ (see the PDF for mass in Figure~\ref{fig:BCD1}(c) as an example).
The values at the $50^{\rm th}$ percentile, $15.87^{\rm th}$ percentile and $84.13^{\rm th}$ percentile of the final PDF yield the estimate, as well as the lower and upper uncertainties, of  $\tau = 7.9_{-2.9}^{+5.2}$\,Myr, $M = 11.17^{+1.39}_{-1.20}~M_{\odot}$ and $R = 4.92^{+1.54}_{-1.06}~R_{\odot}$ for ALS\,8814.

\subsection{Spectral classification from template-matching.}
The spectral properties of ALS\,8814 are also analyzed by comparing its optical spectrum with the spectral library of B-type MK standard stars. Thanks to the recent efforts of the IACOB project, a new grid of Northern standards for the spectral classification of over 150 B-type stars has been developed\cite{2024A&A...690A.176N}. This grid covers the classical classification spectral range from 3,800 $\AA$ to 5,100 $\AA$ and has been artificially degraded to a resolving power \( R = 4,000 \). To ensure consistency, we degrade the high-quality P200/DBSP spectrum of ALS\,8814, collected in December 2022 with a resolving power of 5,000, to match the resolving power of this library.

A \( \chi^2 \) minimization procedure is applied to identify the best-matching template. The \( \chi^2 \) is defined as:

\[
\chi^2 = \sum_{i} \left( \frac{f_{\rm obs}^i - f_{\rm lib,~j}^i}{f_{\rm err}^{i}} \right)^2\text{,}
\]
where \( i \) represents different wavelengths, $f_{\rm lib,~j}$ refers to the spectra of various B-type MK standard stars in the grid, $f_{\rm obs}$ refers to the P200/DBSP spectrum for ALS\,8814, and \( f_{\rm err} \) represents the observed uncertainties. It is important to note that all spectra are normalized to their continuum.

For the calculation, we select the blue spectral range from 4,000 $\AA$ to 4,800 $\AA$, masking the Balmer lines (which are significantly affected by the decretion disk) and the diffuse interstellar bands (DIBs). Based on the resulting \( \chi^2 \) values, the two best-matched templates are HD\,24131 and HD\,144470, both classified as B1V. A comparison of the spectra between ALS\,8814 and HD\,24131 is shown in Figure~\ref{fig:spectra_match}. This result is in excellent agreement with the BCD analysis presented above.

\subsection{ASAS-SN Light curve.} 
ALS\,8814 has been monitoring by the ASAS-SN project.
The $V$-band light curve of ALS\,8814 is downloaded from the ASAS-SN Sky Patrol Photometry Database\cite{Hart2023} (\url{http://asas-sn.ifa.hawaii.edu/skypatrol/}). By selecting photometry with good quality ({\tt flg\_quality = G}), a total of 190 data points are left, which were observed between December, 2014 and December 2018.
As shown in Supplementary Fig.~\ref{fig:lc}, the light curve clearly exhibits Be-type outburst around February, 2017.
The outburst lasted around one year with rising time about 110\,days and decay time about 240\,days. 
The ratio between decay and rising time found in ALS\,8814 is $\sim$2.2, which is consistent with the value found for other early Be-type stars in Ref.\cite{Jonathan2018}.

\subsection{RV curve and fitting.} 
Most of our spectra were obtained using LAMOST/MRS, which offers significantly higher throughput in the red arm compared to the blue arm, resulting in red spectra with a higher signal-to-noise ratio (SNR). 
Among these lines in red-arm, the H$\alpha$ emission line stands out as the strongest, characterized by a stable profile, and is thus selected for measuring RV.
To measure RV, we employed a two-step procedure: 1) determining the relative RV by comparing the observed H$\alpha$ line wing to a specified template using the cross-correlation function (CCF) technique\cite{Tonry1979}, and 2) determining the absolute RV of the template by modeling its full H$\alpha$ line profile for comparison purpose.

For the first step, the wing below half of peak flux of H$\alpha$ emission line of ALS\,8814 between 6555\,$\AA$ and 6575\,$\AA$ are selected to calculate the RV shift by cross-correlating these spectra with a template spectrum taken on November 24, 2018, due to its highest SNR. The selection of the wing of the H$\alpha$ emission line is motivated by its origin in the inner region of the decretion disk, which helps minimize or eliminate RV bias in the centroid caused by contamination from asymmetric disk emissions.
The errors in the measured RVs are calculated as the quadratic sum of the methodological and systematic errors.
The methodological errors are computed using the flux randomization/random subset sampling method\cite{Maoz1990,Peterson1998,Peterson2004}, which involves randomizing the flux at each wavelength point based on its corresponding uncertainty to generate a new spectrum.
For each observed spectrum, we perform 1000 iterations of flux randomization/random subset sampling to generate 1,000 simulated spectra.
Subsequently, we measure the RV of each simulated spectrum relative to the template spectrum, yielding a distribution containing 1,000 RV measurements for each observed spectrum. The $15.87^{\rm th}$ percentile, $50^{\rm th}$ percentile, and $84.13^{\rm th}$ percentile of this distribution represent the $1\sigma$ lower limit, median value, and $1\sigma$ upper limit of the RV, respectively.
The systematic errors are determined utilizing the DIB 6614 feature. We iterate through each observed spectrum, treating it as a template, and measure the relative RVs for all other spectra based solely on the DIB 6614 feature (in range of 6612\,$\AA$ and 6619\,$\AA$). The standard deviation of the measured RVs obtained from other observed spectra for each template is considered as representative of the systematic error.
For a few exposures where the DIB 6614 feature is unclear, we adopt the mean systematic error from all available measurements.

\par For the second step, the two-Gaussian profile is adopted to measure the central wavelength of the H$\alpha$ emission line of the template spectrum:
\begin{equation}\label{eq4}
F_{\rm \lambda} = h + \frac{A_1}{\sqrt{2 \pi} \sigma_1}\exp^\frac{-{(\lambda - {\lambda}_{\rm c})} ^ 2}{2 \sigma_1 ^ 2} - \frac{A_2}{\sqrt{2 \pi} \sigma_2}\exp^\frac{-{(\lambda - {\lambda}_{\rm c})} ^ 2}{2 \sigma_2 ^ 2},
\end{equation}
where $F_{\rm \lambda}$ denotes the normalized flux; $h$ is approximately 1, representing the local continuum; $\lambda_{\rm c}$ is the central wavelength; $A$ is the height; and $\sigma$ is the standard deviation. The subscripts 1 and 2 refer to the two components of the H$\alpha$ line, with one indicating the emission component and the other the shell absorption component.
Using this profile, the H$\alpha$ line of the template spectrum has been successfully fitted, as shown in Supplementary Fig.~\ref{fig:RV_measurement}.
The absolute RV of this exposure is then precisely measured and incorporated into the other relative measurements from the CCF (see Extended Table~\ref{tab:single-epoch RVs}).
As an independent check, this profile is also applied to measure RVs of other observed spectra. 
We note that for the spectra taken by LAMOST/LRS and the second spectrum taken by P200/DBSP, only one Gaussian profile is used for fitting, since low-resolution spectra cannot distinguish the double-peak characteristic (see Supplementary Fig.~\ref{fig:RV_measurement}).
The derived RVs are also listed in Extended Table~\ref{tab:single-epoch RVs}.
The comparison between the RVs obtained from the CCF method and those from the line profile fitting shows excellent consistency,
demonstrating the robustness of our RV measurements.
Furthermore, we utilize the CCF to measure the RVs of the He I 6678 absorption line and the wing of the H$\alpha$ emission line using three additional criteria, thereby validating the reliability of the previously determined RVs. For the wing of the H$\alpha$ emission line, we adopt the following criteria: (1) the full H$\alpha$ emission line, (2) the wing below two-thirds of the peak flux, and (3) the wing below one-third of the peak flux. The results of these measurements are presented in Extended Table~\ref{tab:RVs for contrast} and visualized in Supplementary Fig.~\ref{fig:RV_contrast} for comparison.  
The RVs measured from the H$\alpha$ emission line under different criteria exhibit close agreement, with a slight increase in semi-amplitude observed as fewer wing regions are included. In contrast, the He I 6678 absorption line, which has a complex profile and is subject to variations due to contamination from emission components (as illustrated in Supplementary Fig.~\ref{fig:spectra}b), yields more scattered RV measurements compared to those derived from the wing of the H$\alpha$ emission line. Nevertheless, the RVs obtained from the He I 6678 absorption line still follow the general trend observed in the H$\alpha$ emission line data.  
In conclusion, the RVs measured from the wing of the H$\alpha$ emission line below half of the peak flux are considered as the most reliable for subsequent analysis.

\subsection{Orbital parameters.} The Lomb-Scargle method is employed to determine the orbital period from the RV curve.
The frequency grid is uniformly spaced with an oversampling factor of 500, and the lowest frequency is set to be the reciprocal of the observation time-span. Before setting the highest frequency, we make some estimates for the theoretical lower limit of the orbital period. 
This lower limit can be roughly calculated by the density of star: $P_{\rm min} = 0.369 (\rho_{*}/\rho_{\odot})^{-\frac{1}{2}}$ days\cite{Zheng2019}, where $\rho_{*}$ is the density of the star, while $\rho_{\odot}$ is the density of the Sun.
Therefore, the lower limit of the orbital period of ALS\,8814 is $\sim$1.156 days and the highest frequency is then naturally set to be $\frac{1}{1.156}$. 
The Lomb-Scargle power spectrum reveals three prominent peaks, as displayed in Supplementary Fig.~\ref{fig:power}.The period at the peak location in the power diagram is $\sim$176.8 days, while the periods at the other two peaks are $\sim$120.0 days and $\sim$332.5 days, respectively. Consequently, the real orbital period most likely lies in the range of 100 to 400 days.

\par We then use the python package \texttt{TheJoker}\cite{Adrian2017} to fully model the RV curve of ALS\,8814.
The binary orbital parameters in the model adopted by \texttt{TheJoker} include period $P$, eccentricity $e$, argument of periastron $\omega$, mean anomaly $M_0$, RV semi-amplitude $K$ and systemic velocity of binary system $V_0$. During the fitting process, the prior for P is set to a uniform distribution ranging from 100 to 400 days, as inferred from the Lomb-Scargle analysis mentioned earlier. The best fit parameters (in $1\sigma$ confidence level) are given in Table~\ref{tab:summary}. The posterior distributions of these orbital parameters are shown in Supplementary Fig.~\ref{fig:orbital_corner}.

\subsection{Distance and interstellar extinction.}
We perform a broadband SED fitting to derive interstellar extinction $A_{\rm V}$ and the distance $D$, which is poorly constrained by the \emph{Gaia} DR3 (see Table~\ref{tab:summary}) given its extremely large uncertainty in parallax. This is not surprising for such a close and bright star, given that its wide binary nature inevitably results in imperfect astrometric solutions when modeled as a single star. The high astrometric excess noise and RUWE values (see Table~\ref{tab:summary}) strongly indicate that this is not an astrometric single star, supporting the presence of a high-mass unseen companion, as inferred from RV monitoring. In the SED fitting, photometric measurements from \emph{Swift} (UVM2 band), \emph{Gaia} DR3 (BP, G and RP bands), TYCHO\cite{Høg2000} (B and V bands), \emph{TESS}\cite{Ricker2015} band and the Pan-STARRS1 Surveys\cite{Flewelling2020} (PS1; g, r, z and y bands) are adopted (i band of PS1 is excluded due to potential saturation issues). The photometric information is presented in Extended Data Table~\ref{tab:photometry_data}. 
It is noteworthy that photometric measurements from the infrared band are not included, as the disc surrounding the Be star can contribute significant non-blackbody infrared emission.
\par The SED fitting can provide strong constraints on the ratio $(\frac{R_{\rm Be}}{D})^2$, commonly referred to as the dilution factor (hereafter dilu). In this regard, the distance $D$ can only be accurately determined if the radius $R_{\rm Be}$ of the Be star is known. Fortunately, $R_{\rm Be}$ has been precisely determined using the isochrone matching technique described earlier. On the other hand, there is a strong degeneracy between the effective temperature $T_{\rm eff}$ and the interstellar extinction $A_{\rm V}$ in the SED fitting. While $T_{\rm eff}$ has been well measured using the BCD technique, $A_{\rm V}$ should be precisely determined from the SED fitting. 
The parameters for SED fitting include effective temperature $T_{\rm eff}$, surface gravity $\log g$, extinction $A_{\rm V}$ and dilu. 
We note the value of metallicity [Fe/H] is fixed to solar metallicity, i.e. [Fe/H]~$=0$.
To perform SED fitting, we execute \texttt{emcee}\cite{Foreman2013} with 32 walkers, taking 10,000 steps per walker to sample the parameter space. The initial 5,000 steps of each walker are regarded as the burn-in and consequently discarded. 
The posterior distribution of the fitting parameters $P_{\rm rob}$ is denoted as: 
\begin{equation}
 \ln P_{\rm rob} = \ln \mathcal{L} + \ln \mathcal{P},   
\end{equation}
where $\mathcal{L}$ is likelihood from SED comparison and $\mathcal{P}$ is prior. 
The priors are from our previous estimates of stellar parameters of the Be star: $T_{\rm eff}$ = (26,458$\pm$1,204)\,K and $\log g = 4.09^{+0.09}_{-0.16}$.
We then generate a series of synthetic spectra based on the given parameters ($T_{\rm eff}$, $\log g$) by linearly interpolating the BT-Cond spectra\cite{Allard2012}, which are downloaded from \url{https://lydu.ens-lyon.fr/phoenix/}. This interpolation is using the package implemented in \texttt{Scipy}\cite{Pauli2020}. We convolve the synthetic spectra with the transmission curves of each band used in the SED fitting (accessed from \url{http://svo2.cab.inta-csic.es/theory/fps/}\cite{Rodrigo2012,Rodrigo2020}) to calculate their model fluxes: 
\begin{equation}
F_{\rm mod, band} = {\rm dilu} \times \frac{\int T ({\rm \lambda})~R ({\rm \lambda})~f ({\rm \lambda})~d{\rm \lambda}}{\int T ({\rm \lambda})~d{\rm \lambda}},
\end{equation}
where $T ({\rm \lambda})$ is the transmission curve. Here $R(\lambda)$ is the reddening term:
\begin{equation}
    R(\lambda)=10^{-\frac{k_{\lambda}E(B-V)}{2.5}} \text{,}
\end{equation}
where $k_{\lambda}$ represents the reddening coefficient predicted from the $R_V = \frac{A_{V}}{E (B-V)} = 3.1$\cite{Cardelli1989}, the Fitzpatrick extinction law\cite{Fitzpatrick1999} and it is direcly provided by \texttt{extinction}\cite{Barbary2016} package. Therefore, the likelihood $\mathcal{L}$ is defined as
\begin{equation}
\ln \mathcal{L} = -\frac{1}{2} \sum_\mathrm{band}^{} \left(\frac{F_{\rm obs,band}-F_{\rm mod,band}(\emph{T}_\mathrm{eff}, \log g, {\rm dilu}, \emph{A}_{\rm V})}{\sigma_{\rm obs}}\right)^{2}\,
\end{equation}
where $F_{\rm obs,band}$ is the observed flux and $\sigma_{\rm obs}$ is the error of the observed flux. 
We present the posterior distributions of the parameters and the fitting results in Supplementary Fig.~\ref{fig:sed_corner} and Supplementary Fig.~\ref{fig:sed_plot}, respectively. It is worth noting that the reddening value derived from our analysis is significantly higher than that inferred from the 3D dust extinction map\cite{Green2019} ($A_{\rm V} = 0.91$ mag). The presence of prominent diffuse DIBs in the spectrum supports its highly reddened nature. Quantitatively, we measure the equivalent widths of DIB 6283 and DIB 8620 to be \( 1.09 \pm 0.02 \) $\AA$ and \( 0.42 \pm 0.03 \) $\AA$, respectively. Using empirical relations between extinction and the equivalent widths of these DIBs\cite{Yuan2012,Gaia2023}, we derive an \( E(B-V) \) value of approximately 0.89, corresponding to \( A_V \approx 2.76 \) (see Supplementary Fig.~\ref{fig:DIB_extinction}), which is in excellent agreement with our SED fitting result.

\subsection{Mass of the unseen companion.}
Although the mass of the Be star is known, the absence of detectable emission lines from an accretion disk surrounding the BH in the spectra prevents a direct derivation of the mass ratio between the Be star and the BH. However, given that the mass function of the unseen companion has also been determined, the mass of the unseen companion can be calculated once the orbital inclination angle is known. Determining the orbital inclination angle, however, presents several challenges. Initially, the wide orbit precludes the use of ellipsoidal light curves, which are employed to model and infer orbital inclinations. Moreover, epoch-level astrometric data, which could offer valuable constraints, is not currently available. Although a direct and reliable measurement of the orbital inclination is not feasible, we endeavor to estimate it under the premise of spin-orbit alignment, meaning that we posit the orbital inclination is equivalent to the rotational inclination of the Be star. This assumption is supported by observational evidence\cite{1994AJ....107..306H} and is consistent with the simulation results from our \texttt{POSYDON} binary population synthesis models, which will be elaborated in a subsequent section. The validity of this approach for binaries akin to ALS\,8814 is underscored by these findings, enabling us to approximate the orbital inclination through the estimation the inclination of the stellar rotational plane.
\par We estimate the stellar rotational inclination based on the morphology of the H$\alpha$ profile. Specifically, the H$\alpha$ emission line of a Be star forms in a thin, equatorial, circumstellar disk surrounding the central main sequence B star\cite{Rivinius2013}. The morphology of the H$\alpha$ profile (singly-peaked emission, doubly-peaked emission, or central shell absportion) has long been known to reflect the viewing (or inclination) angle of the system\cite{Porter2003}. Ref.\cite{Sigut2020} quantified this general observation by demonstrating that Be star inclination angles can be robustly extracted from a single observed H$\alpha$ profile by fitting the profile to a library of synthetic profiles generated by the \texttt{Bedisk}/\texttt{Beray} suite of codes\cite{Sigut2018}. Ref.\cite{2024MNRAS.527.2585L} automated the results of this method by using the library of synthetic profiles to train three types of supervised machine learning algorithms: neural networks tasked with regression, neural networks tasked with classification, and support vector regression. The algorithms were trained on grids of relative fluxes from synthetic Be star line profiles, in the vicinity of H$\alpha$. Among the three algorithms, the neural networks tasked with regression consistently outperformed the other methods, so we use it to estimate the stellar inclination for the 28 high-quality LAMOST/MRS spectra for ALS\,8814 H$\alpha$. Each spectral profile is continuum normalized, placed in the stellar rest frame, and re-sampled onto a wavelength grid containing 201 evenly spaced entries extending $\pm1,000\,\rm km\,s^{-1}$ from line centre. The processed spectra are then analyzed using trained neural networks tasked with regression. The final result is obtained by taking a weighted average between the median inclinations derived from two ensembles of five neural networks. As the mass of the Be star in the ALS\,8814 system is estimated to be 11.2 $M_\odot$, the two ensembles correspond to the $M=10\,M_\odot$ and $M=12\,M_\odot$ mass libraries and the weights are calculated using the difference between each library mass and the estimated mass of the observed Be star such that the inclination determinations of the 12 $M_\odot$ models are weighted slightly more heavily than the 10 M$_\odot$ models. The inclination angle of $36.0^{+3.7}_{-6.1}\,^{\circ}$ for ALS\,8814 is derived using the neural networks tasked with regression, based on the distribution of 28 individual inclinations. The results for each LAMOST observation are shown in Supplementary Fig.~\ref{fig:inc}. Additionally, the neural networks tasked with regression exhibited a root mean squared error (RMSE) of 7.6$^{\circ}$ compared to an observational sample of 92 Galactic Be stars, whose inclination angles were determined through direct H$\alpha$ profile fitting\cite{2024MNRAS.527.2585L,Sigut2020}. Taking into account this RMSE, we adopt an inclination angle of $36.0^{+8.5}_{-9.7}\,^{\circ}$ as the final value derived from the machine learning method, which yields a BH mass of $23.51^{+16.75}_{-6.38}\,M_{\odot}$. 

\par Additionally, we validate the inclination measurement by concentrating on the rotational velocity $V$ and the projected rotational velocity $V\sin i$. By estimating both $V$ and $V\sin i$, we can determine their ratio to derive the stellar rotational inclination. The rotational velocity distribution $V/V_{\rm c} = \mathcal{N}(0.73, 0.17)$, derived from a sample of Be stars with $\tau/\tau_{\rm MS} > 0.8$ and $7.5 < M/M_\odot < 15$ (where $\tau_{\rm MS}$ is the main-sequence lifespan of Be stars; Ref.\cite{2016A&A...595A.132Z}), is adopted for ALS\,8814.
The critical rotational velocity $V_{\rm c}$ can be calculated using the following equation:
\begin{equation}
V_{\rm c} = \sqrt{\frac{GM}{R_{\rm e}}} = \sqrt{\frac{2GM}{3R_{\rm p}}},
\end{equation}
where $G$ is the gravitational constant, $M$ is the mass of the star, $R_{\rm e}$ is the equatorial radius and $R_{\rm p}$ is the polar radius. In practice, we can only measure the equivalent radius of the Be star. Therefore, we use \texttt{PHOEBE} to numerically calculate $V_{\rm c}$. 
By modeling a star with the same mass and equivalent radius as ALS\,8814, we determine its critical rotational velocity to be $V_{\rm c} = 580.9$~km~s$^{-1}$.

\par To measure $V\sin i$ from the absorption lines of ALS 8814, we first use the \texttt{spectool} package (\url{https://github.com/zzxihep/spectool}) to generate a series of rotational broadening kernels ranging from 150 to 350 km/s in steps of 0.1~km~s$^{-1}$.
For each rotational broadening kernel, we first convolve the \texttt{TLUSTY} model spectrum (with $T_{\rm eff}$=26,000 K, $\log g =4.0$ and [Fe/H]=0) with the rotational broadening kernel, followed by convolution with the LAMOST/MRS instrumental broadening kernel at a resolving power of $R = 7,500$.
To determine the rotational velocity of ALS\,8814, we calculate the $\chi^2$ values by comparing the broadened spectra to the observed spectrum, specifically, $\chi^2$ is calculated as:
\begin{equation}
\chi^2 = \sum_{i =1}^{n} \left(\frac{F_{\mathrm{obs},i} - F_{\mathrm{mod},i}}{\sigma_{\mathrm{obs},i}}\right)^2,
\end{equation}
where $F_{\mathrm{obs},i}$ and $F_{\mathrm{mod},i}$ are respectively the $i_{\rm th}$ flux points of observed spectrum and broadened model spectrum and $\sigma_{\mathrm{obs},i}$ is the $i_{\rm th}$ flux uncertainty of the observed spectrum. The absorption features analyzed include the He I 6678 line from the LAMOST/MRS spectrum obtained on November 13, 2019, and the He I 5048 line from the LAMOST/MRS spectrum obtained on December 10, 2019. They are picked out due to its relatively high SNR and minimal contamination from emission radiations. Moreover, the earliest spectrum, obtained by P200/DBSP, is analyzed due to its relatively high resolving power ($\sim$5,000) and high SNR. The blue arm of the P200/DBSP spectrum, extending beyond the 4,000 $\AA$ to 5,100 $\AA$, contains a considerable number of spectral lines, including He lines and metallic lines. Therefore, our analysis of this spectrum focuses on this wavelength range, excluding Balmer lines and DIBs, rather than on specific individual lines. The final $\chi^2$ value as a function of rotational broadening kernel for each measurement is shown in Supplementary Fig.~\ref{fig:chisq}. We take the $V\sin i$ value corresponding to the minimum value of $\chi^2$ as the final measurement. The $V\sin i$ values where the $\chi^2$ is greater than the minimum by 1 are considered the $1\sigma$ upper and lower limits of the measurement, resulting in $V\sin i = 240.8^{+1.5}_{-1.4}$ km~s$^{-1}$ for He I 6678 line, $241.9^{+4.8}_{-4.4}$ km~s$^{-1}$ for He I 5048 line and $258.8^{+1.3}_{-2.0}$ km~s$^{-1}$ for the blue arm of the P200/DBSP spectrum, respectively. Supplementary Fig.~\ref{fig:match} demonstrates the agreement between the observed spectra of ALS\,8814 and the \texttt{TLUSTY} model spectra broadened with respective $V\sin i$ and instrumental broadening kernels. All measurement results are close to each other and their weighted average, $248.3^{+2.1}_{-2.2}$ km~s$^{-1}$, is adopted as the final $V\sin i$ measurement. We then estimate the orbital/rotational inclination $i$ to be $36.2^{+13.5}_{-7.7}\,^{\circ}$, which results in the BH mass of $M_{\rm BH} = 22.84^{+11.98}_{-7.88}\,M_\odot$. 
\par The two results obtained by the machine learning method and $V\sin i$ measurement are in excellent agreement, demonstrating the robustness of our inclination estimates. However, when interpreting this result, it is important to wake that it is derived on the assumption of spin-orbit alignment. In reality, there may be a slight spin-orbit misalignment, which could introduce deviations between the measured stellar rotational inclination and the true orbital inclination. All ALS\,8814-like systems can receive a misalignment angle within 10 degrees (see Supplementary Fig.~\ref{fig:kick}) according to our binary population synthesis modeling. Therefore, we consider an additional 10 degree error and pass it into the method error, leading to a final orbit inclination estimate between 22.1$^{\circ}$ and 49.1$^{\circ}$, corresponding to a BH mass of $15.1\pm1.0\,M_\odot$ to $57.8\pm3.9\,M_\odot$. This is a broad estimate, and to truly conduct a precise determination of the orbital inclination will require the epoch-level astrometric data that \emph{Gaia} is expected to publish in the future.

\subsection{Ruling out non-degenerate companions.}
Initially, we employ the spectral disentangling technique, a well-established method in astrophysics, to confirm the absence of a non-degenerate companion in ALS\,8814. Following this confirmation from spectral disentangling, we introduce one novel method to further validate these findings. Additionally, the novel method is also employed to rule out the possibility of an inner binary.
{(1) \it Linear algebra technique to disentangle spectra}
This algorithm, proposed by Ref.\cite{Simon1994}, is designed to disentangle observed spectra. It reconstructs the rest-frame component spectra under the assumption that these spectra are stable over time. We utilize this method to analyze the LAMOST/MRS spectra of ALS\,8814. In the first step, we simulate a mock single-lined spectroscopic binary (SB1) system to validate the method, using the \texttt{TLUSTY} model spectra. The simulated SB1 system has same stellar parameters and orbital parameters as ALS\,8814 system. That is, the visible star is a Be star with an effective temperature $T_{\rm eff} = 26,000$\,K, surface gravity $\log{g}$ = 4.0, metallicity [Fe/H] = 0; the Be star's radius is 4.92 $R_{\odot}$ with a projected rotational velocity $V\sin i=248.3$ km s$^{-1}$. 
We construct 27 mock spectra within the blue band range of 5,000 $\AA$ to 5,200 $\AA$, aligning their orbital phases with those of the observed spectra. During the construction of these mock spectra, we also simulate LAMOST/MRS instrument broadening ($\sim$40 km s$^{-1}$). Three phase points corresponding to observed spectra with low SNRs are excluded. To simulate a scenario with a SNR of $\sim$100, Gaussian-distributed noise is added into the mock spectra. Subsequently, we attempt to disentangle these mock spectra into a single-component spectrum. The algorithm produces a `mean' spectrum (over 27 individual spectra). The residuals of the derived spectrum to the mock spectra (hereafter residual spectra) across 27 phases, are visually inspected (see Supplementary Fig.~\ref{fig:disentangling_mock_SB1} for details). As expected, the residual spectra exhibit characteristics of pure noise, with no discernible patterns indicative of a secondary component, confirming the absence of additional spectral features. 

\par In the second step, we introduce a secondary component to the mock system, making it a double-lined spectroscopic binary (SB2) system. Assuming a 9.8 $M_\odot$ (i.e., the minimum mass of the unseen companion) main-sequence secondary companion, the stellar parameters are set as: $T_{\rm eff} = 24,000$\,K, $\log{g}$ = 4.25, [Fe/H] = 0, $R$ = 4 $R_{\odot}$, and $V\sin i=200$ km s$^{-1}$. 
The spectra of the secondary component are adjusted to match the corresponding RVs according to the orbital phases and the mass ratio and then superimposed onto the primary's spectra to simulate the spectra of an SB2 system. Similarly, Gaussian-distributed noise is added into the SB2's spectra, to simulate a SNR $\sim$ 100 scenario. Similarly, we convolute the composite spectrum with an instrumental broadening kernel of LAMOST/MRS. Again, using the disentangling technique, we disentangle these mock spectra into a single-component spectrum, compute the mean spectrum and thus the residuals of the 27 spectra. In contrast to the previous case, the residual spectra in this SB2 simulation not only include random noise but also exhibit some wiggly patterns near the spectral lines, particularly for the mock spectra at RV extrema. These features are attributable to the contribution of the secondary component (see Supplementary Fig.~\ref{fig:disentangling_mock_SB2} for details).
\par Having established the expected results for mock SB1 and SB2 systems, we proceed to the third step to analyze the multi-epoch spectra of ALS\,8814. We utilize 27 spectra obtained by LAMOST/MRS with SNR $>$ 50 at the blue band (5,000 $\AA$ to 5,200 $\AA$; consistent with the mock spectra described earlier) to compute the mean spectrum and the residual spectra following a same method.
The results indicate that only noise and/or irregular variations, likely due to unstable line profiles, are present in the residual spectra (see Supplementary Fig~\ref{fig:disentangling_observed}). In other words, an SB1 model adequately accounts for the observed data. Consequently, we conclude that the companion of the Be star is not a non-degenerate companion.

\par {(2) \it The impact of companion(s) on the profile of H$\alpha$} 
\par The strong H$\alpha$ emission line of ALS\,8814 suggests that the introduction of absorption lines from a visible companion would alter the H$\alpha$ profile, causing asymmetry, particularly near the RV extrema. In contrast, if no visible companion were present, the H$\alpha$ emission line would remain symmetric. To explore this hypothesis, we select a LAMOST/MRS spectrum to simulate the effect of a visible companion on the H$\alpha$ profile. The selected spectrum was obtained on February 23, 2022, which is near the RV extrema and has a high SNR. The stellar parameters for both the Be star and the companion are consistent with those used in (1). In doing so, we firstly assign RVs to both \texttt{TLUSTY} model spectra respectively representing the Be star and the companion based on the observed data and convolve them with the instrumental broadening kernel of LAMOST/MRS, and calculate the flux ratios at each wavelength for the two spectra. We then normalize the \texttt{TLUSTY} model spectrum of a 9.8 $M_\odot$ secondary, scale it by the flux ratio, and add it to the normalized LAMOST/MRS spectrum to create a composite spectrum. 
As shown in Supplementary Fig.~\ref{fig:mock_inner_ms}, a distinct asymmetry in the H$\alpha$ emission line is clearly present in the composite spectra. However, since the observed H$\alpha$ line remains perfectly symmetric, this suggests that no visible companion is present. Consequently, we conclude that the companion is likely an unseen BH.
\par We also consider the intriguing, albeit extreme, scenario in which the unseen companion is an unresolved binary system consisting of two stars with a combined mass of 9.8 $M_\odot$. If these stars were two equal-mass main-sequence stars, they would be the faintest and most challenging to detect. To explore this possibility, we assume the unseen companion consists of two 4.9 $M_\odot$ main-sequence stars with an orbital period of 10 days. Using Kepler's Third Law, the RV semi-amplitude of the inner binary is calculated to be 108.98 km s$^{-1}$ assuming aligned orbital planes between the inner binary and the outer Be star. The stellar parameters for each 4.9 $M_\odot$ main-sequence star are set as follows: $T_{\rm eff} = 17,000$\,K, $\log{g}$ = 4.00, [Fe/H] = 0, $R$ = 2.66 $R_{\odot}$, and $V\sin i = 200$ km s$^{-1}$. We then follow similar procedures as for the single companion scenario to create composite spectra. Since the orbital phase of the inner binary is unknown, we apply four sets of RV separations, evenly spaced across the calculated RV semi-amplitude (inner RV separations = 0 km s$^{-1}$, 36$\times$2 km s$^{-1}$, 72$\times$2 km s$^{-1}$ and 108$\times$2 km s$^{-1}$, respectively). 
As shown in Supplementary Fig.~\ref{fig:mock_inner_binary}, all four composite spectra display pronounced asymmetry in the H$\alpha$ line, which would be expected from the presence of an inner main-sequence binary. Since no such asymmetry is observed in the real data, we can rule out the presence of an inner main-sequence binary system as a viable scenario.
\par Based on the results of the two aforementioned methods, we have the sensitivity to detect any non-degenerate companions in ALS\,8814 via spectral analysis, even when accounting for the possibility that our orbital inclination estimate may not be reliable and that potential companions could be rapidly rotating. However, no detectable signals from non-degenerate companions are present in our observed spectra, which leads us to conclude that ALS\,8814 does not host any non-degenerate companions in terms of spectral analysis. Furthermore, we conduct a preliminary analysis to assess the potential impact of an inner binary on the SED. In the case of an inner binary consisting two 4.9 $M_\odot$ main-sequence stars, the luminosity ratios relative to the Be star for the Swift\_UVM2, PS1\_r and PS1\_y bands are calculated to be 0.130, 0.284 and 0.324, respectively. These ratios are substantial, indicating that such an inner binary would not be overwhelmed by the luminosity of the Be star. Again due to their significant differences, the presence of such an inner binary would remarkably alter the SED profile. Nevertheless, a single-component SED model provides an excellent fit to the observed photometric data (see Supplementary Fig.~\ref{fig:sed_plot}). This leads us to conclude that the existence of an inner binary in ALS\,8814 is unlikely based on SED analysis. Additionally, the agreement between the observed photometric data and the model in the Swift\_UVM2 band indicates the absence of UV excess, thereby ruling out the presence of a hot hidden companion.
\subsection{X-ray observation and luminosity.}
ALS\,8814 is observed using the \emph{Swift}/XRT with an exposure time of $\sim$680 seconds on November 13, 2023 (ToO ID: 19619; PI: Qian-Yu An). 
The data are analyzed with the packages in \texttt{HEASOFT} version 6.31, including \texttt{XSELECT} and \texttt{XIMAGE}. We extract the X-ray image in 0.3--10 keV band, and estimate the 3$\sigma$ count rate upper limit of $\sim$1.7 $\times 10^{-2}$ cts~s$^{-1}$ at the source position with the task \texttt{sosta}. The absorbed/unabsorbed flux upper limits of $7.6\times10^{-13}$/$1.2\times10^{-12}$ erg~cm$^{-2}$~s$^{-1}$ are derived by assuming a typical power-law model ($N_{\rm H} = 4.9\times10^{21}$ cm$^{-2}$ and $\Gamma$ = 2.0; Ref.\cite{HI4PI2016}), yielding an upper limit of X-ray luminosity of 1.79$\times10^{32}$ erg s$^{-1}$ for ALS\,8814. 
Additionally, we retrieve the archive of eROSITA-DE Data Release 1 and find that ALS\,8814 was observed by eROSITA with a total exposure time of 74.6 seconds between April 1, 2020 and April 5, 2020. 
The absorbed/unabsorbed flux upper limits are, respectively, $1.9\times10^{-13}$/$2.3\times10^{-13}$ erg~cm$^{-2}$~s$^{-1}$, resulting in an upper limit of $L_{\rm X}\leq3.42\times10^{31}$ erg s$^{-1}$, for 02e band (0.2-5.0 keV) by adopting a $N_{\rm H}$ = $3.0\times10^{20}$ cm$^{-2}$. 
Therefore, ALS 8814 can be considered an X-ray faint source based on both observations.

\par We take the result from eROSITA for the subsequent analysis, given its more stringent constraints on the upper limit of luminosity.
In Supplementary Fig.~\ref{fig:X-ray}, we present a comparison of the X-ray luminosity of ALS 8814 with that of Be X-ray binaries, whose orbital periods are taken from Ref.\cite{Reig2011}.
All X-ray flux measurements presented are derived from the 02e band of the eROSITA telescope, while the distances are taken from Ref.\cite{Reig2011}.
It is evident that, for the same orbital period range, the X-ray luminosity of ALS 8814 is significantly lower, by one or two orders of magnitude (or even more), compared to other Be X-ray binaries.
It should be noted that the Be X-ray binaries considered here are all Be star+neutron star systems. 
A previous study\cite{Narayan2008} has indicated that black hole X-ray binaries (BHXBs) exhibit X-ray luminosities that are one or two orders of magnitude lower than those of neutron star X-ray binaries (NSXBs) in their quiescent states.
This result is considered strong evidence for the existence of an event horizon.
However, our case differs somewhat. 
The sample in Ref.\cite{Narayan2008} consists of close binary systems, most of which have orbital periods less than 4 days.
It is reasonable to expect that all Be-neutron star binaries with such short orbital periods would be X-ray sources, implying that there is no bias in the selection process of sources in Ref.\cite{Narayan2008}.
Samples in our analysis are wide binaries with $P_{\rm orb}$ $>$ 30 days, implying the risk of omitting those possible Be-neutron star binaries that do not serve as X-ray sources. 
If not considering this potential selection effect, the faint X-ray luminosity of ALS 8814 should represent another independent piece of evidence for the presence of an event horizon in a BH.
To overcome this selection issue, it is necessary to systematically observe a large sample of Be-neutron star binaries with $P_{\rm orb}$ $<$ 200 days deeply in the X-ray band, to check if there are any such systems that do not exhibit X-ray emission.

\subsection{Kinematics of ALS\,8814.}
The 3D position and velocity of ALS\,8814 are determined 
from its distance from SED fitting, positions, proper motions from \emph{Gaia}, and the systemic velocity from the RV curve fitting. 
The right-handed Cartesian system, with $X$ toward the direction opposite to the Sun, $Y$ in the direction of Galactic rotation, and $Z$ in the direction of north Galactic pole (NGP), is adopted during the calculations. The Sun is placed at $(X, Z) = (-8.178, 0.025)$\,kpc\cite{GRAVITY2019,Joss2016} and its peculiar velocities respect to the Local Standard of Rest are $(U_{\odot}, V_{\odot}, W_{\odot}) = (7.01, 10.13, 4.95)$\,km\,s$^{-1}$\cite{Huang2015}.
The circular velocity at the solar position is set to $V_{\rm c} (R_0) = 234.04$\,km\,s$^{-1}$\cite{Zhou2023}.
Follow the method described in Ref.\cite{Zhao2023}, the 3D peculiar velocities at Galactic plane crossing $V_{\rm pec}^{Z = 0}$ are calculated by backward orbit integration in time of 10\,Gyr under the Milky way potential {\tt milkypotential}\cite{Bovy2015} via python package {\tt gala}\cite{Price-Whelan17}.

\subsection{BSE Population synthesis of ALS\,8814$-$like binaries.}
We employ the BSE code\cite{Hurley2002} to model the formation of Galactic Be-BH binaries including ALS\,8814. This code has been significantly modified involving the processes of stellar evolution and binary interaction, in particular the treatments of mass transfer and its stability\cite{2014ApJ...796...37S}. Evolution begins from a primordial binary containing two zero-age main sequence  stars, the primary star first fills its Roche lobe and transfers its envelope to the secondary star via Roche lobe overflow. Mass and angular momentum accretion onto the secondary star causes it to expand and spin up. We assume that the secondary star accretes at a rate being equal to the mass transfer rate multiplied by a factor of $(1-\Omega/\Omega_{\rm cr})$, where $\Omega$ is the angular velocity of the secondary star and $\Omega_{\rm cr}$ is its critical value. Under this assumption, the secondary star rotating at $\Omega_{\rm cr}$ will no longer accrete any mass. When dealing with the rotational evolution of the secondary star, we include the effect from stellar winds, tides and mass transfer\cite{Mink2013}. In addition, mass accretion may drive the secondary star to get out of thermal equilibrium and expand, possibly resulting in the formation of a contact binary as the secondary star has also filled its own Roche lobe\cite{Nelson2001}. In our simulations, we assume all contact binaries undergo common envelope evolution. For the primordial binaries with primary masses less than $60~M_\odot$, detailed binary evolution calculations have been used to obtain the parameter space that determining whether the binaries experience stable mass transfer or common envelope evolution. The calculations showed that the maximal mass ratio of the primary star to the secondary star for avoiding the formation of contact binaries is about 6, and these calculated results had been incorporated in the BSE code\cite{2014ApJ...796...37S}. For more massive binaries with donor masses larger than $60~M_\odot$, recent investigations indicated that the maximal mass ratio for stable mass transfer can be reach as high as about 10\cite{Pavlovskii2017,Shao2021}. So for the primordial binaries with primary masses larger than $60~M_\odot$, we assume that only the binaries with mass ratios of $>10$ undergo common envelope evolution. 

For stable mass transfer, we assume the  transferred matter that is not accreted by the secondary star escapes from the binary system and carries the specific orbital angular momentum of the secondary star. During common envelope evolution, we adopt the energy conservation equation to deal with the
orbital evolution of the binary system\cite{Webbink1984}. For stellar winds, we use the prescription of Belczynski et al.\cite{2010ApJ...714.1217B} to treat mass-loss rates for various types of stars. Recent observations of Cyg X$-$1 with a $\sim21M_\odot$ BH suggested a significant reduction in mass loss via stellar winds relative to commonly used prescriptions for helium stars (Wolf–Rayet)\cite{2021Sci...371.1046M,Neijssel2021}. For helium stars, the mass-loss rate is given as a function of stellar luminosity $L$ and metallicity $Z$, 
\begin{equation}
\dot{M}_{\rm WR} = f_{\rm WR} \times 10^{-13}(L/L_\odot)^{1.5} (Z/Z_\odot)^{0.86}\,M_\odot\,\rm yr^{-1},
\label{bse_mdotWR}
\end{equation}
where $f_{\rm WR}$ is set by default to $ 1.0$\cite{2010ApJ...714.1217B}. In our simulations, we decrease $f_{\rm WR}$ from 1.0 to 0.1/0.3 to test its effect on the formation of ALS\,8814 with a $\sim 15-36M_\odot$ BH (see Supplementary Fig.~\ref{fig:model}).  

Each of our simulations evolves $\sim 10^7$ primordial binaries. For the parameter configurations of primordial binaries, the masses of the primary stars are assumed to follow the initial mass function\cite{Kroupa1993}, the mass ratios of the secondary to the primary stars obey a flat distribution between 0 and 1 (Ref.\cite{Kobulnicky2007}), the orbital separations are uniformly distributed in the logarithmic space\cite{Abt1983}, and the eccentricities have a thermal-equilibrium distribution between 0 and 1 (Ref.\cite{Duquennoy1991}). In our simulations, we take the primary masses in the range of 15--100 $M_{\odot}$, the secondary masses in the range of 1--100 $M_{\odot}$, and the orbital separations in the range of 3--10,000 $R_{\odot}$. For the stars formed in the Milky Way, we assume a constant star formation rate of 3\,$M_{\odot}\,{\rm yr}^{-1} $ with a metallicity of $Z = 0.02$.

Following previous works\cite{2014ApJ...796...37S}, we identify Be stars as main-sequence stars with rotational velocities exceeding 80\% of their Keplerian limits. In our simulations, we pick out all Be-BH binaries and record their relevant parameters. After analyzing our recorded data, we estimate that $\sim380-450$ of Be-BH binaries exist in the Milky Way. This number is not strongly dependent on the options of $f_{
\rm WR}$. We find that the formation of ALS\,8814 favors a relatively low mass-loss rate with $f_{
\rm WR}\lesssim 0.3$. In the case of $f_{
\rm WR}= 0.1$, Supplementary Fig.~\ref{fig:formation} shows an evolutionary track from a primordial binary to the formation of a system similar to ALS\,8814. After mass transfer, the primary evolves to be a helium star and the secondary becomes a Be star. Subsequently, the He star directly collapses into a $\sim 16.4~M_\odot$ BH without any kick. 

\subsection{\texttt{POSYDON} population synthesis of ALS\,8814$-$like binaries.}
In a recent study\cite{2024ApJ...971..133R}, the \texttt{POSYDON} binary population synthesis (BPS) code was used to explore the Galactic Be-XRB population, highlighting the success of self-consistent evolution of rotation during mass transfer at reproducing many features in the observed properties of the Be-XRB population. In this work, we explore the aforementioned population synthesis simulation without any adjustments, to see if any ALS\,8814$-$ like binaries were produced in their models.
Here we briefly reiterate the relevant BPS modeling assumptions of Ref.\cite{2024ApJ...971..133R} and refer the reader for a complete description.

The \texttt{POSYDON} version 1 (solar metallicity with $Z=0.0142$ only) BPS models incorporate detailed 1D stellar and binary evolution simulations with \texttt{MESA}, which assume fully conservative Roche-lobe Overflow (RLO) mass transfer, while limiting accretors to sub-critical rotational with a rotationally enhanced wind. In practice, this leads to a mass transfer efficiency near $5\%$ (the accretor accretes, on average, 5\% of the donor's mass loss) for an astrophysical population of detached binaries harboring compact objects\cite{2024ApJ...971..133R}. For the Wolf–Rayet wind prescription, we employ Equation 1 from Ref.\cite{2000A&A...360..227N}, which differs from the \texttt{BSE} code.

Binary models which evolve into dynamically unstable mass transfer are assumed to enter common envelope evolution which we model using the $\alpha-\lambda$ formalism\cite{Webbink1984}, where we take $\alpha=1$ and $\lambda$ is calculated from stellar structures.
The resulting compact objects formed through core collapse supernovae (SNe) are modeled using Ref.\cite{Patton2020} as implemented in \texttt{POSYDON} version 1, where BHs receive isotropically oriented natal kicks with magnitude randomly drawn from a Maxwellian distribution with standard deviation $\sigma_\mathrm{CCSN} = 265 ~\mathrm{km~s}^{-1}$, scaled by $1.4\,M_\odot/M_\mathrm{BH}$.
If the binary model remains bound post-SNe, \texttt{POSYDON} matches the secondary to an equivalent single star model to continue evolving the binary in a detached configuration.

To model a constant star formation history of the Milky Way, we draw randomly distributed time samples of the time evolving parameters of binaries during their detached evolution.
In addition to this time sampling, each sample is weighted by the lifetime of the BH and Be star detached phase, as defined in Equation 2 of Ref.\cite{2024ApJ...971..133R}, to account for the relative lifetime of different binaries. The purpose of this weighting is to reflect the higher probability of observing longer-lived systems. The panel c of Supplementary Fig.~\ref{fig:model} displays the mass distribution of BH and Be star systems from the \texttt{POSYDON} population synthesis modeling. In this study, we do not aim to compare codes with \texttt{BSE} due to significant differences in physics and methodologies. Three primary factors contributing to this disparity in number distribution are: (a) consideration of weighted number distribution; (b) variation in initial metallicity; (c) differences in wind prescriptions.

From the time sampled evolution, we further select binaries which match simultaneously a reasonable range of measured properties of ALS\,8814: in orbital period ($P_\mathrm{orb}\in[100-300]~\mathrm{day}$), eccentricity ($e\in[0.15-0.25]$), component masses ($M_1 = M_\mathrm{Be}\in[10-12]~M_\odot$; $M_2 = M_\mathrm{BH}\in[15-58]~M_\odot$), and surface rotation ($\Omega/\Omega_\mathrm{cr} > 0.5$), we find 19 candidate binaries.

All of our ALS\,8814$-$like binary models evolve through a phase of RLO mass transfer initiated at terminal-age main sequence (Case B mass transfer) which spins up the secondary to high fractions of its critical rotation rate.
We display the evolution of one such system in Supplementary Fig.~\ref{fig:evolution_model}. This sample system contains a $\sim$50 $M_{\odot}$ BH progenitor and a $\sim$11 $M_{\odot}$ companion in a circular orbit with a period of 73.5 days at the zero-age main sequence phase. The strong stellar wind removes a significant fraction of the donor's hydrogen envelope, leaving behind a helium enriched star with a surface helium abundance around $\gtrsim 50\%$. At 4.34 Myr, the partially stripped donor finishes the main sequence evolution and evolves into a RLO phase lasting a few $10^4$ years, during which mass transfer accelerates the removal of the hydrogen-rich envelope. At 4.47 Myr, the donor evolves into a helium star with a mass of 25.7 $M_{\odot}$. At the same time, the WR wind continues to reduce its mass, leading to the exposure of a high-mass C/O core around 4.7 Myr, just before the iron core collapse.

During the donor's SN at 4.73 Myr, the newly formed BHs receive modest natal kicks in the range $10-50~\mathrm{km~s}^{-1}$ resulting in low systemic velocities in the range $5-30~\mathrm{km~s}^{-1}$.
Additionally, the BH natal kick induces an eccentricity of 0.18 and tilts the new orbital angular momentum vector (relative to the pre-SN configuration), which we find varies from $i \simeq 1-20$ degrees.
Assuming the secondary's stellar spin is aligned with the pre-SN orbital angular momentum, the aforementioned range in inclination angles would translate directly to the inclination between the spin axis of the Be star and the orbital inclination.
Furthermore, when comparing our models with systemic velocities similar to the measured peculiar velocity of $8.0\pm6.8~\mathrm{km~s}^{-1}$, we find that most have spin-orbit misalignment angles smaller than 6 degrees (see Supplementary Fig.~\ref{fig:kick}).
This suggests the assumption of spin-orbit alignment for ALS\,8814 is quite reasonable. However, a complete exploration of different SNe models is necessary to draw more robust conclusions. 

Following the formation of the BH, our binary model's orbital properties do not significantly evolve until the end of their detached evolution due to their wide orbital separation.
However, the Be star's $\Omega/\Omega_\mathrm{cr}$ increases due to the relatively little angular momentum lost from winds on the main sequence evolution, combined with radial expansion as it approaches the terminal age of the main sequence.
The equatorial rotational rate of the secondary in our binary \texttt{MESA} models is $\simeq 500 ~\mathrm{km~s}^{-1}$.
As the Be star evolves further, all models initiate a second phase of RLO between the now post-main sequence Be star and BH, resulting in 12 models with stripped helium stars and 7 hydrogen rich models which all undergo a core collapse SN which ultimately disrupts the binaries due to our model imparting significantly stronger kicks to NSs. 
Therefore, we predict ALS\,8814$-$like binaries are unlikely to be progenitors of NS-BH mergers observed by LIGO.
However, other Be-BH binaries with less massive Be companions may undergo an electron-capture SNe (ECSN), and are more likely to remain bound to form NS-BH mergers due to the smaller natal kicks predicted in ECSN.

\subsection{Comparison to other known dormant BHs.}
We compare ALS\,8814 with other known dormant BHs (NGC\,3201\#21859\cite{Giesers2019}, NGC\,1850\,BH1\cite{Saracino2023}, VTFS\,243\cite{Shenar2022}, HD\,130298\cite{Mahy2022}, MWC\,656\cite{2014Natur.505..378C}, J0521\cite{Thompson2019}, AS\,386\cite{Khokhlov2018}, NGC\,3201\#12560\cite{Giesers2018}, Gaia\,BH1\cite{El-Badry2023}, G3425\cite{Wang2024}, Gaia\,BH2\cite{El-Badry2023B} and Gaia\,BH3\cite{Gaia2024}) in terms of BH mass-orbital period relationship in Supplementary Fig.~\ref{fig:m_p}. Among the dormant BHs identified by RV monitoring, ALS\,8814 stands out as hosting the most massive BH while also having the longest orbital period. In contrast, the orbital period of ALS\,8814 is shorter than those of all BH binaries identified by astrometry, suggesting a distinct boundary between the two populations and highlighting the benefits of different search methods targeting each population. Furthermore, while the visible star in AS\,386 exhibits the B[e] phenomenon, it does not qualify as a classical Be star because its emission lines stem from forbidden lines rather than a decretion disk. Consequently, in terms of the nature of the visible stars, ALS 8814 remains unique as the only confirmed Be star-BH binary.
\end{methods}

\renewcommand{\tablename}{Extended Data Table}
\renewcommand\thetable{\arabic{table}}  
\setcounter{MaxMatrixCols}{40}
\setcounter{table}{0} 
\begin{table}
\centering
\renewcommand{\arraystretch}{0.95}	
\centering
  \caption{Spectral observations of ALS\,8814 for the purpose of RV measurements. The spectrum marked by a star is used as a template in CCF.}
  \centering
  \tiny
  \begin{tabular}{@{}cccccccc@{}}
  \hline
  \thead{\textbf{Facility}\\~} &
  \thead{\textbf{Resolving~power}\\~} & \thead{\textbf{(Observation~time)$_{\rm mid}$}\\(YYYY-MM-DD hh:mm:ss)} &
  \thead{\textbf{HJD$_{\rm mid}$}\\~} & 
  \thead{\textbf{Exposure time}\\(s)} &
  \thead{\textbf{Phase}\\~} & 
  \thead{\textbf{RV\_fit}\\($\mathrm{km~s^{-1}}$)} &
  \thead{\textbf{RV\_CCF}\\($\mathrm{km~s^{-1}}$)} \\
  \multicolumn{1}{c}{(1)}&
  \multicolumn{1}{c}{(2)}& 
  \multicolumn{1}{c}{(3)}& 
  \multicolumn{1}{c}{(4)}& 
  \multicolumn{1}{c}{(5)}&
  \multicolumn{1}{c}{(6)}&
  \multicolumn{1}{c}{(7)}&
  \multicolumn{1}{c}{(8)}\\  
  \hline
  \multirow{30}{*}{LAMOST/MRS} &
  \multirow{30}{*}{$\sim$7500}
   &2017-11-05 20:25:18 & 2458063.355 & 6$\times$600 & 0.119 & 56.67$\pm$4.16 & 58.68$\pm$4.18\\
   &  &2017-11-07 20:01:28 & 2458065.338 & 7$\times$600 & 0.131 & 51.19$\pm$4.22 & 54.40$^{+4.28}_{-4.25}$\\
   &  &2017-11-26 18:00:55 & 2458084.256 & 2$\times$1200 & 0.238 & 16.17$\pm$4.11 & 17.22$^{+4.11}_{-4.12}$\\
   &  &2017-11-28 17:47:08 & 2458086.246 & 6$\times$1200 & 0.249 & 14.39$\pm$4.26 & 14.75$\pm$4.26\\
   &  &2017-11-29 19:19:14 & 2458087.310 & 3$\times$600 & 0.255 & 11.78$\pm$3.46 & 12.03$\pm$3.46\\
   &  &2017-11-30 17:18:24 & 2458088.226 & 5$\times$1200 & 0.260 & 10.69$\pm$4.35 & 10.38$\pm$4.36\\
   &  &2017-12-01 16:49:43 & 2458089.206 & 3$\times$600 & 0.266 & 8.40$\pm$3.88 & 7.68$^{+3.90}_{-3.88}$\\
   &  &2017-12-02 16:38:09 & 2458090.198 & 3$\times$600 & 0.271 & 8.17$\pm$4.01 & 8.10$\pm$4.02\\
   &  &2017-12-26 15:38:26 & 2458114.157 & 4$\times$1200 & 0.407 & -13.65$\pm$3.86 & -13.63$\pm$3.87\\
   &  &2018-01-03 16:15:24 & 2458122.183 & 5$\times$1200 & 0.453 & -20.55$\pm$3.77 & -22.00$\pm$3.78\\
   &  &2018-01-25 13:53:36 & 2458144.084 & 5$\times$1200 & 0.577 & -22.18$\pm$4.11 & -25.59$\pm$4.11\\
   &  &2018-01-30 12:33:42 & 2458149.028 & 3$\times$600 & 0.605 & --19.70$\pm$4.34 & -21.63$^{+4.34}_{-4.35}$\\
   &  &2018-02-03 14:39:44 & 2458153.115 & 2$\times$1200 & 0.628 & -16.60$\pm$4.09 & -19.75$^{+4.14}_{-4.15}$\\
   &  &2018-03-02 12:07:00 & 2458180.007 & 3$\times$1200 & 0.780 & 8.72$\pm$4.47 & 3.66$^{+4.58}_{-4.59}$\\
   &  &2018-03-08 12:29:43 & 2458186.022 & 1200 & 0.814 & 11.06$\pm$4.22 & 8.74$^{+4.25}_{-4.24}$\\
   &  &2018-11-24 18:51:44* & 2458447.291 &	5$\times$1200 & 0.294 & 6.21$\pm$4.11 & 6.21$\pm$4.11\\
   &  &2019-03-16 11:55:57 & 2458558.998 & 3$\times$1200 & 0.927 & 52.59$\pm$4.71 & 55.57$\pm$4.71\\
   &  &2019-11-13 18:52:37 & 2458801.291 & 3$\times$1200 & 0.299 & 5.71$\pm$3.77 & 7.51$\pm$3.77\\
   &  &2019-12-10 16:51:53 & 2458828.208 & 4$\times$1200 & 0.452 & -19.17$\pm$4.14 & -22.20$^{+4.16}_{-4.35}$\\
   &  &2020-01-13 14:26:40 & 2458862.107 & 3$\times$1200 & 0.644 & -20.94$\pm$3.92 & -23.04$\pm$3.92\\
   &  &2020-02-11 13:33:23 & 2458891.069 & 3$\times$1200 & 0.808 & 16.76$\pm$4.22 & 17.66$\pm$4.22\\
   &  &2020-11-27 16:55:19 & 2459181.210 & 3$\times$1200 & 0.451 & -17.03$\pm$3.64 & -18.15$\pm$3.66\\
   &  &2021-01-22 14:12:19 & 2459237.097 & 3$\times$1200 & 0.768 & 2.21$\pm$3.93 & 2.05$^{+4.15}_{-4.14}$\\
   &  &2021-02-19 11:44:42 & 2459264.992 & 3$\times$1200 & 0.926 & 54.03$\pm$4.23 & 58.16$\pm$4.23\\
   &  &2021-02-26 12:21:52 & 2459272.018 & 4$\times$1200 & 0.965 & 68.88$\pm$3.48 & 72.72$\pm$3.48\\
   &  &2021-11-18 17:53:52 & 2459537.250 & 3$\times$1200 & 0.468 & -18.04$\pm$3.92 & -17.99$\pm$3.93\\
   &  &2022-01-12 14:20:15 & 2459592.103 & 3$\times$1200 & 0.778 & 4.06$\pm$3.88 & 3.74$\pm$3.88\\
   &  &2022-02-23 12:56:03 & 2459634.042 & 3$\times$1200 & 0.016 & 74.12$\pm$4.37 & 79.54$\pm$4.37\\
   &  &2023-01-02 15:06:33 & 2459947.135 & 4$\times$1200 & 0.789 & 7.80$\pm$3.99 & 8.58$\pm$3.99\\
   &  &2023-03-07 11:19:27 & 2460010.974 & 3$\times$1200 & 0.151 & 43.88$\pm$4.44 & 46.32$\pm$4.50\\   
  \hline
  \multirow{6}{*}{LAMOST/LRS} &
  \multirow{6}{*}{$\sim$1800}
   &2017-11-18 17:40:19 & 2458076.241 & 8$\times$600 & 0.192 & 34.71$\pm$4.09 & 35.09$\pm$3.38\\
   &  &2017-12-17 15:59:53 & 2458105.172 & 8$\times$600 & 0.356 & -1.39$\pm$4.08 & -0.55$^{+3.37}_{-3.38}$\\
   &  &2017-12-21 15:48:50 & 2458109.165 & 8$\times$600 & 0.379 & -3.93$\pm$4.08 & -3.23$\pm$3.38\\
   &  &2018-01-16 14:05:00 & 2458135.092 & 8$\times$600 & 0.526 & -21.44$\pm$4.09 & -20.35$\pm$3.38\\
   &  &2018-02-12 12:39:38 & 2458162.031 & 8$\times$600 & 0.678 & -8.78$\pm$4.09 & -7.89$\pm$3.38\\
   &  &2018-12-04 17:37:05 & 2458457.239 & 3$\times$600 & 0.350 & -1.83$\pm$4.12 & -1.13$\pm$3.41\\
  \hline
  \multirow{3}{*}{P200/DBSP} & 
  $\sim$5000 &2022-12-18 12:03:42 & 2459932.008 & 300 & 0.704 & -4.87$\pm$4.62 & -4.84$\pm$3.94\\
  &$\sim$2800 &2023-04-11 03:23:53 & 2460045.640 & 60 & 0.347 & -8.58$\pm$4.14 & -7.80$^{+3.58}_{-3.63}$\\
  &$\sim$4500 &2023-09-11 12:42:39 & 2460199.028 & 180 & 0.216 & 28.34$\pm$3.27 & 28.34$\pm$2.91\\
  \hline
  \multirow{1}{*}{2.4-m Telescope/HiRES} &
  \multirow{1}{*}{$\sim$28000}
  &2023-02-27 15:47:25& 2460003.160 & 2000 & 0.107 & 59.31$\pm$4.91 & 59.11$^{+4.05}_{-4.06}$\\
  \hline  
  \end{tabular}\label{tab:single-epoch RVs}
  \vspace{1ex}
  {\raggedright Notes: column(1): facility and instrument; column(2): the resolving power of the spectra; column(3): the UTC time in the mid-time of each exposure; column(4): the Heliocentric Julian Date in the mid-time of each exposure; column(5): total exposure time of each spectrum; column(6): the orbital phase; column(7): the measurement results of RVs by Gaussian-like model fitting; column(8): the measurement results of RVs by CCF (adopted for the following study). 
 \par} 
\end{table}
\clearpage

\begin{table}
\centering
\renewcommand{\arraystretch}{0.95}	
\centering
  \caption{RVs respectively measured by the wing below half of peak flux of H$\alpha$ emission line (adopted), the entire H$\alpha$ emission line, the wing below two-thirds of peak flux of H$\alpha$ emission line and the wing below one-third of peak flux of H$\alpha$ emission line of ALS\,8814.}
  \centering
  \tiny
  \begin{tabular}{@{}cccccccc@{}}
  \hline
  \thead{\textbf{Facility}\\~} &
  \thead{\textbf{Resolving~power}\\~} & 
  \thead{\textbf{HJD$_{\rm mid}$}\\~} & 
  \thead{\textbf{Phase}\\~} &   \thead{\textbf{RV\_(H$\bm{\alpha})_\mathrm{\frac{1}{2}}$}\\($\mathrm{km~s^{-1}}$)}& 
  \thead{\textbf{RV\_H\bm{$\alpha$}}\\($\mathrm{km~s^{-1}}$)} &  \thead{\textbf{RV\_(H$\bm{\alpha})_\mathrm{\frac{2}{3}}$}\\($\mathrm{km~s^{-1}}$)}& \thead{\textbf{RV\_(H$\bm{\alpha})_\mathrm{\frac{1}{3}}$}\\($\mathrm{km~s^{-1}}$)}
  \\
  \multicolumn{1}{c}{(1)}&
  \multicolumn{1}{c}{(2)}& 
  \multicolumn{1}{c}{(3)}& 
  \multicolumn{1}{c}{(4)}& 
  \multicolumn{1}{c}{(5)}&
  \multicolumn{1}{c}{(6)}&
  \multicolumn{1}{c}{(7)}&  
  \multicolumn{1}{c}{(8)}\\  
  \hline
  \multirow{30}{*}{LAMOST/MRS} &
  \multirow{30}{*}{$\sim$7500}
   & 2458063.355 & 0.119 & 58.68$\pm4.18$ & 56.65$\pm$4.16 & 58.38$\pm4.17$ & 60.73$^{+4.31}_{-4.21}$ \\
   &  & 2458065.338 & 0.131 & 54.40$^{+4.28}_{-4.25}$ & 51.26$\pm$4.23 & 53.00$\pm4.24$ & 56.02$\pm4.29$ \\
   &  & 2458084.256 & 0.238 & 17.22$^{+4.11}_{-4.12}$ & 15.83$\pm$4.11 & 17.24$\pm4.11$ & 18.19$^{+4.12}_{-4.13}$ \\
   &  & 2458086.246 & 0.249 & 14.75$\pm$4.26 & 14.19$\pm$4.26 & 14.93$\pm4.26$ & 14.64$\pm4.27$ \\
   &  & 2458087.310 & 0.255 & 12.03$\pm$3.46 & 11.37$\pm$3.46 & 12.38$\pm3.46$ & 11.83$^{+3.46}_{-3.50}$ \\
   &  & 2458088.226 & 0.260 & 10.38$\pm$4.36 & 10.49$\pm$4.35 & 10.40$\pm4.35$ & 9.22$\pm4.36$ \\
   &  & 2458089.206 & 0.266 & 7.68$^{+3.90}_{-3.88}$ & 8.22$\pm$3.88 & 8.44$\pm3.88$ & 7.08$\pm3.89$ \\
   &  & 2458090.198 & 0.271 & 8.10$\pm$4.02 & 7.89$\pm$4.01 & 8.37$\pm4.01$ & 7.54$\pm4.02$ \\
   &  & 2458114.157 & 0.407 & -13.63$\pm$3.87 & -13.95$\pm$3.86 & -13.12$\pm3.87$ & -14.07$\pm3.89$ \\
   &  & 2458122.183 & 0.453 & -22.00$\pm$3.78 & -20.77$\pm$3.77 & -21.17$\pm3.18$ & -23.53$\pm3.78$ \\
   &  & 2458144.084 & 0.577 & -25.59$\pm$4.11 & -22.39$\pm$4.11 & -23.92$\pm4.11$ & -27.57$^{+4.17}_{-4.12}$ \\
   &  & 2458149.028 & 0.605 & -21.63$^{+4.34}_{-4.35}$ & -19.75$\pm$4.34 & -20.39$^{+4.35}_{-4.36}$ & -22.45$\pm4.35$ \\
   &  & 2458153.115 & 0.628 & -19.75$^{+4.14}_{-4.15}$ & -16.79$^{+4.10}_{-4.09}$ & -18.17$^{+4.12}_{-4.11}$ & -20.27$^{+4.15}_{-4.17}$ \\
   &  & 2458180.007 & 0.780 & 3.66$^{+4.58}_{-4.59}$ & 7.87$^{+4.47}_{-4.48}$ & 6.56$^{+4.51}_{-4.52}$ & 2.99$^{+4.62}_{-4.63}$ \\
   &  & 2458186.022 & 0.814 & 8.74$^{+4.25}_{-4.24}$ & 10.93$\pm$4.22 & 10.21$\pm$4.22 & 8.52$^{+4.25}_{-4.26}$ \\
   &  & 2458447.291 & 0.294 & 6.21$\pm$4.11 & 6.21$\pm$4.11 & 6.21$\pm$4.11 & 6.21$\pm$4.11 \\
   &  & 2458558.998 & 0.927 & 55.57$\pm$4.71 & 52.56$\pm$4.71 & 55.44$\pm$4.71 & 57.63$\pm$4.71 \\
   &  & 2458801.291 & 0.299 & 7.51$\pm$3.77 & 5.63$\pm$3.77 & 6.86$\pm$3.77 & 7.53$\pm$3.77 \\
   &  & 2458828.208 & 0.452 & -22.20$^{+4.16}_{-4.35}$ & -20.08$\pm$4.14 & -21.44$^{+4.15}_{-4.16}$ & -25.36$^{+4.16}_{-4.29}$ \\
   &  & 2458862.107 & 0.644 & -23.04$\pm$3.92 & -21.45$\pm3.92$ & -22.54$\pm3.92$ & -25.65$\pm3.92$ \\
   &  & 2458891.069 & 0.808 & 17.66$\pm$4.22 & 16.46$\pm$4.22 & 17.33$\pm4.22$ & 17.34$\pm4.22$ \\
   &  & 2459181.210 & 0.451 & -18.15$\pm$3.66 & -17.06$^{+3.65}_{-3.64}$ & -16.79$^{+3.65}_{-3.66}$ & -17.73$^{+3.65}_{-3.86}$ \\
   &  & 2459237.097 & 0.768 & 2.05$^{+4.15}_{-4.14}$ & 1.63$\pm$3.94 & 2.26$^{+4.04}_{-4.09}$ & 1.11$^{+4.33}_{-4.24}$ \\
   &  & 2459264.992 & 0.926 & 58.16$\pm$4.23 & 54.44$\pm$4.23 & 57.04$\pm$4.23 & 60.51$^{+4.24}_{-4.50}$ \\ 
   &  & 2459272.018 & 0.965 & 72.72$\pm$3.48 & 68.69$\pm$3.48 & 72.55$^{+3.48}_{-3.68}$ & 76.40$^{+3.79}_{-3.50}$ \\
   &  & 2459537.250 & 0.468 & -17.99$\pm$3.93 & -18.32$\pm$3.92 & -18.27$^{+3.93}_{-3.92}$ & -19.91$^{+4.10}_{-3.95}$ \\
   &  & 2459592.103 & 0.778 & 3.74$\pm$3.88 & 3.99$\pm$3.88 & 4.31$^{+3.88}_{-3.90}$ & 3.51$\pm$3.89 \\
   &  & 2459634.042 & 0.016 & 79.54$\pm$4.37 & 73.97$\pm4.37$ & 77.15$^{+5.51}_{-4.37}$ & 82.98$\pm$4.37 \\
   &  & 2459947.135 & 0.789 & 8.58$\pm$3.99 & 8.02$\pm$3.99 & 8.14$\pm$3.99 & 8.21$^{+4.02}_{-3.99}$ \\
   &  & 2460010.974 & 0.151 & 46.32$\pm$4.50 & 43.56$^{+4.45}_{-4.44}$ & 45.19$^{+4.46}_{-4.47}$ & 47.98$^{+4.56}_{-4.61}$ \\
  \hline
  \multirow{6}{*}{LAMOST/LRS} &
  \multirow{6}{*}{$\sim$1800}
   & 2458076.241 & 0.192 & 35.92$\pm$4.09 & 35.09$\pm$4.09 & 35.91$\pm$4.09 & 42.13$\pm$4.11 \\
   &  & 2458105.172 & 0.356 & -2.47$^{+4.09}_{-4.11}$ & -0.55$\pm$4.08 & -0.53$\pm$4.08 & -2.24$\pm$4.09 \\
   &  & 2458109.165 & 0.379 & -3.63$\pm$4.09 & -3.23$^{+4.08}_{-4.09}$ & -3.07$\pm$4.09 & -3.58$\pm$4.09 \\
   &  & 2458135.092 & 0.526 & -21.81$\pm$4.09 & -20.35$\pm$4.09 & -21.22$\pm$4.09 & -26.25$\pm$4.09 \\
   &  & 2458162.031 & 0.678 & -7.87$\pm$4.09 & -7.89$\pm$4.09 & -7.89$\pm$4.09 & -7.89$\pm$4.09 \\
   &  & 2458457.239 & 0.350 & -1.12$\pm$4.11 & -1.13$\pm$4.11 & -1.13$\pm$4.11 & -1.13$^{+4.11}_{-4.12}$ \\
  \hline
  \multirow{3}{*}{P200/DBSP} & 
  $\sim$5000 & 2459932.008 & 0.704 & -3.83$\pm$4.63 & -4.84$^{+4.63}_{-4.62}$ & -4.08$\pm$4.63 & -4.27$^{+4.66}_{-4.74}$ \\
  &$\sim$2800 & 2460045.640 & 0.347 & -4.91$^{+4.37}_{-4.34}$ & -7.80$^{+4.25}_{-4.30}$ & -6.59$^{+4.28}_{-4.35}$ & -10.58$^{+4.26}_{-4.36}$ \\
  &$\sim$4500 & 2460199.028 & 0.216 & 28.68$\pm$3.28 & 28.34$\pm$3.27 & 30.38$^{+3.27}_{-3.28}$ & 28.76$^{+3.43}_{-3.31}$ \\
  \hline
  \multirow{1}{*}{2.4-m Telescope/HiRES} &
  \multirow{1}{*}{$\sim$28000}
  & 2460003.160 & 0.107 & 62.40$^{+4.92}_{-4.93}$ & 59.11$^{+4.91}_{-4.92}$ & 60.60$\pm$4.92 & 63.97$^{+4.95}_{-4.94}$ \\
  \hline  
  \end{tabular}\label{tab:RVs for contrast}
  \vspace{1ex}
  \\
  {\raggedright Notes: columns(1)-(4): refer to Extended Data Table~\ref{tab:single-epoch RVs}; column(5): the measurement results of RVs by the wing below half of peak flux of H$\alpha$ emission line (adopted); column(6): the measurement results of RVs by the entire H$\alpha$ emission line; column(7): the measurement results of RVs by the wing below two-thirds of peak flux of H$\alpha$ emission line; column(8): the measurement results of RVs by the wing below one-third of peak flux of H$\alpha$ emission line. 
 \par} 
\end{table}
\clearpage

\begin{table}
\centering
\renewcommand{\arraystretch}{1.6}	
\centering
  \caption{Spectral observations of ALS\,8814 and seven BCD standard stars. Targets marked by stars are spectroscopic standard stars for flux calibration purpose.}
  \centering
  \scriptsize
  \begin{tabular}{@{}cccccc@{}}
  \hline
  \thead{\textbf{Target}\\~} &
  \thead{\textbf{SpT/LC}\\~} &
  \thead{\bm{{$\lambda_{1}$}}\\($\mathrm{\AA}$)} &
  \thead{\textbf{D}\\(dex)}&
  \thead{\textbf{Slit width}\\(")} &	  
  \thead{\textbf{Exposure time}\\(s)}\\
  \multicolumn{1}{c}{(1)}&
  \multicolumn{1}{c}{(2)}& 
  \multicolumn{1}{c}{(3)}& 
  \multicolumn{1}{c}{(4)}& 
  \multicolumn{1}{c}{(5)}& 
  \multicolumn{1}{c}{(6)}\\
  \hline
  \multirow{2}{*}{HD\,13267} &
  \multirow{2}{*}{B6~Ia}& 
  \multirow{2}{*}{22.61} &
  \multirow{2}{*}{0.124}& 
  7.0 & 2$\times$10\\
  & & & &1.8 & 2$\times$15 + 2$\times$30\\
  \hline
  \multirow{2}{*}{HD\,23850} &
  \multirow{2}{*}{B7~III}& 
  \multirow{2}{*}{40.80} &
  \multirow{2}{*}{0.372}& 
  7.0 & 2$\times$2\\
  & & & &1.8 & 2$\times$8 + 2$\times$4\\
  \hline
  \multirow{2}{*}{HD\,30614} &
  \multirow{2}{*}{B0.5~Ia}& 
  \multirow{2}{*}{32.79} &
  \multirow{2}{*}{0.070}& 
  7.0 & 2$\times$5\\
  & & & &1.8 & 2$\times$8 + 2$\times$6\\
  \hline
  \multirow{2}{*}{HD\,3360} &
  \multirow{2}{*}{B2~IV}& 
  \multirow{2}{*}{40.88} &
  \multirow{2}{*}{0.167}& 
  7.0 & 2$\times$1 + 2$\times$0.5\\
  & & & &1.8 & 2$\times$4 + 2$\times$2\\
  \hline
  \multirow{2}{*}{HD\,3369} &
  \multirow{2}{*}{B5~V}& 
  \multirow{2}{*}{49.70} &
  \multirow{2}{*}{0.291}& 
  7.0 & 2$\times$4\\
  & & & &1.8 & 2$\times$10 + 2$\times$5\\
  \hline
  \multirow{2}{*}{HR\,1122} &
  \multirow{2}{*}{B5~IIIe}& 
  \multirow{2}{*}{40.16} &
  \multirow{2}{*}{0.293}& 
  7.0 & 2$\times$1\\
  & & & &1.8 & 2$\times$6 + 2$\times$3\\
  \hline
  \multirow{2}{*}{HR\,533} &
  \multirow{2}{*}{B2~Ve}& 
  \multirow{2}{*}{61.93} &
  \multirow{2}{*}{0.152}& 
  7.0 & 2$\times$5\\
  & & & &1.8 & 2$\times$15 + 2$\times$10\\
  \hline
  \multirow{2}{*}{ALS\,8814} &
  \multirow{2}{*}{B1V}& 
  \multirow{2}{*}{62.43} &
  \multirow{2}{*}{0.122}& 
  7.0 & 400\\
  & & & &1.8 & 800 + 400\\
  \hline
  \multirow{1}{*}{HR\,753*} &
  \multirow{1}{*}{-}& 
  \multirow{1}{*}{-} &
  \multirow{1}{*}{-}& 
  7.0 & 2$\times$3\\
  \hline
  \multirow{1}{*}{BD+25d4655*} &
  \multirow{1}{*}{-}& 
  \multirow{1}{*}{-} &
  \multirow{1}{*}{-}& 
  7.0 & 300\\
  \hline
  \multirow{1}{*}{BD+75d325*} &
  \multirow{1}{*}{-}& 
  \multirow{1}{*}{-} &
  \multirow{1}{*}{-}& 
  7.0 & 300\\
  \hline
  \end{tabular}\label{tab:BCD_obs}
  \vspace{1ex}
  
  {\raggedright Notes: column(1): target name; column(2): spectral type and luminosity classification; column(3): position of the Balmer discontinuity; column(4): depth of the Balmer discontinuity; column(5): slit width used in exposure; column(6): total exposure time. 
 \par}  
\end{table}
\clearpage

\begin{table}
\centering
\renewcommand{\arraystretch}{1.6}	
\centering
  \caption{Photometric data imported for SED. Photometric data from \emph{Gaia} DR3, \emph{TESS} and Panoramic Survey Telescope and and Rapid Response System 1 (PS1) are extracted from \texttt{Vizier}\cite{Ochsenbein2000} (I/355/gaiadr3\cite{Gaia2022}, IV/38/tic\cite{Stassun2019} and II/349/ps1\cite{Chambers2017}, respectively), while photometric data from TYCHO and \emph{Swift} are extracted from VOSA\cite{Bayo2008} (\url{http://svo2.cab.inta-csic.es/theory/vosa/}) and specialized observation by \emph{Swift} by our proposal, respectively. Besides, Z$_{\rm p}$ can be found at \url{http://svo2.cab.inta-csic.es/theory/fps/}.} 
  \centering
  \scriptsize
  \begin{tabular}{@{}cccccc@{}}
  \hline
  \thead{\textbf{Survey}\\~} &
  \thead{\textbf{Band}\\~} &
  \thead{\textbf{Magnitude}\\~} &	  
  \thead{\textbf{Flux}\\(erg s$^{-1}$ cm$^{-2}$ \AA$^{-1}$)} &	
  \thead{\textbf{Magnitude system}\\~} &
  \thead{\textbf{Z\bm{$_{\rm p}$}}\\(erg s$^{-1}$ cm$^{-2}$ \AA$^{-1}$)}\\
  \multicolumn{1}{c}{(1)}&
  \multicolumn{1}{c}{(2)}& 
  \multicolumn{1}{c}{(3)}& 
  \multicolumn{1}{c}{(4)}& 
  \multicolumn{1}{c}{(5)}& 
  \multicolumn{1}{c}{(6)}\\
  \hline
  \multirow{1}{*}{\emph{Swift}/XRT} &
   UVM2& - & $(7.67\pm0.11)\times10^{-14}$ & - & -\\  
  \hline
  \multirow{3}{*}{\emph{Gaia} DR3} &
   BP&$10.33\pm0.01$& - & Vega & $4.08\times10^{-9}$\\
   &G&$9.96\pm0.01$& - & Vega & $2.50\times10^{-9}$\\ 
   &RP & $9.38\pm0.02$ & - & Vega & $1.27\times10^{-9}$\\
  \hline
  \multirow{2}{*}{TYCHO} &
   B&$11.01\pm0.06$& $(2.56\pm0.14)\times10^{-13}$ & Vega & $6.51\times10^{-9}$\\
   &V&$10.33\pm0.05$& $(2.93\pm0.12)\times10^{-13}$ &	Vega & $4.00\times10^{-9}$\\
  \hline
  \multirow{1}{*}{{TESS}} & 
  T&$9.62\pm0.25$& - & Vega & $1.33\times10^{-9}$\\
  \hline
  \multirow{4}{*}{PS1} &
   g & red$10.36\pm0.04$ & - & AB & $4.63\times10^{-9}$\\
   & r & $9.99\pm0.03$ & - & AB & $2.83\times10^{-9}$\\ 
   & z & $9.72\pm0.02$ & - & AB & $1.45\times10^{-9}$\\
   & y & $9.71\pm0.03$ & - & AB & $1.17\times10^{-9}$\\
   \hline
  \end{tabular}\label{tab:photometry_data}
  \vspace{1ex}

  {\raggedright Notes: column(1): survey name; column(2): band name; column(3): magnitude of the corresponding band; column(4): flux of the corresponding band; column(5): magnitude system; column(6): zero point of flux of the corresponding magnitude system. 
 \par} 
\end{table}
\clearpage

\renewcommand{\figurename}{Supplementary Figure}
\setcounter{figure}{0}
\begin{figure*}
\begin{center}
\includegraphics[width=1\textwidth]{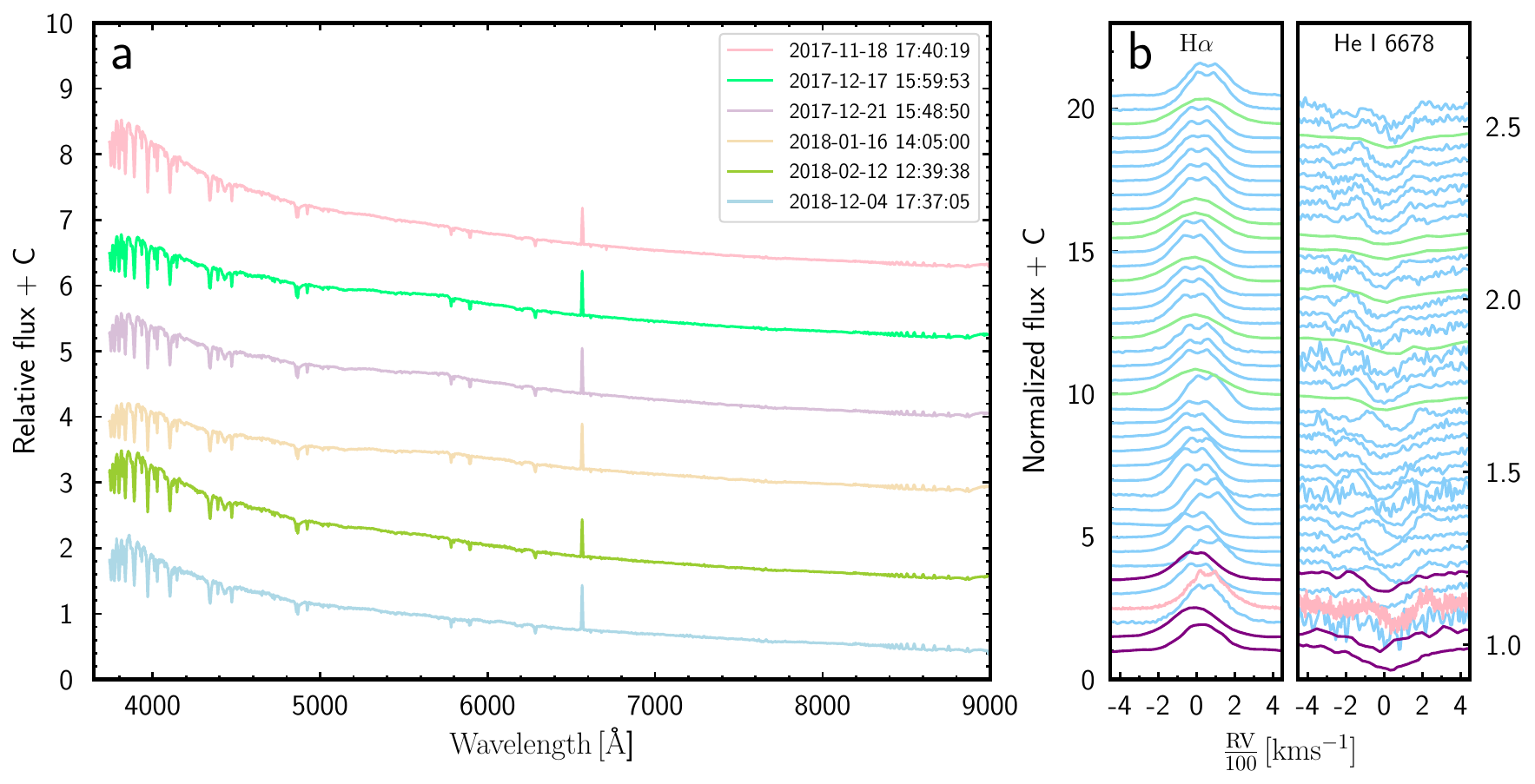}
\caption{{\bf Optical Spectra of ALS\,8814.} {\bf a,} Six spectra from LAMOST/LRS ($R\approx1800$). Mid-time of each observation is listed in the top-right corner. Spectral profiles suggest that the visible star is hot B-type star with significant emission in H$\alpha$. {\bf b,} All the 40 spectra (light sky blue for LAMOST/MRS, light green for LAMOST/LRS, purple for P200/DBSP and light pink for 2.4-m Telescope/HiRES) for H$\alpha$ emission lines and He I 6678 stellar absorption lines. These spectra are organized chronologically, with the upper ones representing earlier observations. Significant RV variations occur in the same direction for both the H$\alpha$ emission lines and the He I 6678 stellar absorption lines.
To avoid overlap, arbitrary offsets are applied in both panels. All observed wavelengths are corrected to the heliocentric vacuum reference frame.}
\label{fig:spectra}
\end{center}
\end{figure*}
\clearpage
\begin{figure*}
\begin{center}
\includegraphics[width=1\textwidth]{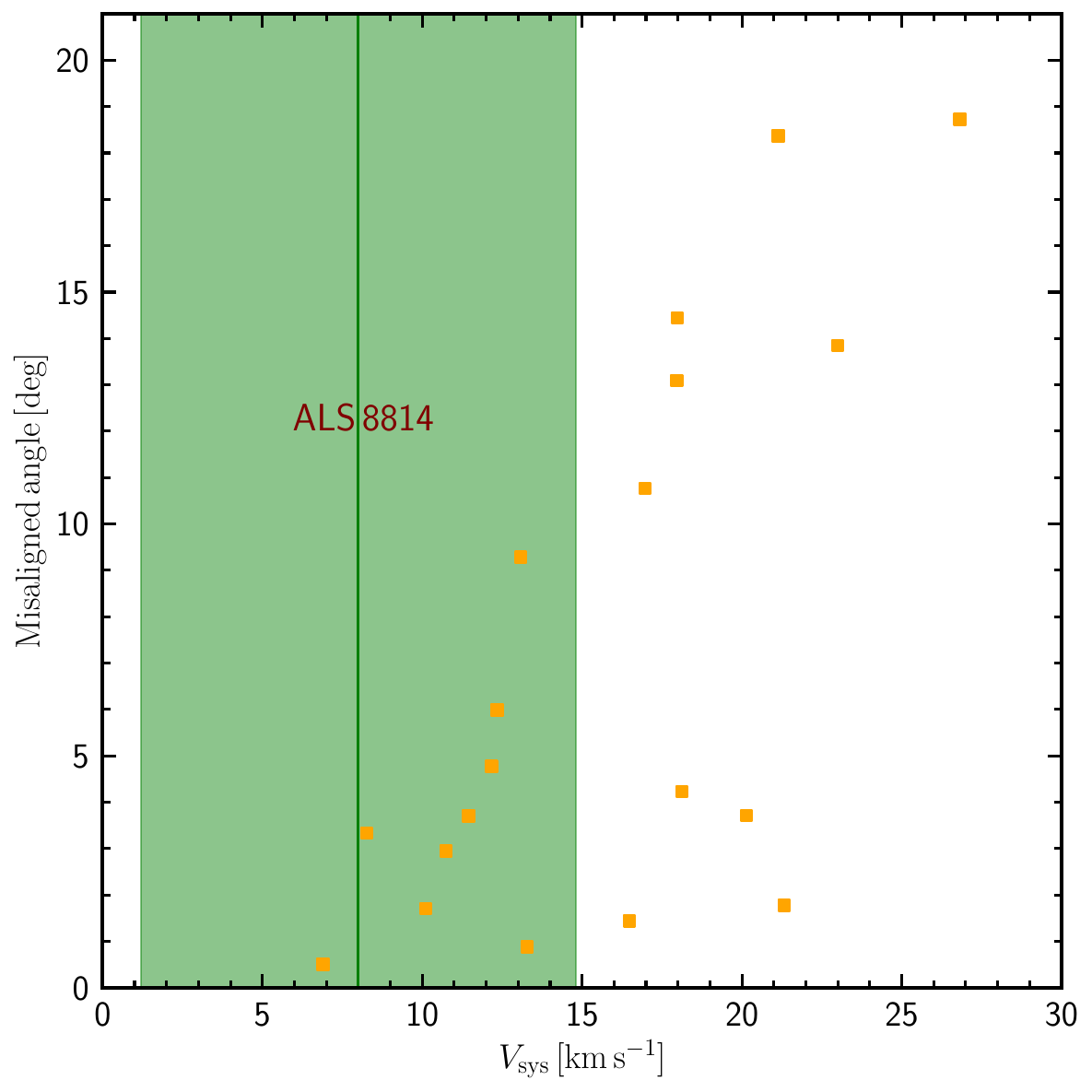}
\caption{{\bf Misaligned angle versus systematic velocity in binary population synthesis modeling.} The orange points represent the ALS\,8814-like binaries (selected by mass and orbital properties) in our {\tt POSYDON} BPS model. Green vertical line and green shaded region indicate the median value and 68.26$\%$ confidence level of systemic velocity for ALS\,8814, respectively.}
\label{fig:kick}
\end{center}
\end{figure*}
\clearpage
\begin{figure*}
\begin{center}
\includegraphics[width=1\textwidth]{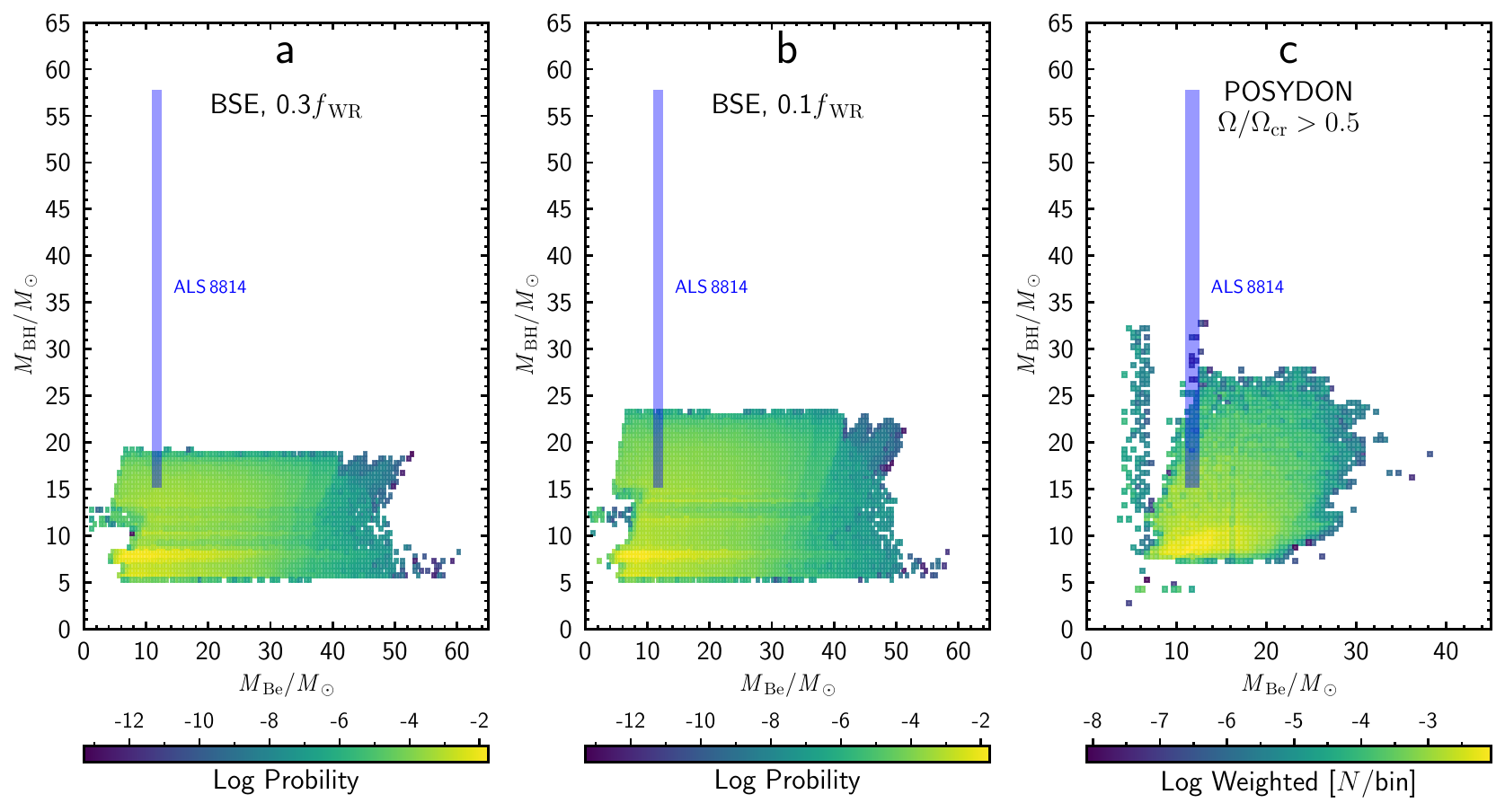}
\caption{{\bf Number distributions of Galactic detached BH–Be binaries in BH mass and companion mass plane.} \textbf{a,} This is predicted by BSE simulations using a 0.3$f_{\rm WR}$, where $f_{\rm WR}$ is the usually adopted  mass-loss rate of Wolf–Rayet stars\cite{2010ApJ...714.1217B}. The colours in each pixel are scaled according to the corresponding numbers. \textbf{b,} The same as \textbf{a}, but predicted for 0.1$f_{\rm WR}$. \textbf{c}, Modeled with the \texttt{POSYDON} BPS code selecting systems with rapidly rotating Be companions ($\Omega/\Omega_\mathrm{cr}>0.5$); here, colours in each pixel show systems weighted by their lifetime in the detached phase as defined in Ref.\cite{2024ApJ...971..133R}. In all the three panels, the blue shaded regions represent the mass range estimates of ALS\,8814.}
\label{fig:model}
\end{center}
\end{figure*}
\clearpage
\begin{figure*}
\begin{center}
\includegraphics[width=1\textwidth]{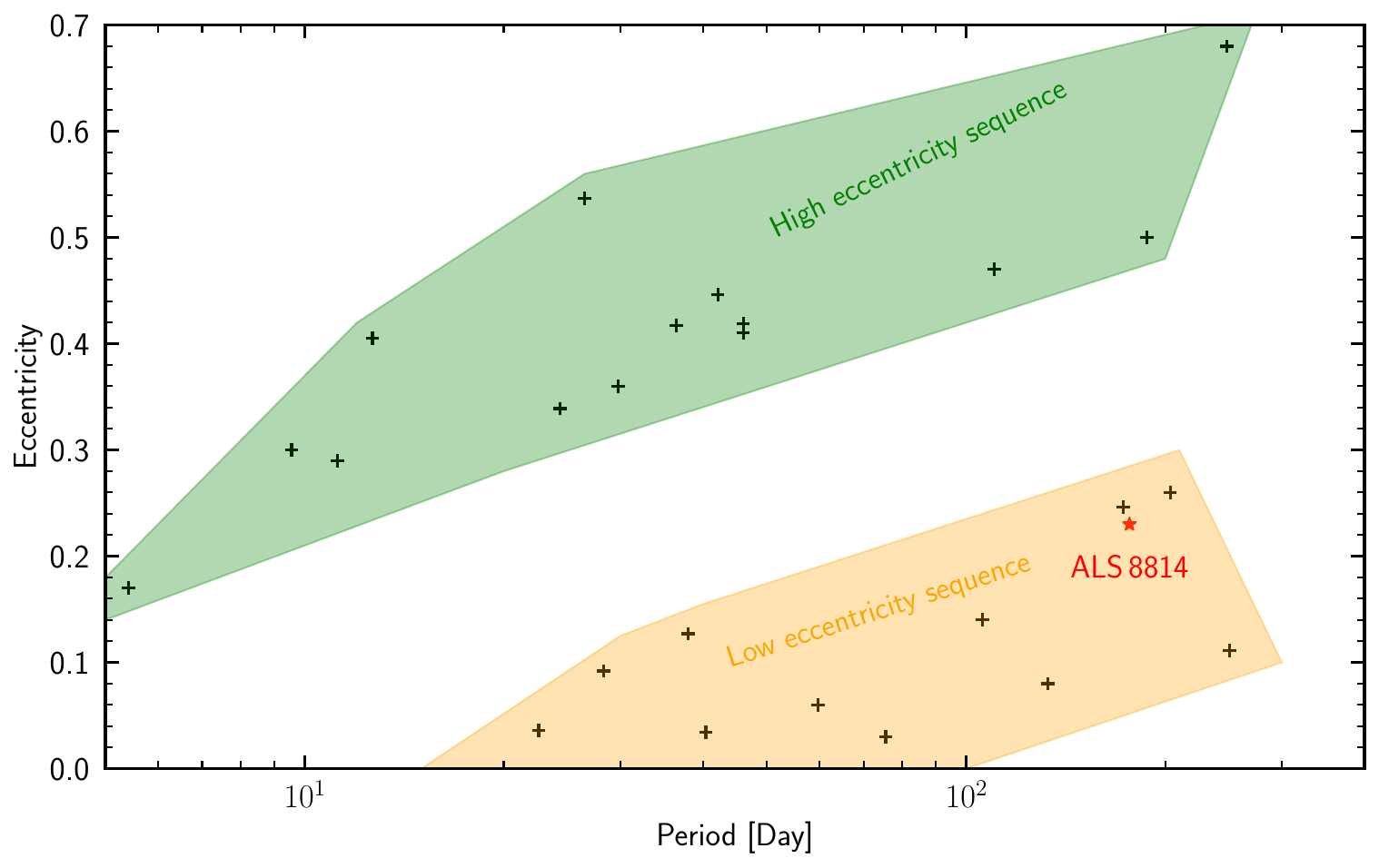}
\caption{{\bf Eccentricity -- orbital period for the known Be high-mass X-ray binaries\cite{2023A&A...671A.149F}.} The orange and green shaded regions respectively represent the low-eccentricity and high-eccentricity sequences, defined by the known Be high-mass X-ray binaries (marked by crosses). ALS\,8814 falls on the low-eccentricity sequence.}
\label{fig:periodecc}
\end{center}
\end{figure*}
\clearpage

\begin{figure*}
\begin{center}
\includegraphics[width=1\textwidth]{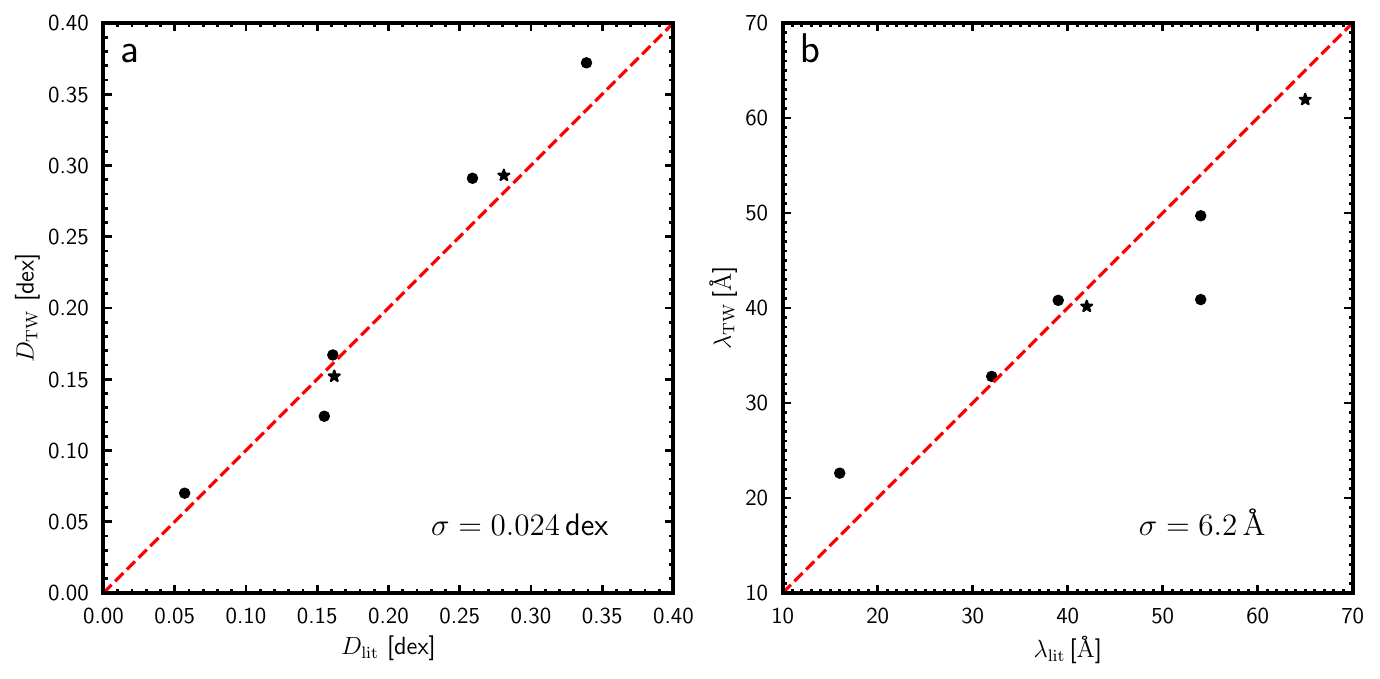}
\caption{\textbf{Comparison of BCD parameters between this work and the literature\cite{2009A&A...501..297Z, Gkouvelis2016}}. {\bf a,} Comparison for the  Balmer jump depth $D$. {\bf b,} Comparison for the  position of the Balmer discontinuity $\lambda_1$. The scatters of the differences between this work and the literature are marked in each panel. In both panels, five circles represent B-type stars, while the two pentagrams represent Be stars.}
\label{fig:bcd_comp}
\end{center}
\end{figure*}
\clearpage

\begin{figure*}
\begin{center}
\includegraphics[width=1\textwidth]{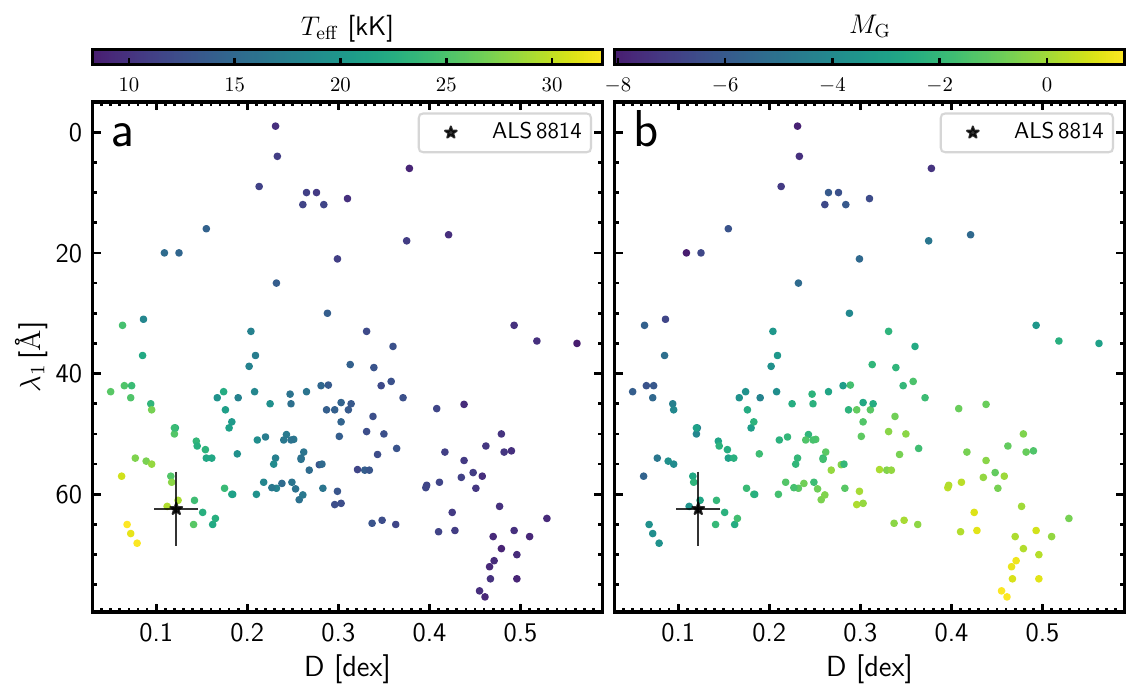}
\caption{{\bf a,} Comparison of ALS\,8814 (black star and solid lines) to empirical spectra in the plane of the position of the Balmer discontinuity $\lambda_{1}$ versus the depth of the Balmer discontinuity $D$. Colorful data points calculated by empirical spectra indicate the effective temperature. {\bf b,} The same as {\bf a} but colorful data points exhibit different absolute $G$ magnitude.}
\label{fig:BCD2}
\end{center}
\end{figure*}
\clearpage

\begin{figure*}
\begin{center}
\includegraphics[width=1\textwidth]{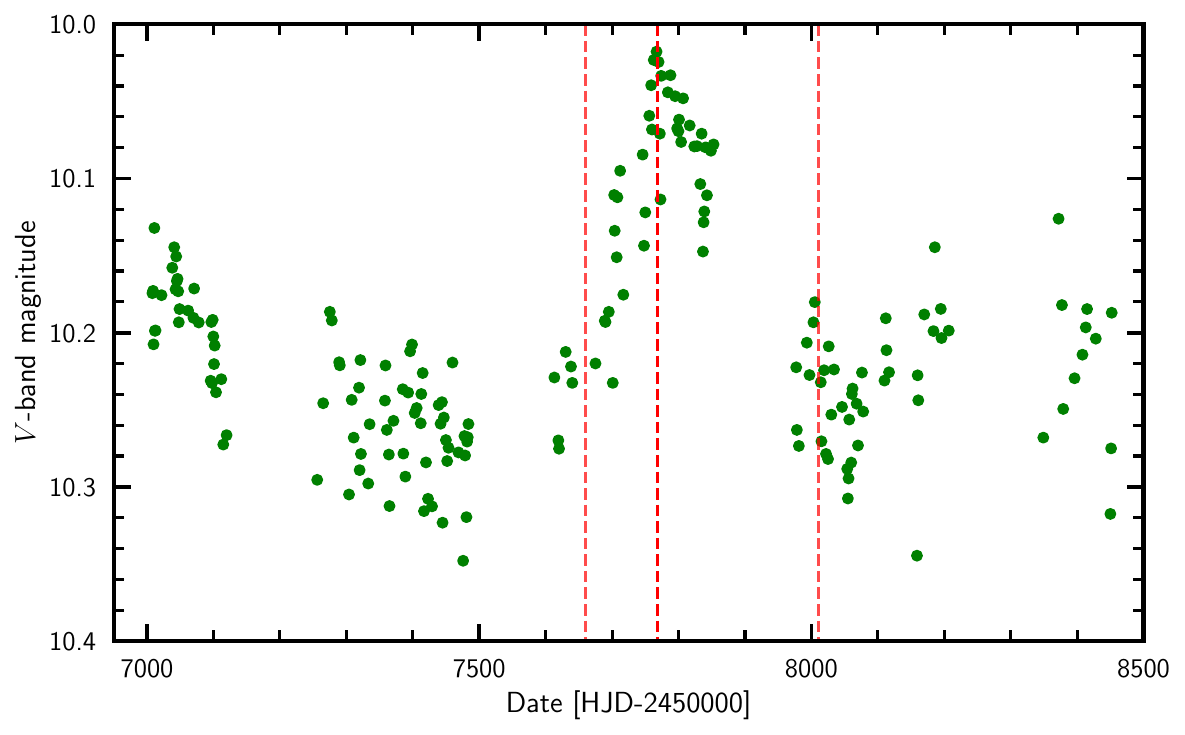}
\caption{{\bf The $V$-band light curve of ALS\,8814 from ASAS-SN.} The deep red dashed line indicates the peak of the outburst, while the two light red dashed lines indicate the start and end of the outburst, respectively.}
\label{fig:lc}
\end{center}
\end{figure*}
\clearpage

\begin{figure*}
\begin{center}
\includegraphics[width=1\textwidth]{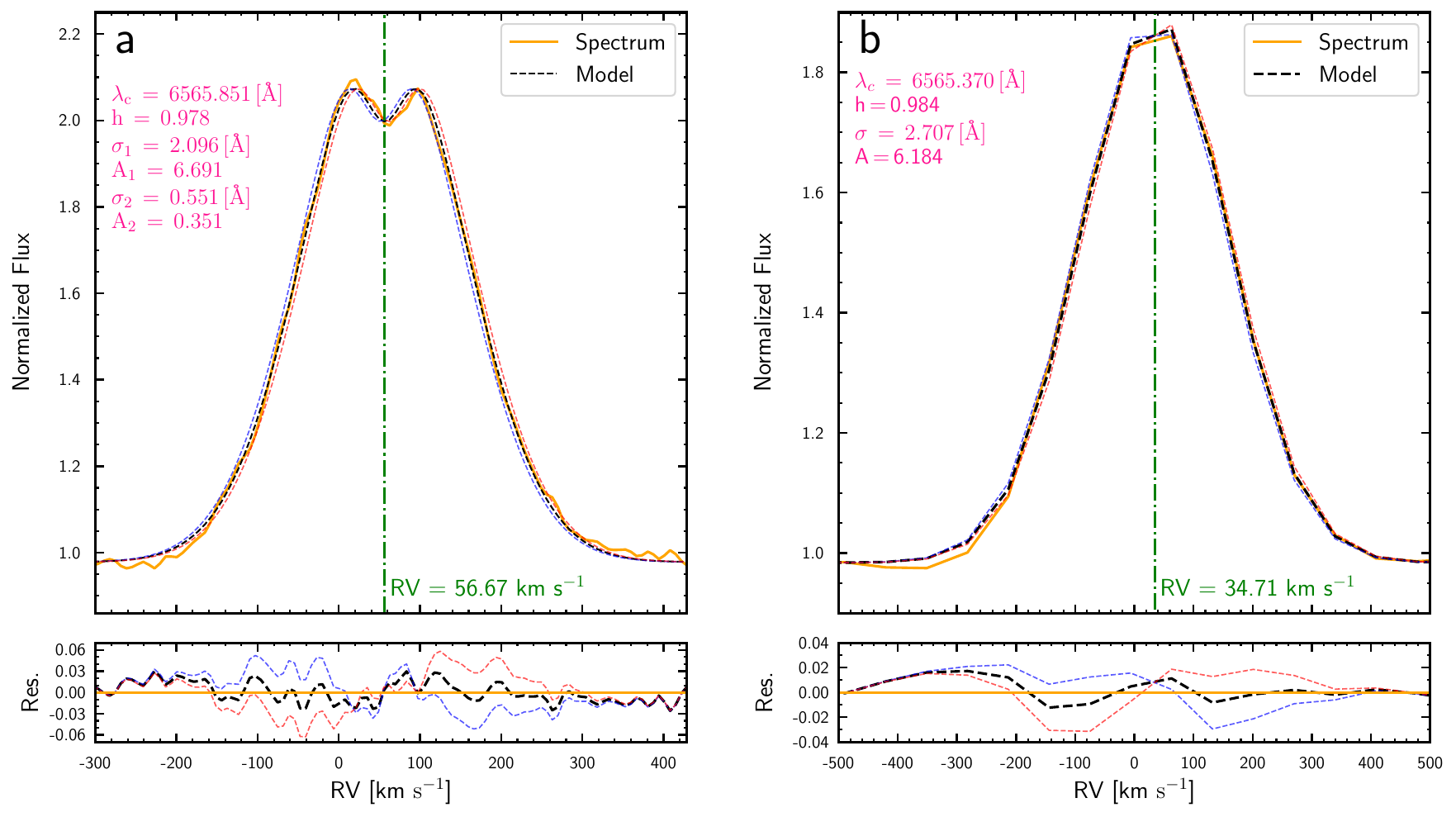}
\caption{{\bf a,} Upper: the comparison between the reference spectrum used by CCF (orange line) from the LAMOST/MRS taken on November 5, 2017 and model (black dashed line) described in Eq.~\ref{eq4}. The fitting results of each parameter are listed in the upper-left corner. The green vertical dashed line indicates the RV measured by this spectrum with value marked in the bottom. 
The blue dashed and red dashed lines correspond, respectively, to the best-fit models incorporating an additional blue/redshift of 5 km s$^{-1}$. 
Lower: the fitting residuals for the three models. It is obvious that residuals for two blue/redshited models are significantly higher than that of the best-fit model. {\bf b,} The same as the left panel, but for the spectrum taken from the LAMOST/LRS observed on November 18, 2017. For this low-resoltion spectrum, only one Gaussian component of Eq.~\ref{eq4} is adopted.}
\label{fig:RV_measurement}
\end{center}
\end{figure*}
\clearpage

\begin{figure*}
\begin{center}
\includegraphics[width=0.95\textwidth]{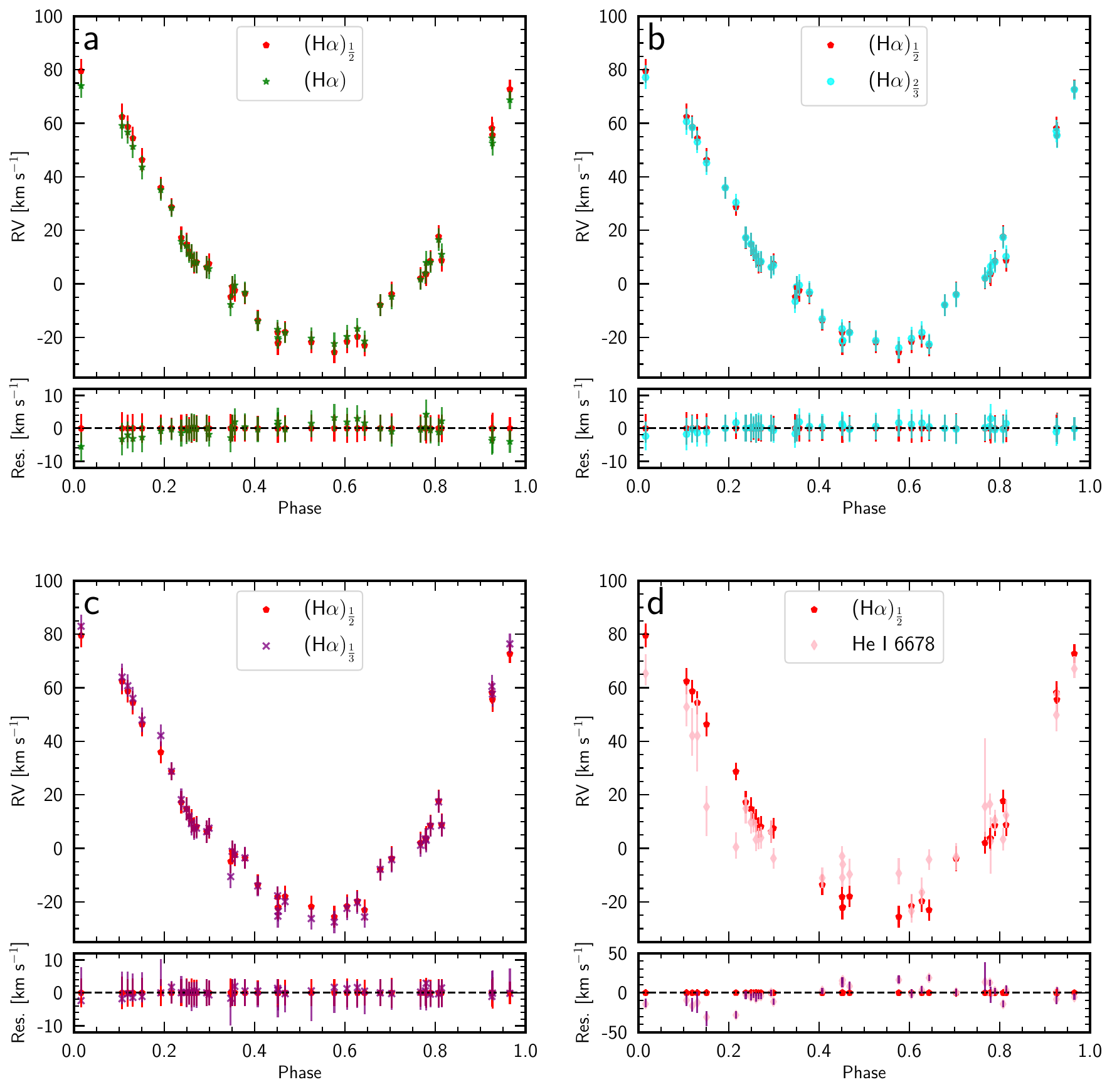}
\caption{{\bf Contrast of RVs measured by different methods.} {\bf a,} Upper: RVs measured by the wing below half of peak flux of H$\alpha$ emission line and the full H$\alpha$ emission line, respectively. Each set of RV measurements is represented by points with unique styles, as indicated in the legend for reference. Lower: the discrepancy between two sets of RVs. {\bf b-d,} The same as {\bf a}, but for contrast of RVs between other sets. Please note: He I 6678 absorption line is not as strong or has as high a SNR as the H$\alpha$ emission line. Consequently, RV measurements derived from low-resolution spectra (with R$<$3,000) using the He I 6678 absorption line are considered unreliable and have therefore not been included in this plot.}
\label{fig:RV_contrast}
\end{center}
\end{figure*}
\clearpage

\begin{figure*}
\begin{center}
\includegraphics[width=1\textwidth]{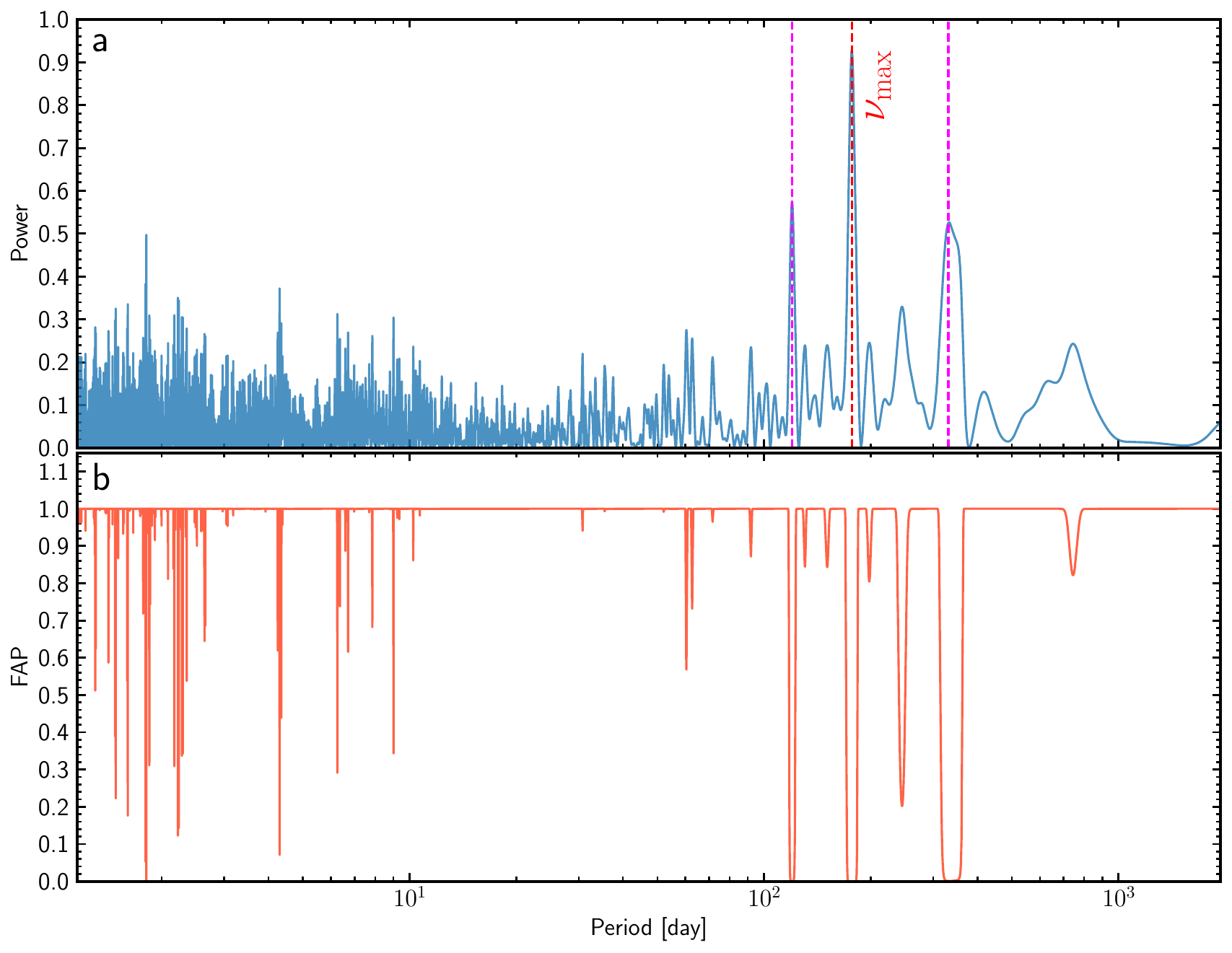}
\caption{{\bf a,} The Lomb-Scargle powers of RV curve. The highest power, marked by red vertical dashed line, indicating the most likely orbital period of $\sim$176.8 days. Additionally, we also mark the second and third highest power, respectively corresponding to $\sim$120.0 days and $\sim$332.5 days, by magenta dashed lines. {\bf b,} The false alarm probability for each sampling points.}
\label{fig:power}
\end{center}
\end{figure*}
\clearpage
\begin{figure*}
\begin{center}
\includegraphics[width=1\textwidth]{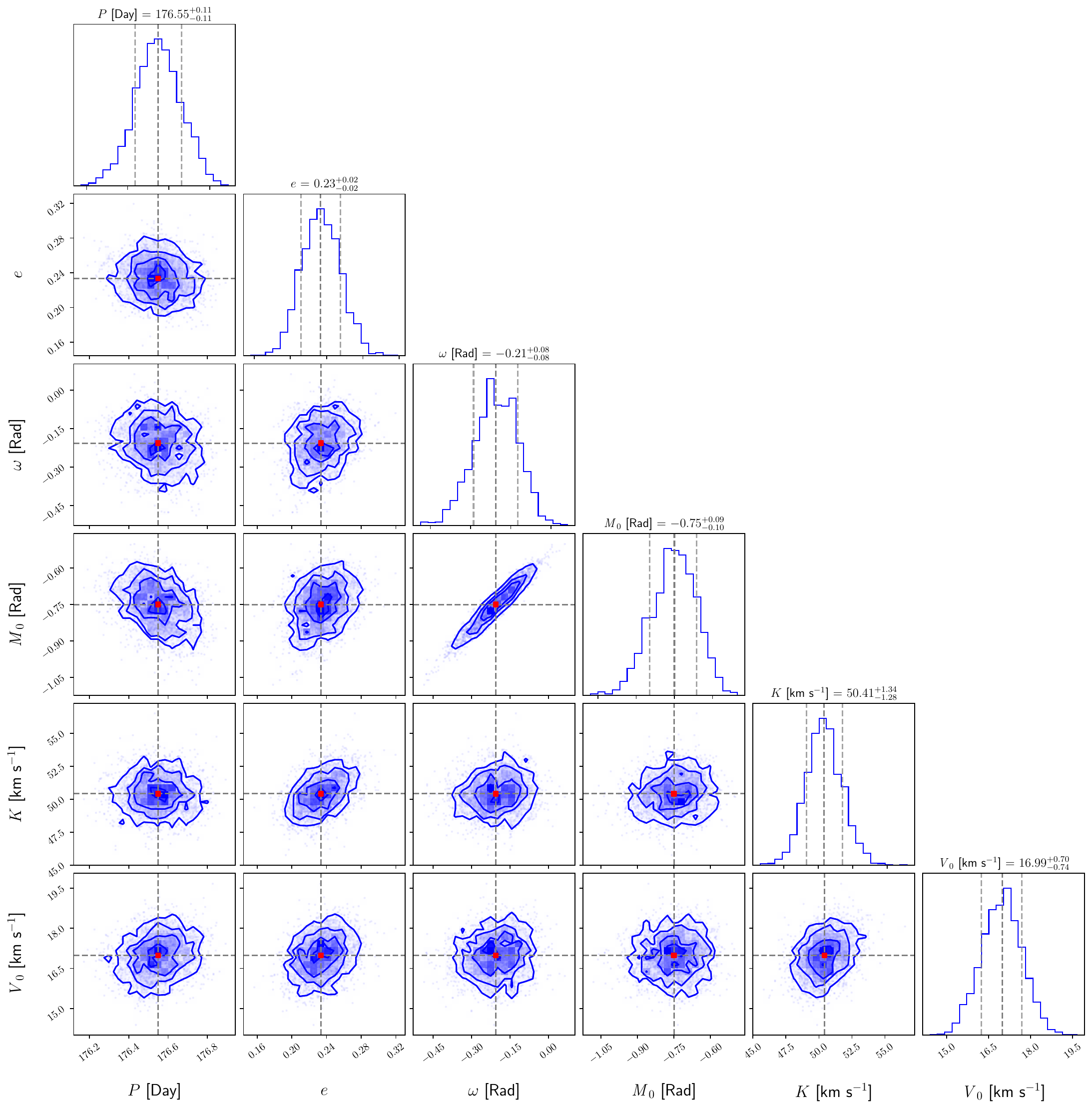}
\caption{{\bf Corner plot for the MCMC sampling of the RV curve.} The histograms indicate the marginal distributions for the fitted orbital parameters and the density plots indicate the joint distributions. In the histograms, median of each parameter and their 1 $\sigma$ uncertainties are indicated by vertical dark gray dashed lines and vertical light gray dashed lines, respectively.}
\label{fig:orbital_corner}
\end{center}
\end{figure*}
\clearpage

\begin{figure*}
\begin{center}
\includegraphics[width=1\textwidth]{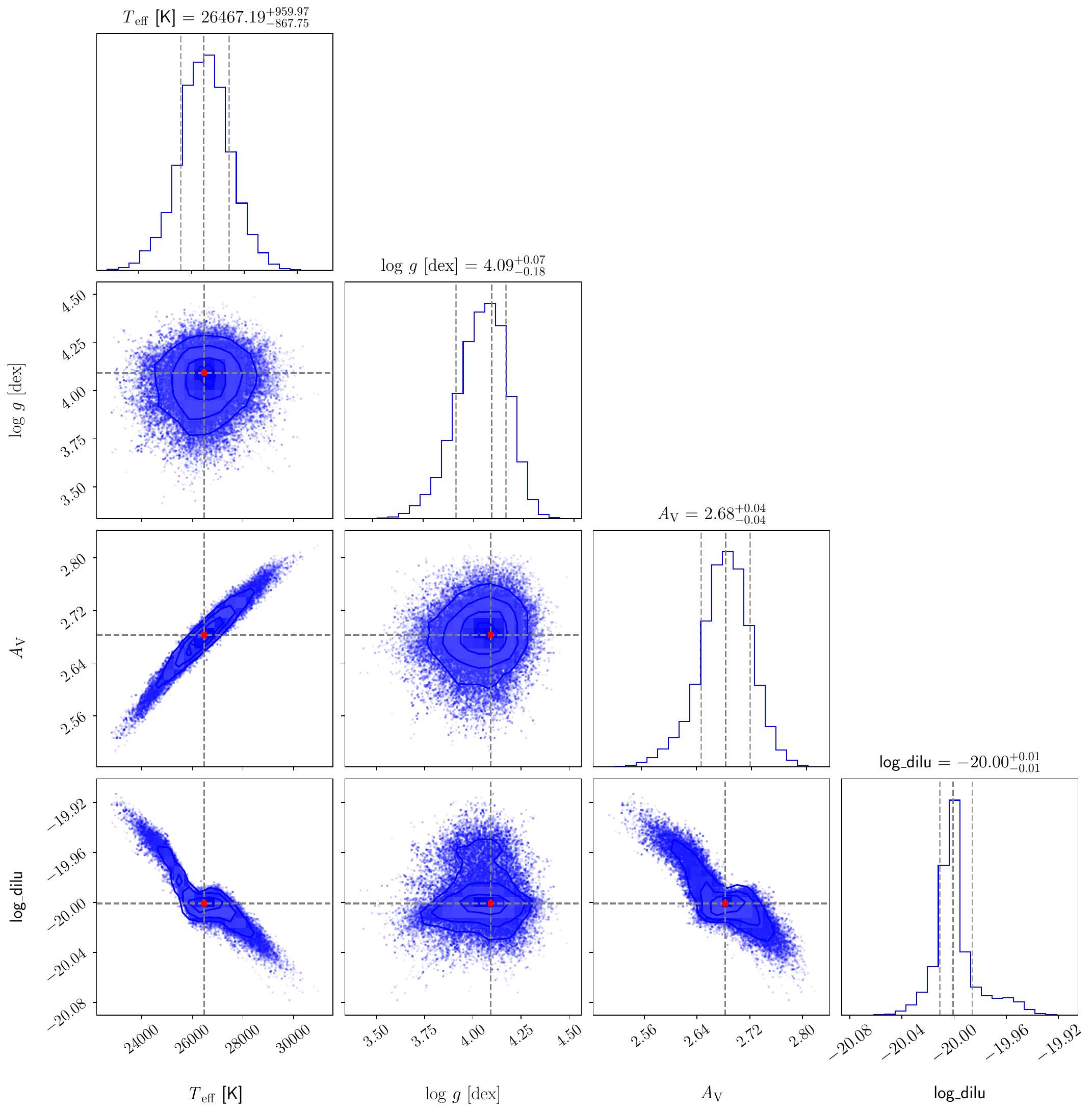}
\caption{{\bf The corner plot same as Supplementary Fig.~\ref{fig:orbital_corner}, but for SED fitting.} Here, we utilize the logarithm of the dilu, log\_dilu, instead of the dilu itself for visibility. [Fe/H] is fixed at 0.}
\label{fig:sed_corner}
\end{center}
\end{figure*}
\clearpage
\begin{figure*}
\begin{center}
\includegraphics[width=1\textwidth]{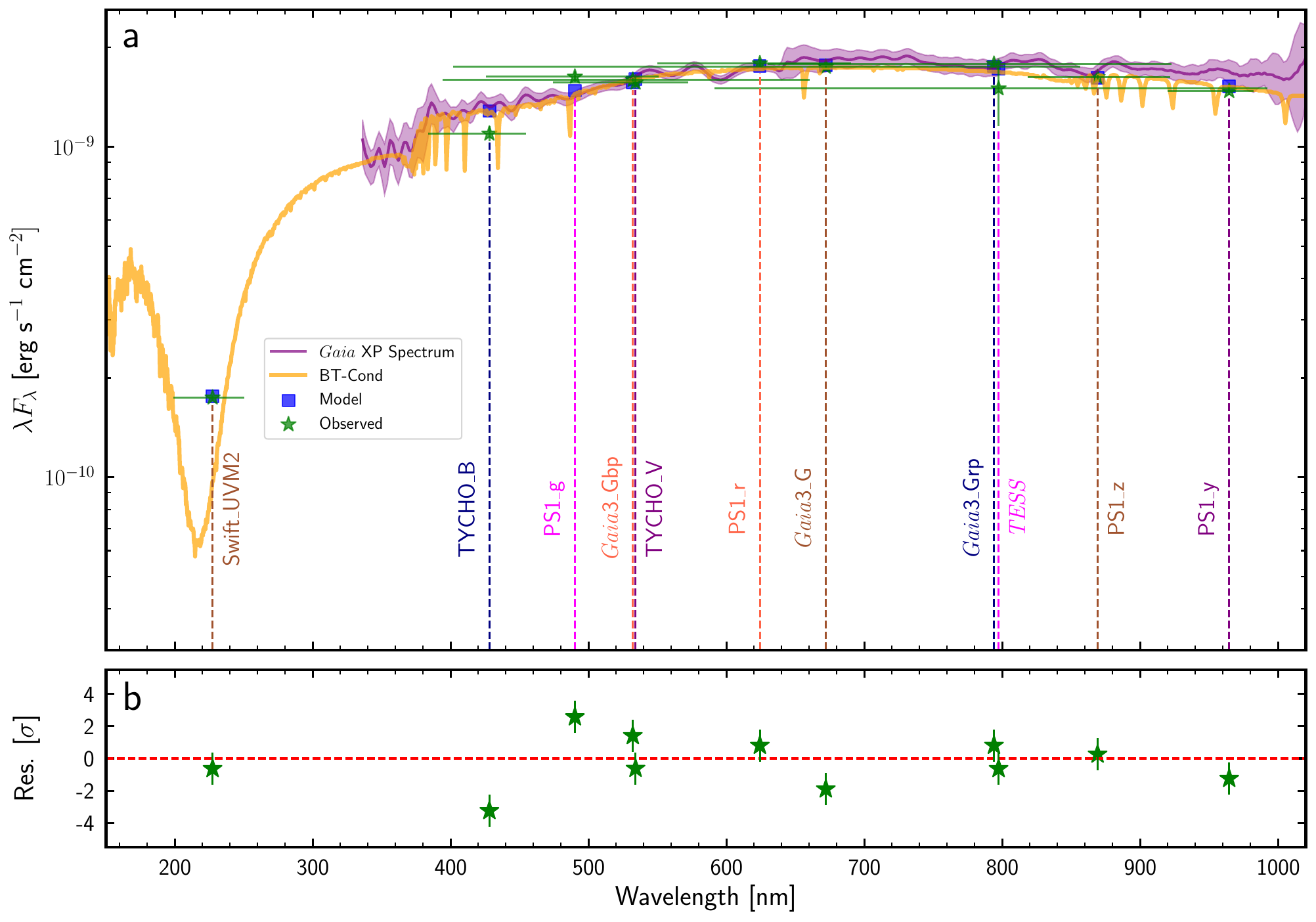}
\caption{{\bf The broaband SED of ALS\,8814.} In both panels, green stars display the observed photometric points. {\bf a,} The orange curve represents the best-fit spectrum for the Be star, while the blue squares depict the model photometric data points. Green horizontal and vertical lines indicate the effective widths of the transmission curves and the 1 $\sigma$ uncertainties associated with the observed photometric data, respectively. Additionally, there are 11 dashed lines in various colours, each corresponding to the central wavelength of a specific filter. The names of the filters are labeled next to the dashed lines in matching text colours. Moreover, the \emph{Gaia} XP Spectrum\cite{Carrasco2021,DeAngeli2023,Montegriffo2023} (purple line) with 1 $\sigma$ uncertainty associated with the flux (light purple shaded region) is also plotted here for comparison. {\bf b,} The fitting residuals, indicated as $\sigma$, i.e., $\frac{{\rm Observed} - {\rm Model}}{\sigma_{\rm observed}}$.}
\label{fig:sed_plot}
\end{center}
\end{figure*}
\clearpage
\begin{figure*}
\begin{center}
\includegraphics[width=1\textwidth]{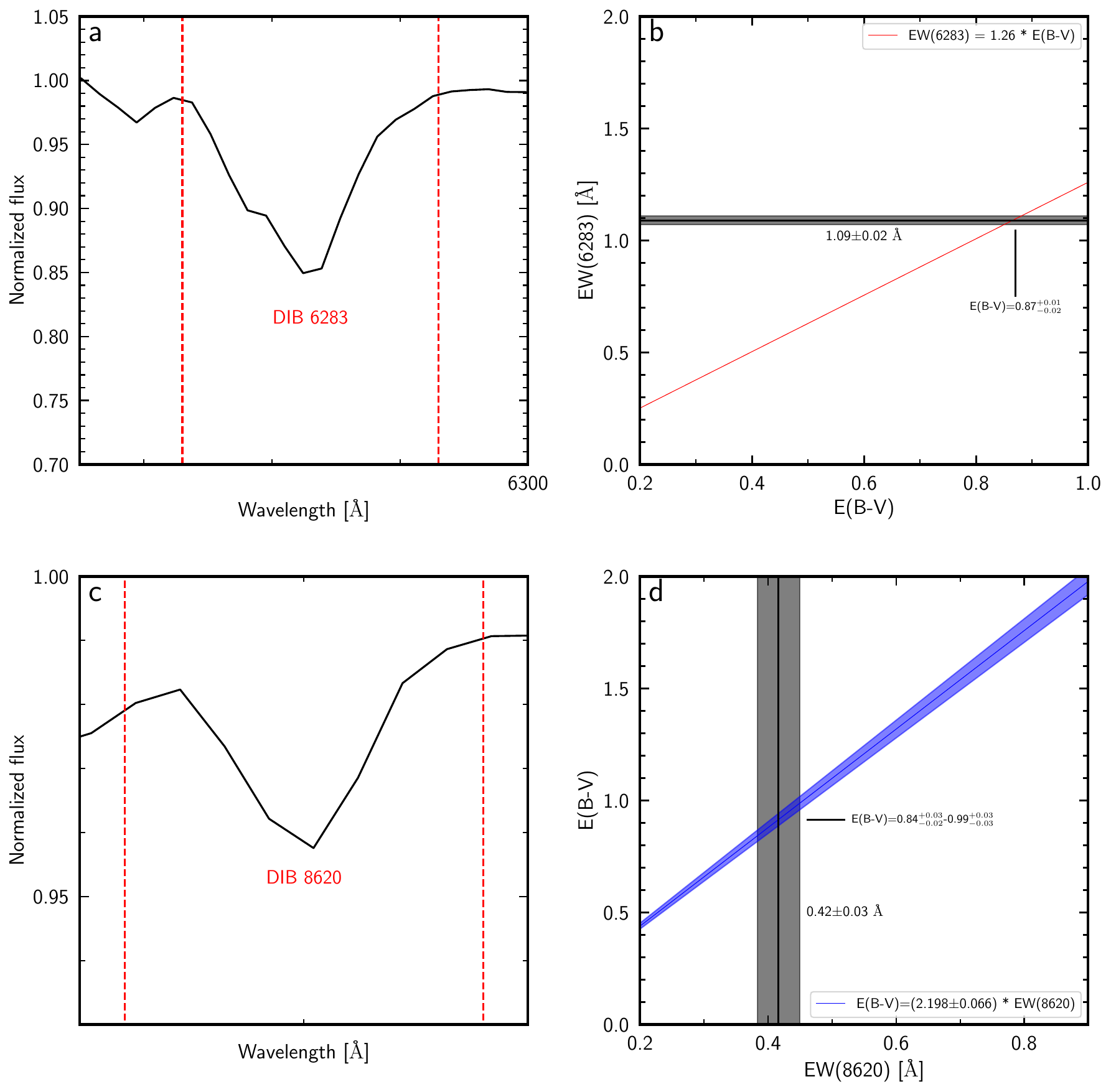}
\caption{{\bf The equivalent widths of DIB 6283 and DIB 8620 versus extinction.} {\bf a,} The LAMOST/LRS spectrum for DIB 6283. DIB 6283 are indicated by two vertical red dashed lines. {\bf b,} The equivalent widths of DIB 6283 versus extinction denoted as E(B-V). Red line is the empirical relation between extinction and the equivalent width of DIB 6283 given by Ref.\cite{Yuan2012}. The black shaded region indicates the equivalent widths of DIB 6283 with 1 $\sigma$ uncertainty. {\bf c-d,} The similar to {\bf a-b}, but for DIB 8620.}
\label{fig:DIB_extinction}
\end{center}
\end{figure*}
\clearpage
\begin{figure*}
\begin{center}
\includegraphics[width=1\textwidth]{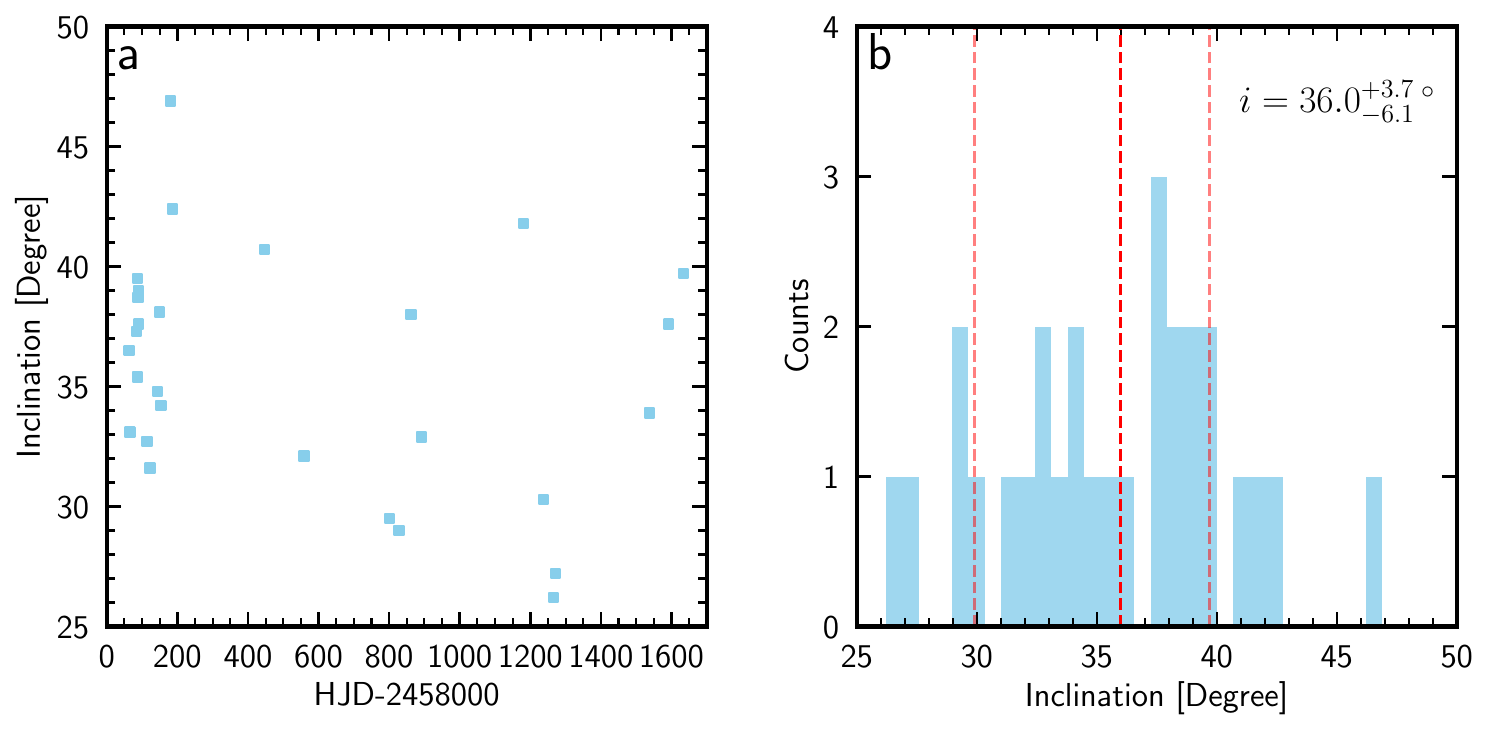}
\caption{{\bf Inclination derived by machine learning method.} {\bf a,} Inclination as a function of Heliocentric Julian date. {\bf b,} Histogram of inclination. Vertical dark red dashed lines and vertical light red dashed lines indicate the median and the 1 $\sigma$ uncertainties of the inclination distribution.}
\label{fig:inc}
\end{center}
\end{figure*}
\clearpage
\begin{figure*}
\begin{center}
\includegraphics[width=1\textwidth]{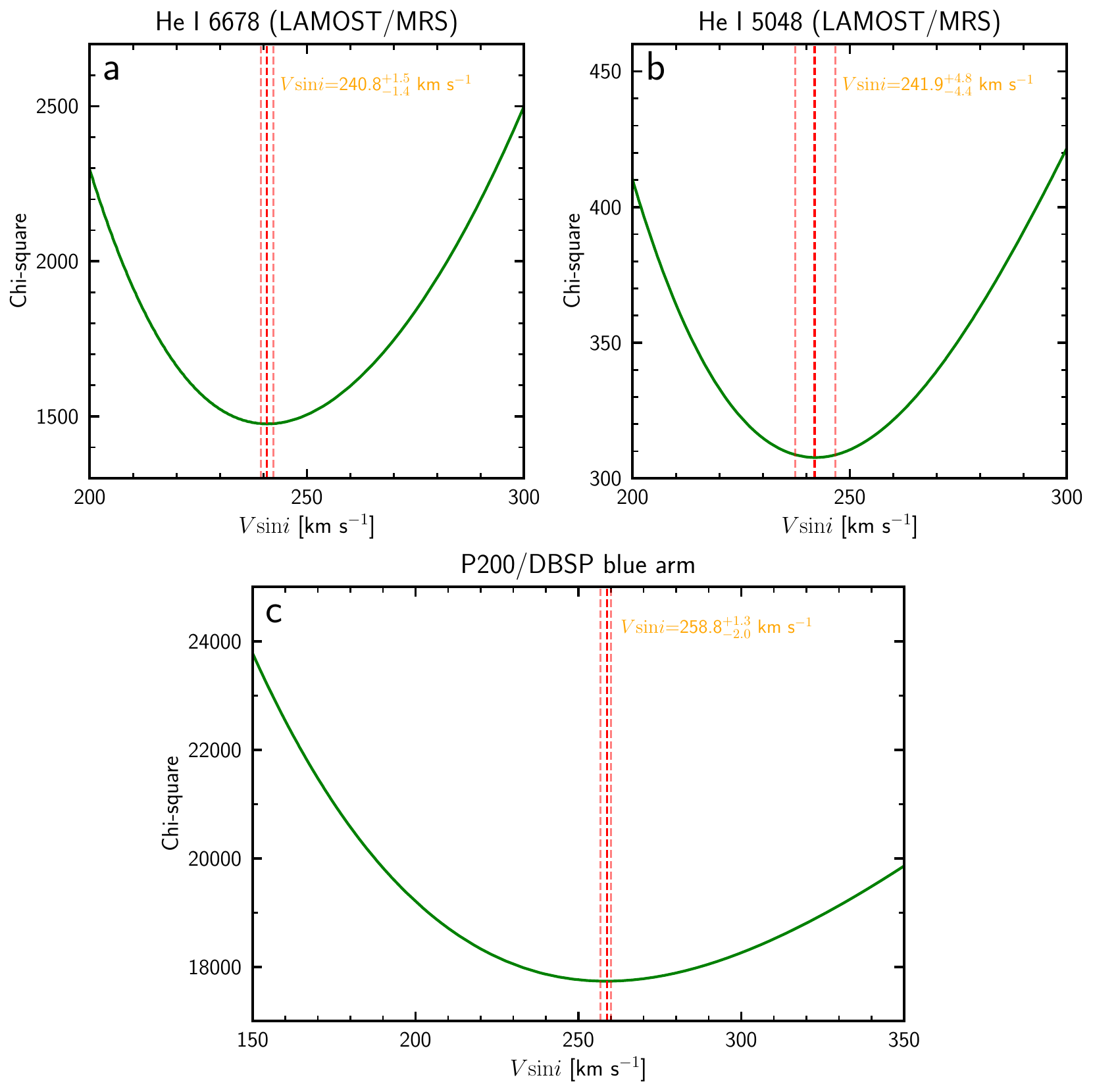}
\caption{{\bf The fitting $\chi^2$ versus \bm{$V\sin i$}.} {\bf a,} The median and the 1 $\sigma$ uncertainties of the measurement result by He I 6678 line of $V\sin i$ are indicated by vertical dark red dashed lines and vertical light red dashed lines, respectively. {\bf b-c,} The same as {\bf a}, but for measurement result by He I 5048 line and the blue arm of P200/DBSP, respectively.}
\label{fig:chisq}
\end{center}
\end{figure*}
\clearpage
\begin{figure*}
\begin{center}
\includegraphics[width=0.95\textwidth]{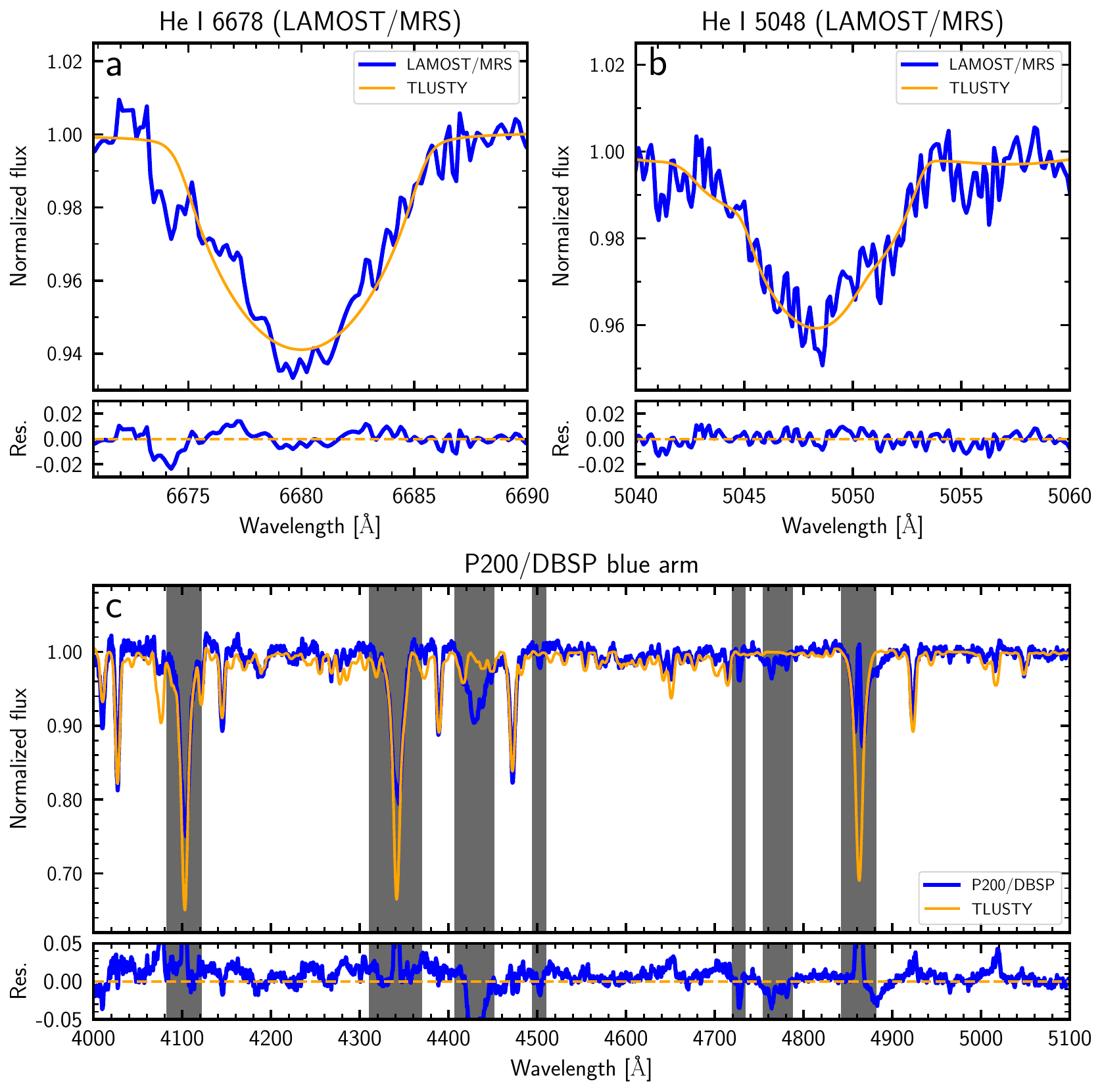}
\caption{{\bf Observed spectra of ALS\,8814 compared with the best broadened \texttt{TLUSTY} model spectrum.} {\bf a,} Upper: the blue spectrum corresponds to the observed spectrum (LAMOST/MRS) of He I 6678 absorption line, while the orange spectrum corresponds to the \texttt{TLUSTY} model spectrum, broadened by rotation with the $V\sin i$ corresponding to the minimum $\chi^2$ in panel {\bf a} in Supplementary Fig.\ref{fig:chisq}, along with the R $\sim$ 7500 LAMOST/MRS instrumental broadening. Lower: the fitting residuals for He I 6678 absorption line. {\bf b-c,} The same as {\bf a}, but for He I 5048 line and the blue arm of P200/DBSP, respectively. The gray regions in {\bf c} cover Balmer lines and DIBs and are excluded from chi-square calculations.}
\label{fig:match}
\end{center}
\end{figure*}
\clearpage
\begin{figure*}
\begin{center}
\includegraphics[width=0.8\textwidth]{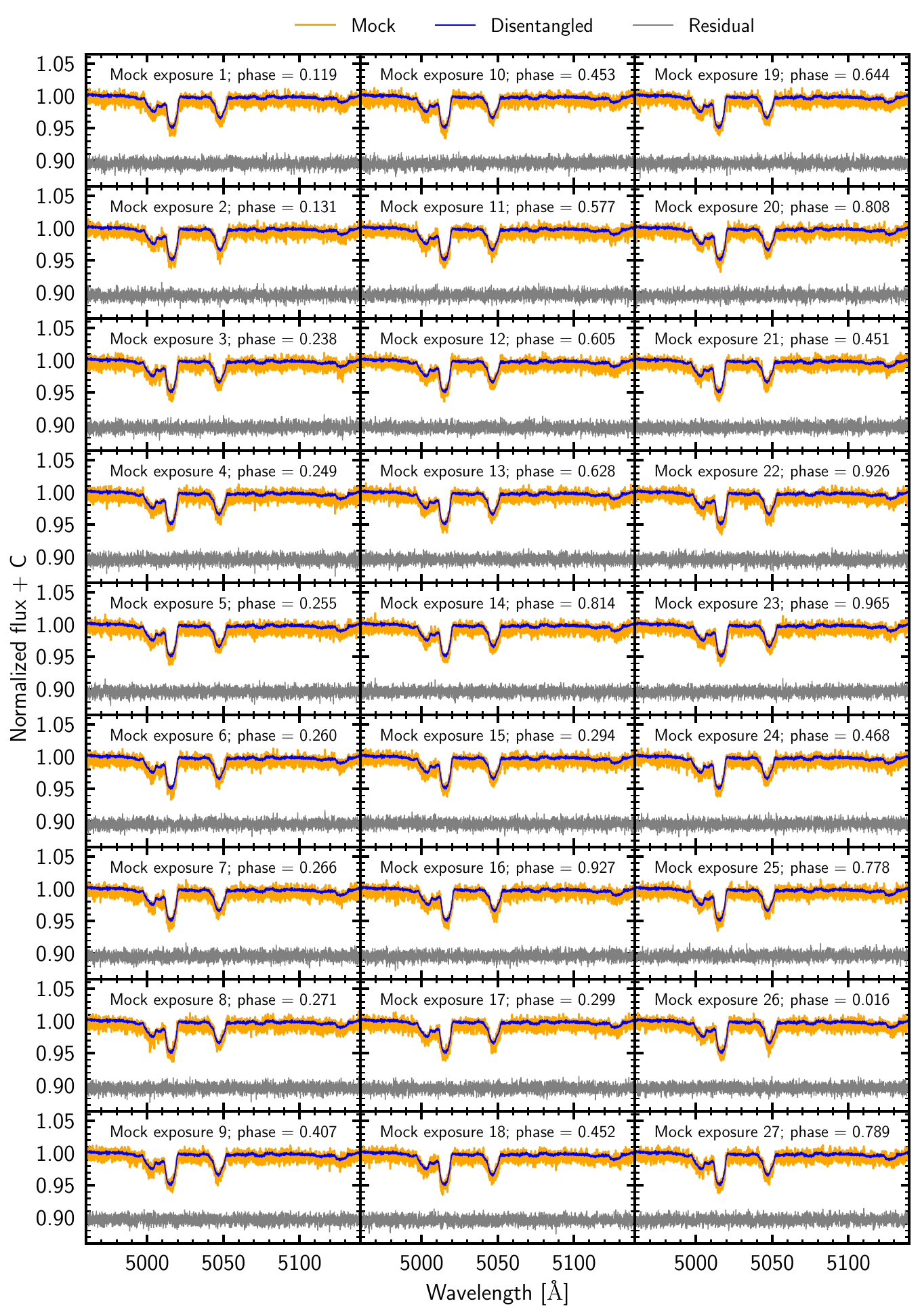}
\caption{{\bf Spectral disentanglement experiment simulating SB1 system for ALS\,8814.} Mock (orange), disentangled (blue) and residual spectra (gray) are shown in each panel.}
\label{fig:disentangling_mock_SB1}
\end{center}
\end{figure*}
\clearpage
\begin{figure*}
\begin{center}
\includegraphics[width=0.8\textwidth]{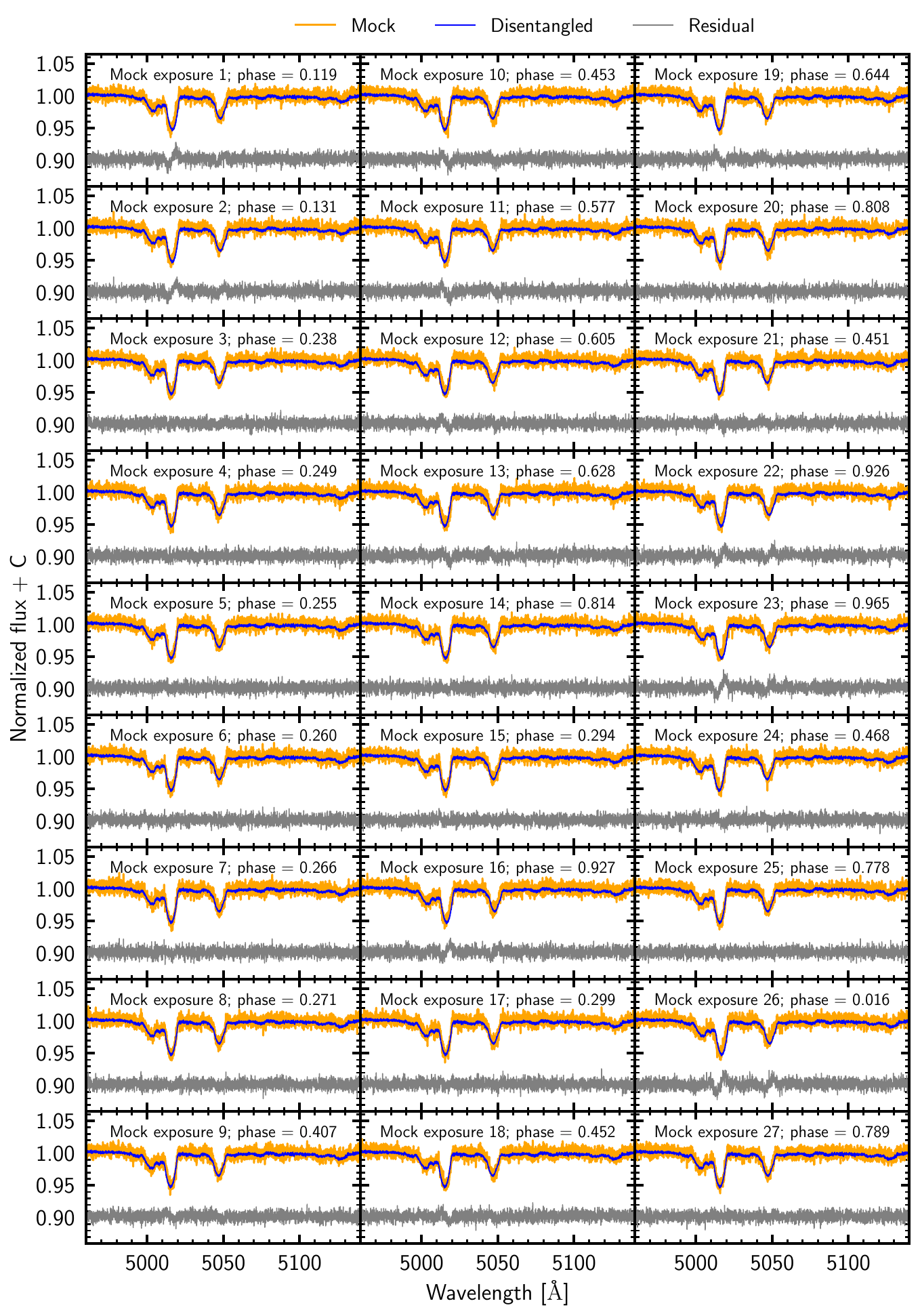}
\caption{{\bf The plot same as Supplementary Fig.~\ref{fig:disentangling_mock_SB1} but for simulating SB2 system.}}
\label{fig:disentangling_mock_SB2}
\end{center}
\end{figure*}
\clearpage
\begin{figure*}
\begin{center}
\includegraphics[width=0.8\textwidth]{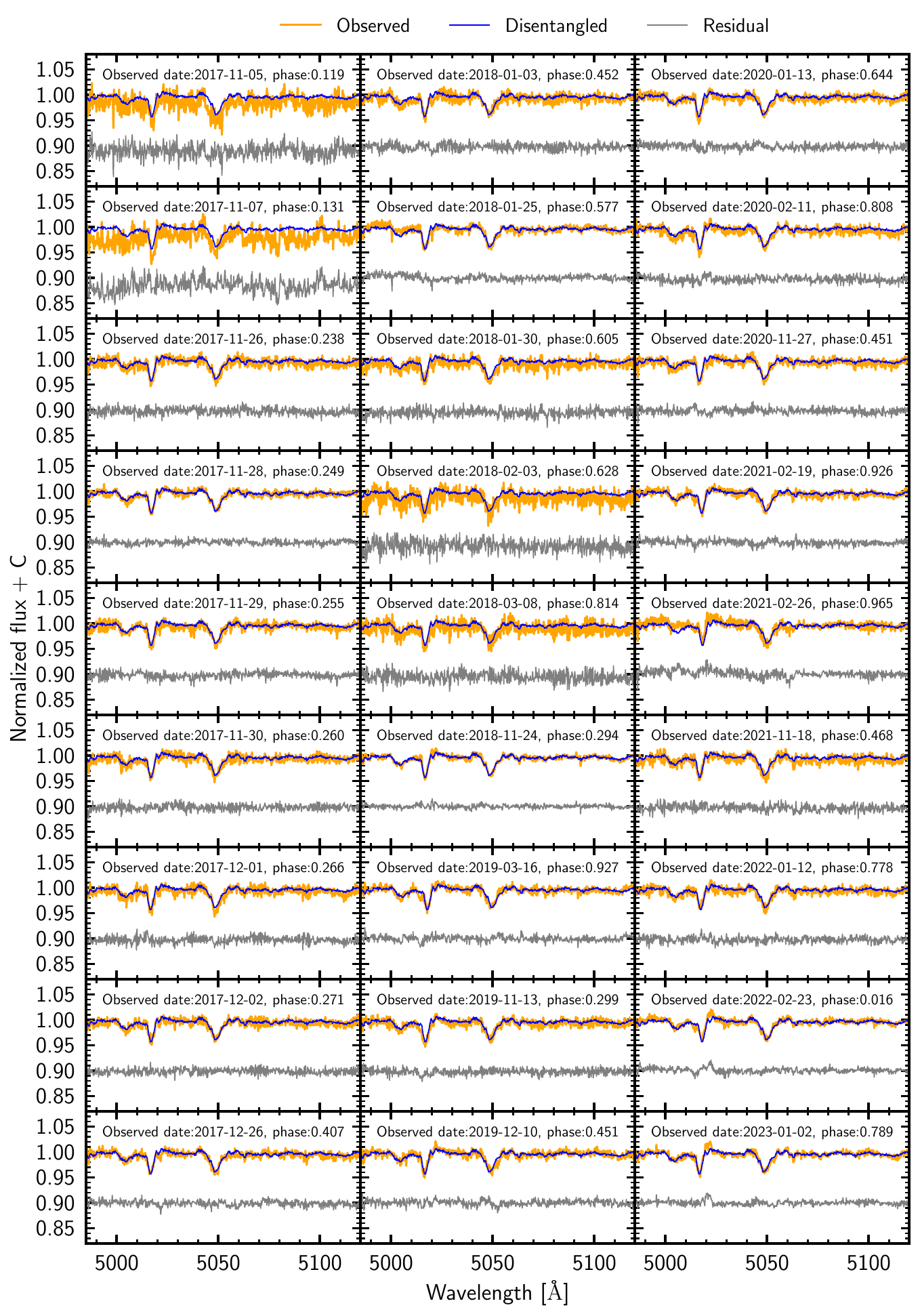}
\caption{{\bf Spectral disentanglement for ALS\,8814.} The orange, blue, and gray colors represent the mock, disentangled, and residual spectra, respectively.}
\label{fig:disentangling_observed}
\end{center}
\end{figure*}
\clearpage

\begin{figure*}
\begin{center}
\includegraphics[width=1\textwidth]{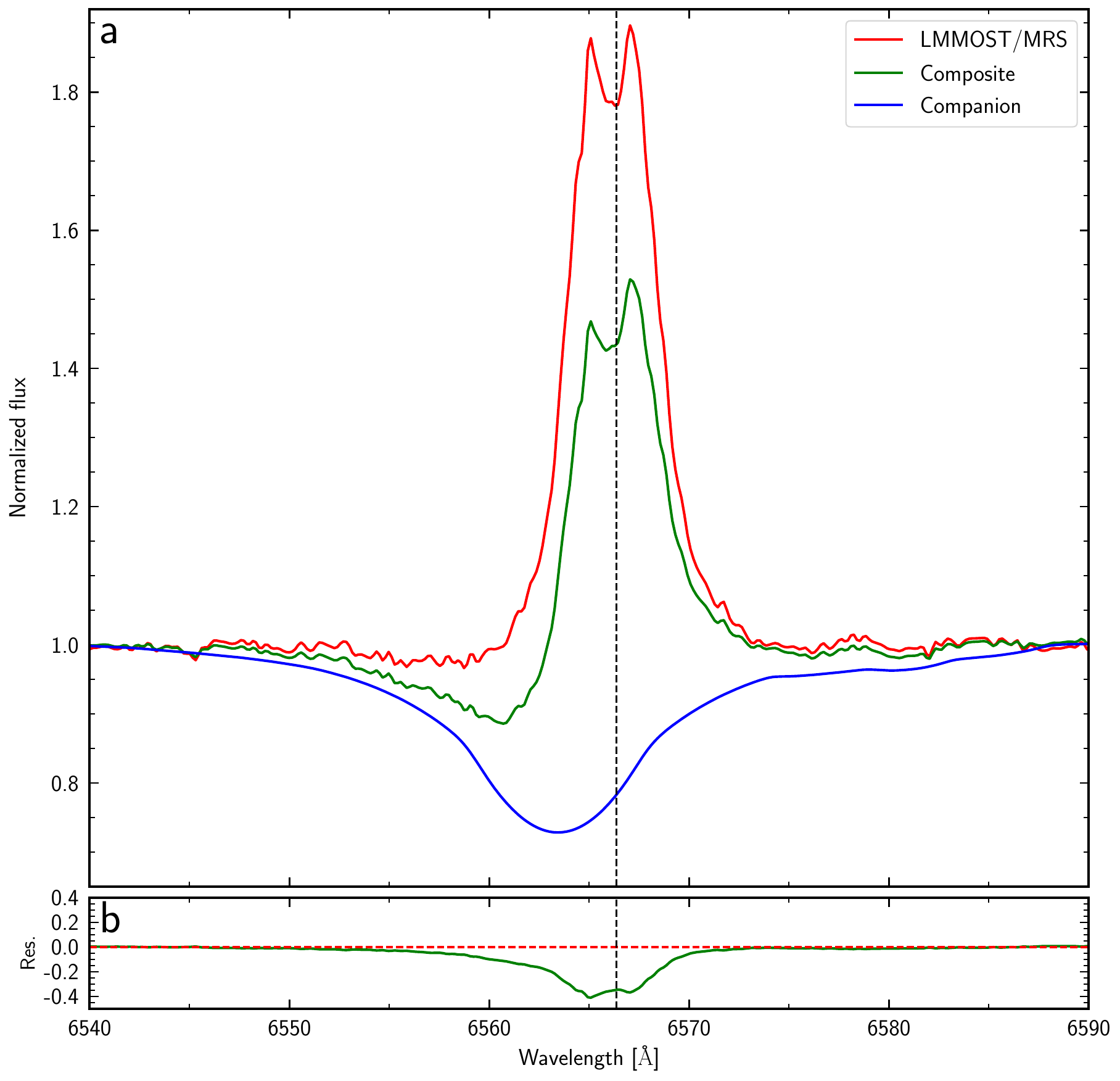}
\caption{{\bf H$\alpha$ emission line comparison between LAMOST/MRS spectrum and composite spectra with an introduced main-sequence companion.} {\bf a,} The red line, green line and blue line respectively represent the observed LAMOST/MRS spectrum, the composite spectrum and the main-sequence companion spectrum. {\bf b,} The discrepancy between the LAMOST/MRS spectrum and the composite spectrum.}
\label{fig:mock_inner_ms}
\end{center}
\end{figure*}
\clearpage
\begin{figure*}
\begin{center}
\includegraphics[width=1\textwidth]{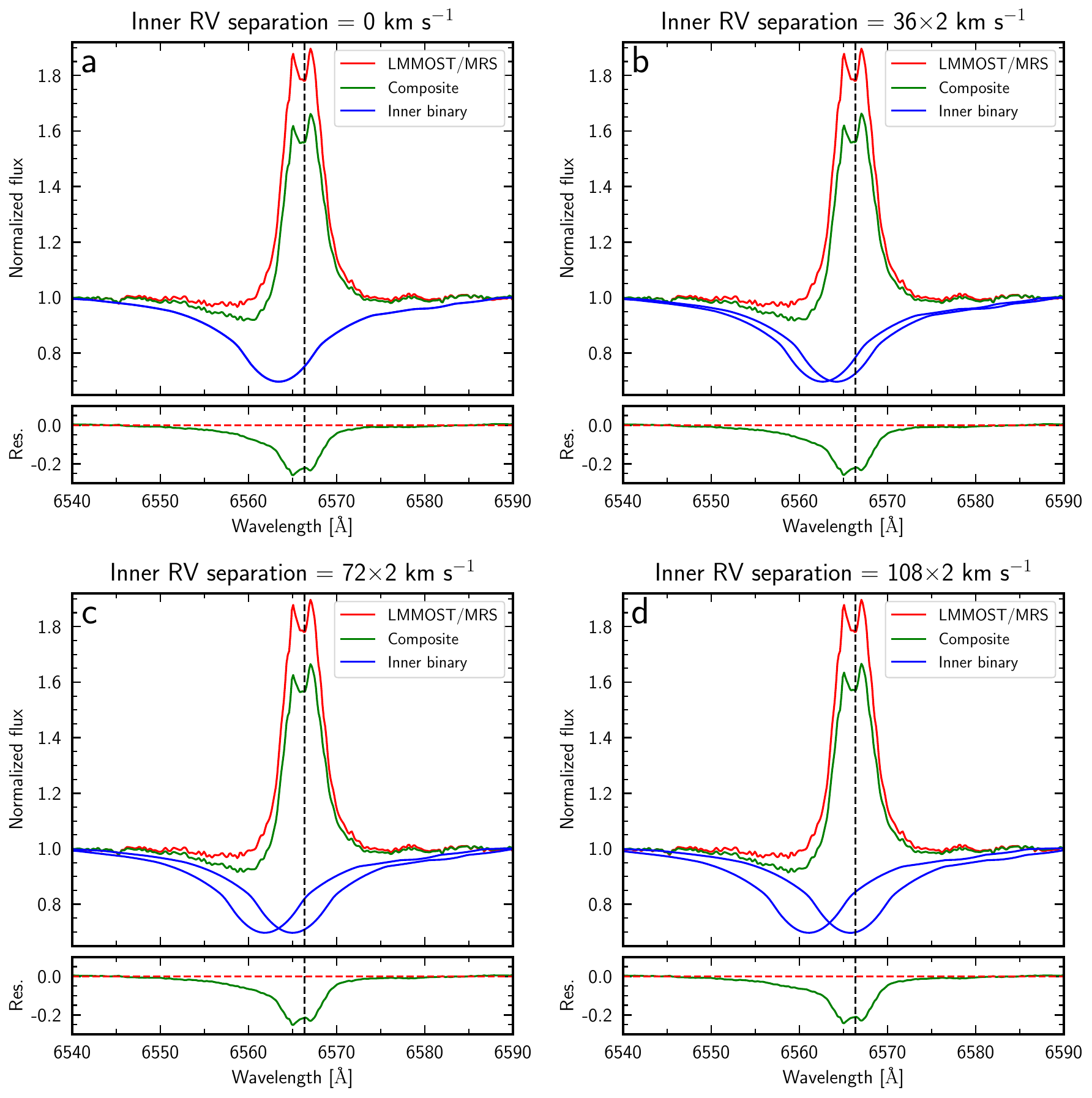}
\caption{{\bf H$\alpha$ emission line comparison between LAMOST/MRS spectrum and composite spectra with an introduced inner main-sequence binary.} {\bf a,} The case for an inner RV separation = 0 km s$^{-1}$. Upper: the red line, green line and blue line respectively represent the observed LAMOST/MRS spectrum, the composite spectrum and the inner binary spectra. Lower: the discrepancy between the LAMOST/MRS spectrum and the composite spectrum. {\bf b-d,} The same as {\bf a}, but for inner RV separations = 36$\times$2 km s$^{-1}$, 72$\times$2 km s$^{-1}$ and 108$\times$2 km s$^{-1}$, respectively.}
\label{fig:mock_inner_binary}
\end{center}
\end{figure*}
\clearpage
\begin{figure*}
\begin{center}
\includegraphics[width=1\textwidth]{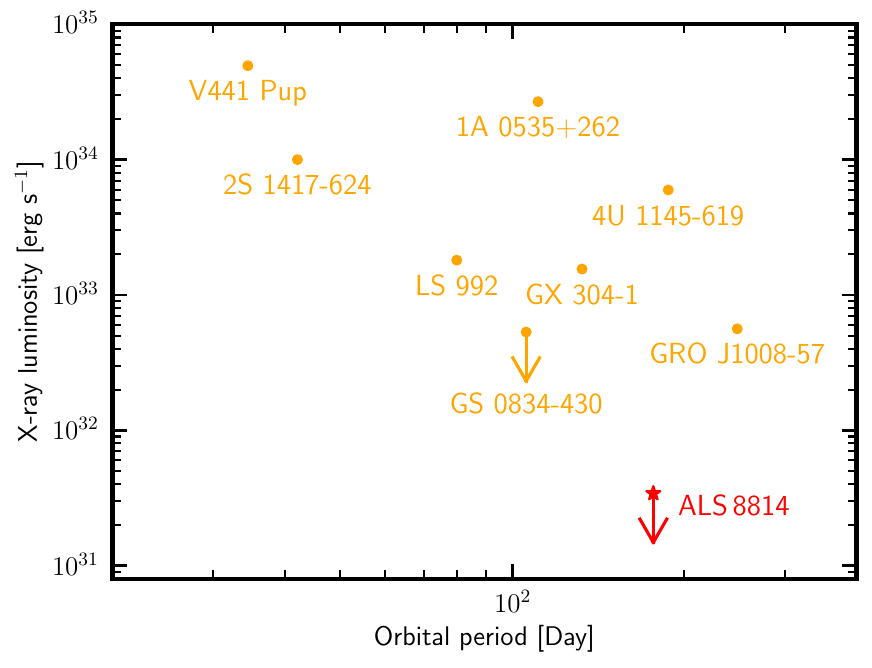}
\caption{{\bf The X-ray luminosity versus orbital period of Be-NS X-ray binaries observed by eROSITA.} The orange points represent Be-NS X-ray binaries and ALS\,8814 is also shown as a red pentagram for comparsion. The downward-pointing arrows represent binaries with only an upper limit provision for X-ray luminosity.}

\label{fig:X-ray}
\end{center}
\end{figure*}
\clearpage
\begin{figure*}
\begin{center}
\includegraphics[width=0.84\textwidth]{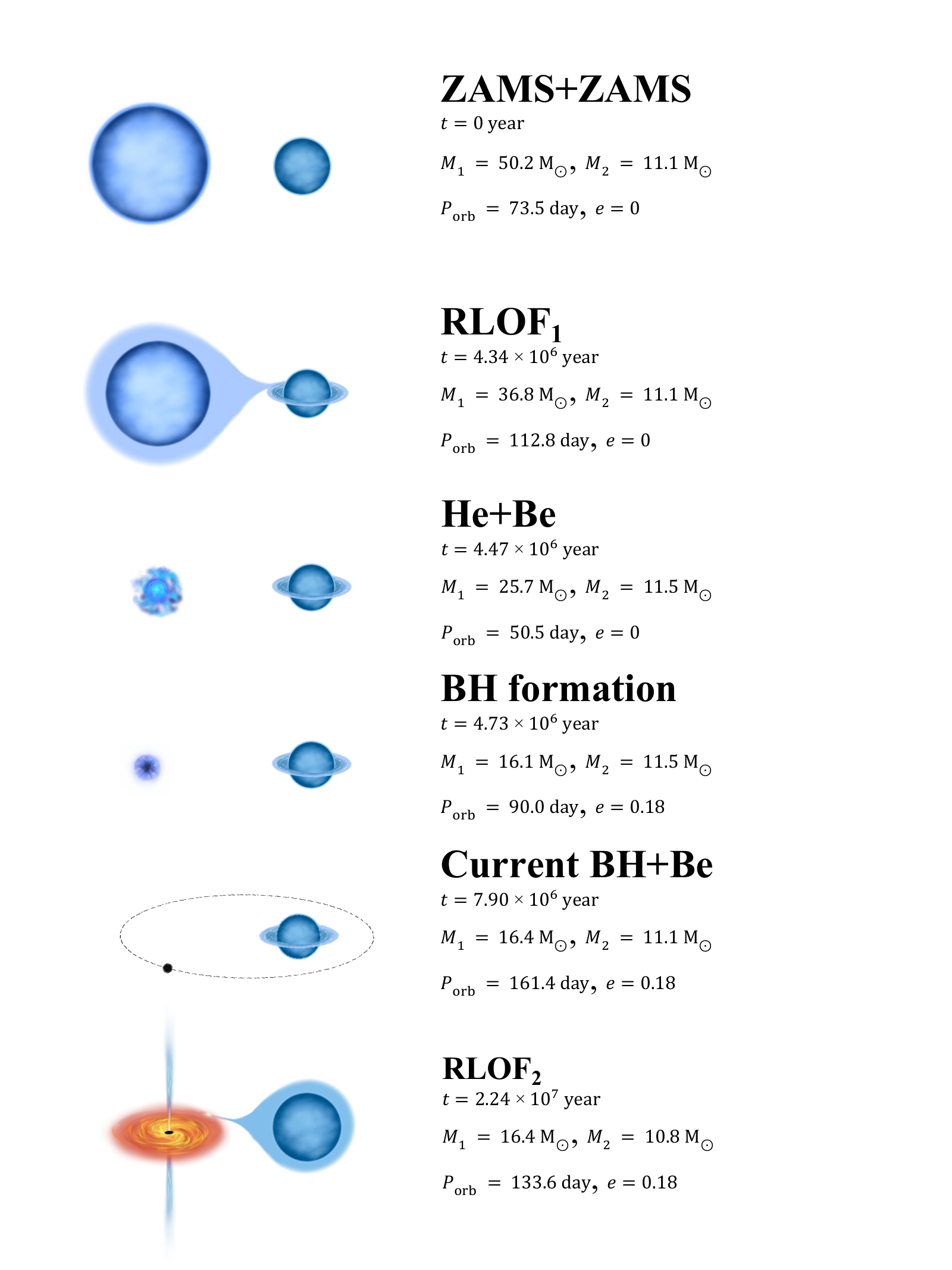}
\caption{{\bf Schematic plot depicts the formation of a binary system similar to ALS\,8814.} The primordial binary contains two zero-age main sequence (ZAMS) stars. The evolution involves mass transfer through Roche-lobe overflow (RLOF), the formation of a helium (He) star and then a BH.}
\label{fig:formation}
\end{center}
\end{figure*}
\clearpage
\begin{figure*}
\begin{center}
\includegraphics[width=0.63\textwidth]{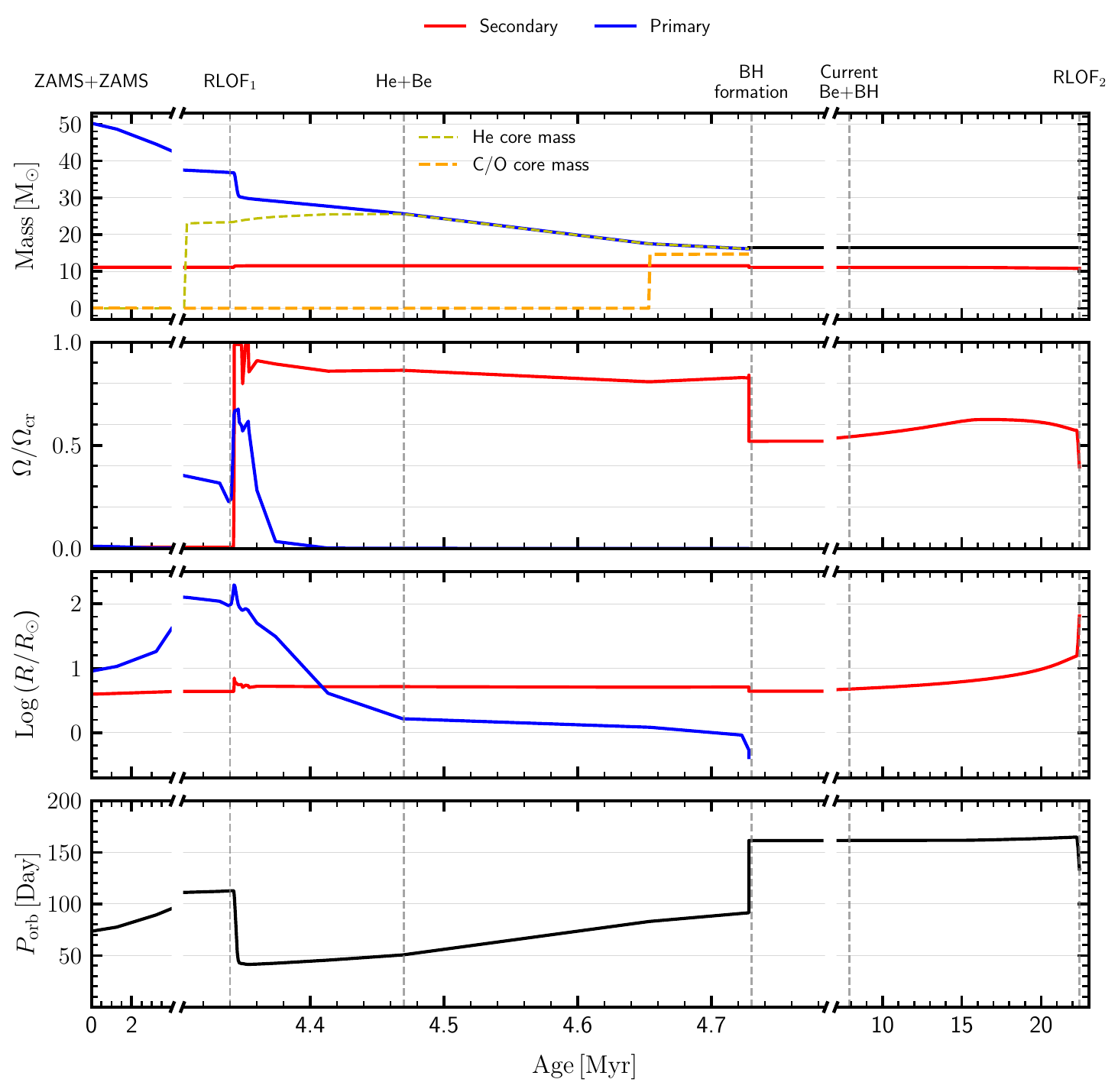}
\caption{{\bf Time evolution of one of the \texttt{POSYDON} binary population synthesis models reproducing the observed properties of ALS\,8814.} In the upper three panels, the blue line indicates the primary star (initially the more massive star, the BH progenitor) and the red line indicates the secondary star (the Be star). The topmost panel displays the masses of the primary star and the secondary star as a function of age. The helium core and C/O core masses are represented by beige and orange dashed lines, respectively. The second top panel provides information on the ratio between the surface and critical rotation (break-up) rate of the two stars. The second bottom panel shows the photosphere radius of the two stars. The most bottom panel displays the orbital period. 
The key events during binary evolution are indicated by vertical dashed lines: at 0 Myr, the binary is in the zero-age main-sequence stage (labeled as `ZAMS + ZAMS'); at 4.34 Myr, the progenitor of the BH reaches the end of the main sequence and triggers the first mass transfer with a stable Roche-lobe overflow phase (labeled as `RLOF$_{1}$'); near 4.47 Myr, the progenitor of the BH becomes a He star (labeled as `He + Be'); near 4.73 Myr, the progenitor of the BH experiences a SN event and becomes a BH (labeled as BH formation); the current configuration of the system is denoted as `Current Be + BH' at 7.9 Myr; near 22.42 Myr, the Be star becomes an evolved star and initiates the second mass transfer phase (labeled as `${\rm RLOF}_2$').}
\label{fig:evolution_model}
\end{center}
\end{figure*}
\clearpage
\begin{figure*}
\begin{center}
\includegraphics[width=1\textwidth]{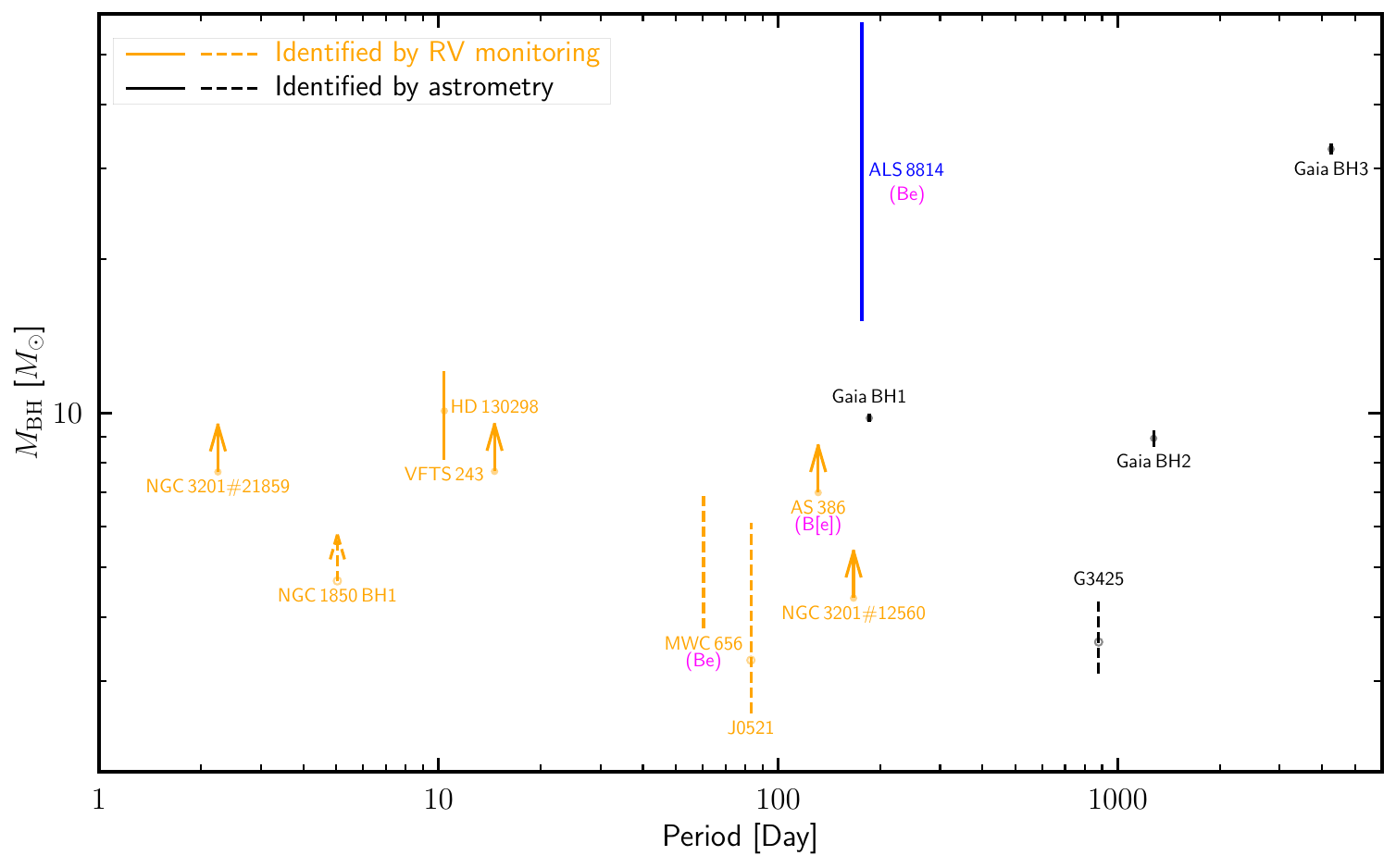}
\caption{{\bf Comparison of ALS\,8814 with other known dormant BHs.} Orange symbols denote other known dormant BHs identified by RV monitoring, while black symbols denote those identified by astrometry. The source names are labeled next to corresponding symbols. Open circles and dashed lines indicate that their corresponding sources are controversial or uncertain on whether they host BHs or not. The upward-pointing arrows denote sources with
only a lower limit provision for BH mass.}
\label{fig:m_p}
\end{center}
\end{figure*}

\begin{addendum}
\item[Acknowledgements]
W.M.G. acknowledges support from the National Key R\&D Program of China under grant Nos. 2023YFA1607901 and 2021YFA1600401, and the National Natural Science Foundation of China (NSFC) under grants 11925301, 12033006 and 12433007. Y.H. acknowledges the NSFC with Nos. of 11903027 and 11833006 and the National Key RD Program of China No. 2019YFA0405503. JFL acknowledges support the NSFC through grant Nos. of
11988101 and 11933004, and support from the New Cornerstone Science Foundation through the New Cornerstone Investigator Program and the XPLORER PRIZE. W.M.G., J.F. Wang, and J.F. Wu acknowledge support from NSFC under grant No. 12221003. Bryan Lailey acknowledges support from the University of Western Ontario’s Physics and Astronomy department. T.\ A.\ A.\ Sigut is grateful for support from the Natural Sciences and Engineering Research Council of Canada (NSERC) through the Discovery Grant program. L.X.D. acknowledges support from the National Key Research and Development Program of China (2021YFA0718500), the Natural Science Foundation of China under grant Nos. 12041301, 12121003. K.A.R. and M.S. are supported by the Gordon and Betty Moore Foundation (PI Kalogera, grant award GBMF8477). K.A.R.also thanks the Riedel Family Fellowship and the LSSTC Data Science Fellowship Program, which is funded by LSSTC, NSF Cybertraining Grant No.\ 1829740, the Brinson Foundation, and the Moore Foundation. S.G. acknowledges the funding support by the CIERA Postdoctoral Fellowship. S.J.G. acknowledges support from the Natural Science Foundation of China under grant No. 123B2045. Guoshoujing Telescope (the Large Sky Area Multi-Object Fiber Spectroscopic Telescope LAMOST) is a National Major Scientific Project built by the Chinese Academy of Sciences. Funding for the project has been provided by the National Development and Reform Commission. LAMOST is operated and managed by the National Astronomical Observatories, Chinese Academy of Sciences. This research uses data obtained through the Telescope Access Program (TAP), which has been funded by the TAP member institutes. We acknowledge the support of the staff of the Lijiang 2.4m telescope. Funding for the telescope has been provided by Chinese Academy of Sciences and the People's Government of Yunnan Province. We acknowledge the support of the staff of the Xinglong 2.16-m telescope. We thank Las Cumbres Observatory and its staff for their continued support of ASAS-SN. ASAS-SN is funded in part by the Gordon and Betty Moore Foundation through grants GBMF5490 and GBMF10501 to the Ohio State University, and also funded in part by the Alfred P. Sloan Foundation grant G-2021-14192. We are grateful to the \emph{Swift} team for approving and rapid scheduling of the monitoring campaign. This work made use of data supplied by the UK \emph{Swift} Science Data Centre at the University of Leicester and data obtained with NuSTAR mission, a project led by Caltech, funded by NASA and managed by JPL. The \texttt{POSYDON} project is supported primarily by two sources: a Swiss National Science Foundation Professorship grant (PI Fragos, project number PP00P2\_176868) and the Gordon and Betty Moore Foundation (PI Kalogera, grant award GBMF8477). \texttt{IRAF} is distributed by the National Optical Astronomy Observatory, which is operated by the Association of the Universities for Research in Astronomy, Inc. (AURA) under cooperative agreement with the National Science Foundation. Funding for the TESS mission is provided by NASA’s Science Mission directorate. This research has made use of the Spanish Virtual Observatory (\url{https://svo.cab.inta-csic.es}) project funded by MCIN/AEI/10.13039/501100011033/ through grant PID2020-112949GB-I00. VOSA has been partially updated by using funding from the European Union's Horizon 2020 Research and Innovation Programme, under Grant Agreement No. 776403 (EXOPLANETS-A).

\item[Author contributions]
W.M.G., Y.H., and J.F.L. led the project. Q.Y.A contributed to the early discovery of ALS\,8814. W.M.G. and Y.H. proposed the follow-up optical observations and coordinated the spectroscopic and photometric data reduction and analysis, with significant inputs from Z.X.Z.. W.M.G., Z.X.Z and T.Y. contributed to their expertise of the P200 proposals and observations, and S.Y.Q and B.W.Z contributed their P200 observation time, and Y.H. contributed 2.4-m Telescope observation time, and Z.X.Z. contributed to the P200 and 2.4-m Telescope data reduction and analysis. Q.Y.A., T.Y. and W.M.G. proposed the follow-up X-ray observations, and S.S.W. contributed to the expertise of Swift’s data reduction and analysis. Y.H derived out the stellar parameters from flux-calibrated spectrum and analyzed the kinematics and ASAS-SN light curve of ALS\,8814. Q.Y.A. measured radial velocities and solved the orbital solution, estimated the orbital inclination and BH mass with contributions from B.D.L., and conducted the SED fitting; Q.Y.A., T.Y. and Z.X.Z. contributed to rule out non-degenerate companions. Y.S. and S.J.G. contributed the BSE evolution analysis. K.A.R., M.S. and S.G. contributed to the \texttt{POSYDON} evolutionary modeling analysis, as well as the writing of the \texttt{POSYDON} section. Q.Y.A., Y.H., W.M.G., Z.X.Z., T.Y., S.S.W., S.J.G., Y.S., M.S., K.A.R., S.G. and B.D.L. wrote the manuscript. All authors reviewed and contributed to the manuscript.

\item[Competing interests] 
The authors declare no competing interests.

\item[Author information] 
Readers are welcome to comment on the online version of the paper. Correspondence and requests for materials should be addressed to W.M.G. (email: guwm@xmu.edu.cn) and/or J.F.L. (email: jfliu@nao.cas.cn).

\item[Data availability]
The data supporting the plots in this paper and other findings of this study are available from the corresponding authors upon reasonable request.

\item[Code Availability]
We use standard data analysis tools in the Python 3.10. We refer the reader to the \texttt{astropy} (\url{https://www.astropy.org/}), \texttt{BSE} (\url{http://astronomy.swin.edu.au/~jhurley/}), \texttt{emcee} (\url{https://emcee.readthedocs.io/en/stable/index.html}), \texttt{extinction} (\url{https://zenodo.org/records/804967}), \texttt{MESA} (\url{https://docs.mesastar.org/}), \texttt{Parsec} (\url{http://stev.oapd.inaf.it/cgi-bin/cmd}), \texttt{PHOEBE} (\url{http://phoebe-project.org/}), \texttt{POSYDON} (\url{https://posydon.org/}), \texttt{spectool} (\url{https://github.com/zzxihep/spectool}), \texttt{TheJoker} (\url{https://github.com/adrn/thejoker}), \texttt{VOSA} (\url{http://svo2.cab.inta-csic.es/theory/vosa/}) and web pages for access and availability policies.
\end{addendum}
\end{document}